\documentclass[10pt]{article}
\usepackage{a4wide}
\usepackage{amsmath,amssymb,mathrsfs,accents}
\usepackage{verbatim}

\title{\textbf{Thermal equilibrium states of a linear scalar quantum field in stationary
spacetimes}}
\author{Ko Sanders\thanks{E-mail: kosanders@uchicago.edu}\\
Enrico Fermi Institute, University of Chicago\\
5640 South Ellis Avenue, Chicago, IL 60637, USA}
\date{11 April 2013}

\numberwithin{equation}{section}

\newtheorem{definition}{Definition}[section]
\newtheorem{theorem}{Theorem}[section]
\newtheorem{proposition}{Proposition}[section]
\newtheorem{corollary}{Corollary}[section]
\newtheorem{lemma}{Lemma}[section]
\newenvironment{proof*}{\smallskip\par\noindent\emph{Proof: }
 \ignorespaces}{\hfill$\Box$\smallskip\par\ignorespaces}
\newtheorem{remark}{Remark}[section]
\newtheorem{example}{Example}[section]

\newcommand{\map}[3]{\ensuremath{#1\!:\!#2\!\rightarrow\!#3}}
\newcommand{\alg}[1]{\ensuremath{\mathcal{#1}}}

\begin{document}

\markboth{Ko Sanders}{Thermal equilibrium states in stationary spacetimes}

\maketitle

\begin{abstract}
The linear scalar quantum field, propagating in a globally hyperbolic
spacetime, is a relatively simple physical model that allows us to
study many aspects in explicit detail. In this review we focus on the
thermal equilibrium (KMS) states of such a field in a stationary
spacetime. Our presentation draws on several existing sources and
aims to give a unified exposition, while weakening certain technical
assumptions. In particular we drop all assumptions on the behaviour
of the time-like Killing field, which is important for physical
applications to the exterior region of a stationary black hole.

Our review includes results on the existence and uniqueness of ground
and KMS states, as well as an evaluation of the evidence supporting
the KMS-condition as a characterization of thermal equilibrium. We
draw attention to the poorly understood behaviour of the temperature
of the quantum field with respect to locality.

If the spacetime is standard static, the analysis can be done more
explicitly. For compact Cauchy surfaces we consider Gibbs states
and their properties. For general Cauchy surfaces we give a
detailed justification of the Wick rotation, including the explicit
determination of the Killing time dependence of the quasi-free KMS
states.
% \keywords{KMS condition; stationary spacetimes.}
\end{abstract}

% \ccode{PACS numbers: 04.62.+v, 11.10.Wx, 05.30.Jp}
% 04.62.+v Quantum fields in curved spacetime
% 11.10.Wx Finite-temperature field theory
% 05.30.Jp Boson systems (in quantum statistical mechanics)

\tableofcontents

\section{Introduction}\label{Sec_Intro}

For a quantum mechanical system with a Hilbert space $\mathcal{H}$, a
thermal equilibrium state can be described by the density matrix for the
Gibbs grand canonical ensemble,
\begin{equation}\label{Eqn_GibbsIntro}
\rho^{(\beta,\mu)}:=Z^{-1}e^{-\beta(H-\mu N)},
\end{equation}
where $H$ is the Hamiltonian operator of the system, $N$ the particle number
operator, $\beta$ the inverse temperature and $\mu$ the chemical
potential.\footnote{We work in natural (Planck) units throughout:
$c=G=\hbar=k_B=1$.} $Z$ is a normalization factor, which ensures that the
trace $\mathrm{Tr}\ \rho^{(\beta,\mu)}=1$. For this to be well defined we need
to know that $e^{-\beta(H-\mu N)}$ is a trace-class operator, a condition
which can often be established in explicit models, especially when the
system is confined to a bounded region of space.

For physical purposes it is of some interest to study thermal equilibrium
in much more general situations than for quantum mechanical systems, such as
for a quantum field propagating in a given gravitational background field. In
these cases one immediately encounters three well known problems: in a
general curved spacetime there is no clear notion of particle, no clear
choice of a Hamiltonian operator and, even if there were, the exponentiated
operator in Eq.\ (\ref{Eqn_GibbsIntro}) might not be of trace-class. Additional
problems arise if one wants to use the technique of Wick rotation, which has
important computational advantages in the quantum mechanical case, but which
requires a preferred choice of a well behaved time coordinate.

In this review paper we treat the problems above for the explicit example
of a linear scalar quantum field propagating in a globally hyperbolic spacetime.
We combine results and arguments from several sources into a unified
exposition and we take the opportunity to show that some of the technical
conditions made in the earlier literature may be dropped or weakened.

It is well known how to formulate a linear scalar quantum field theory in
all globally hyperbolic spacetimes \cite{Dimock1980,Baer+2009,Wald1994,Brunetti+2003}.
A notion of particle and Hamiltonian can be introduced whenever the
spacetime is also stationary \cite{Wald1994}. We will therefore focus on
stationary spacetimes, in which case the notion of global thermal equilibrium
is (in principle) well understood \cite{Fulling+1987,Kay1985_1}. Under
suitable positivity assumptions on the field equation we first give a full
characterization of all ground states on the Weyl algebra and we
describe in detail a uniquely preferred ground state \cite{Kay1978}.
More precisely, our assumptions are that the field should satisfy the
(modified) Klein-Gordon equation
\[
-\Box\phi+V\phi=0
\]
with a smooth, real-valued potential $V$ which is stationary and
strictly positive everywhere. Unlike Ref.\ \cite{Kay1978} we do not insist
that the ground state should have a mass gap, which allows us to drop
the restrictions that the norm and the lapse function of the time-like
Killing field be suitably bounded away from zero. This is of some
importance in certain physical applications, e.g.\ when the stationary
spacetime is the exterior region of a stationary black
hole \cite{Kay+1991,SandersIP}. In that case the norm of the Killing field
may become arbitrarily small.

Gibbs states as in Eq.\ (\ref{Eqn_GibbsIntro}) have a certain property,
first noticed by Kubo \cite{Kubo1957} and Martin and
Schwinger \cite{Martin+1959} and now known as the KMS-condition. This
property was proposed as a defining characteristic for thermal equilibrium
states by Ref.\ \cite{Haag+1967}, even when the Gibbs state is no longer
defined, on the grounds that it survives the thermodynamic (infinite volume)
limit under general circumstances for systems in quantum statistical
mechanics in Minkowski spacetime. Further support for this proposal comes
from an investigation of the second law of thermodynamics for general
$C^*$-dynamical systems \cite{Pusz+1978} and from the study of explicit
models in quantum statistical mechanics \cite{Bratteli+}. In addition to
its physical context, the KMS-condition has also become important in the
abstract theory of operator algebras, where it is related to Tomita's
modular theory \cite{Kadison+}.

In the case of a standard static spacetime (see Sec.\ \ref{Sec_Geometry}
for the definition) with a compact Cauchy surface we will see that the
Gibbs state of Eq.\ (\ref{Eqn_GibbsIntro}) makes sense. In the case of
a general stationary spacetime we will give a full characterization of
all KMS states on the Weyl algebra and we describe uniquely preferred
KMS states at any temperature \cite{Kay1985_1}.
Unfortunately, the arguments of Ref.\ \cite{Haag+1967} concerning the
thermodynamic limit fail to work for quantum field theories. This
indicates that the behaviour of the temperature of a quantum field,
with respect to locality, is presently rather poorly understood, even
in a spacetime with a favourable background geometry. With a view to
physical applications, e.g.\ in cosmology, an improved understanding
would be highly desirable. (At this point we would also like to point
out that Refs.\ \cite{Buchholz+2007,Buchholz+2002} have recently proposed
a notion of local thermal equilibrium in general curved spacetimes, but
the full merit of this new approach is as yet unclear and a review of
these recent developments is beyond the scope of this paper.)

When we study the Wick rotation we will restrict attention to
spacetimes which are standard static. Under these geometric
circumstances there is a preferred Killing time coordinate and
it is well understood how KMS states can be obtained
from a Wick rotation \cite{Fulling+1987,Birke+2002}. We show that any
technical assumptions are automatically verified for the systems
under consideration. After complexifying the Killing time coordinate
we obtain an associated Riemannian manifold and we compactify the
imaginary time coordinate to a circle of radius $R$. We then show
that there exists a uniquely distinguished Euclidean Green's function,
which can be analytically continued back to the Lorentzian
spacetime. We will find the explicit Killing time dependence of this
Green's function and on the Lorentzian side we recover the two-point
distribution of the preferred KMS state with inverse temperature
$\beta=2\pi R$.

The contents of this paper are organised as follows.
Section \ref{Sec_Alg} below considers some basic features of
thermal equilibrium states in an abstract, algebraic setting. The
main aim is to elucidate the structure of the spaces of all ground and
KMS states on the Weyl algebra under minimal assumptions. Section
\ref{Sec_Geometry} provides a review of recent geometric results
on stationary, globally hyperbolic spacetimes and the subclass of
standard static ones. In addition, it introduces the spacetime
complexification procedure needed to perform the Wick rotation. After
these algebraic and geometric preliminaries we describe in Section
\ref{Sec_Field} the linear scalar field under consideration, with an
emphasis on those results that depend on the presence of
the time-like Killing field. This section also contains a discussion
of the two-point distributions of thermal equilibrium states. Section
\ref{Sec_GroundProp} considers the space of ground states and the
GNS-representation of the uniquely preferred ground state. It also
includes a discussion of the renormalised stress-energy-momentum tensor.
Section \ref{Sec_Thermal} considers thermal equilibrium states at
non-zero temperature, from several perspectives. It contains existence
results of Gibbs states, under suitable assumptions, and
it discusses the motivations to use the KMS-condition to characterize
thermal equilibrium. Furthermore, it characterizes all KMS states,
including a uniquely preferred one, and in the static
case it provides a rigorous justification of the Wick rotation.
A number of useful results from functional analysis, needed for
Sections \ref{Sec_Alg}, \ref{Sec_Field} and \ref{Sec_Thermal}, are
collected in \ref{App_SP}, so as not to hamper the flow
of the presentation. These results concern strictly positive
operators and the relation between operators in Hilbert spaces and
distributions.

\section{Equilibrium states in algebraic dynamical systems}\label{Sec_Alg}

Much of the structure of dynamical systems can be conveniently
described in an abstract algebraic setting, which subsumes a
great variety of physical applications. In this section we provide a
brief overview of a number of notions and results relating to
equilibrium states for such systems and some more specialised
results pertaining to Weyl $C^*$-algebras. (For a detailed
treatment of Weyl $C^*$-algebras we refer to Ref.\ \cite{Binz+2004}
and references therein.)

Note that we generally do not assume any continuity of the time
evolution, so our results must remain more limited than those
for $C^*$-dynamical systems or $W^*$-dynamical systems \cite{Sakai1991,Bratteli+}.
This is in line with our physical applications later on, where
we will consider the Weyl $C^*$-algebra of certain pre-symplectic
spaces. As it turns out, for these systems the time evolution
will not be norm continuous in the given algebra, but there will
be continuity at the level of the symplectic space. To
accommodate for such situations, the results in this section
will only make ad hoc continuity assumptions in suitable
representations.

\subsection{Algebraic dynamical systems and equilibrium states}\label{SSec_AlgDynSys}

We begin with the following basic definition:
\begin{definition}\label{Def_AlgDSys}
An \emph{algebraic dynamical system} $(\mathfrak{A},\alpha_t)$
consists of a $^*$-algebra $\mathfrak{A}$ with unit $I$,
together with a one-parameter group of $^*$-isomorphisms
$\alpha_t$ on $\mathfrak{A}$.
\end{definition}
The algebra $\mathfrak{A}$ is interpreted as the algebra of
observables and $\alpha_t$ describes the time evolution. A state
$\omega$ on $\mathfrak{A}$ is a linear functional
$\map{\omega}{\mathfrak{A}}{\mathbb{C}}$ which
is normalised, $\omega(I)=1$, and positive, $\omega(A^*A)\ge 0$
for all $A\in\mathfrak{A}$. Every state gives rise to a unique
(up to unitary equivalence) GNS-triple \cite{Bratteli+}
$(\pi_{\omega},\mathcal{H}_{\omega},\Omega_{\omega})$, where
$\mathcal{H}_{\omega}$ is a Hilbert space and $\pi_{\omega}$ is a
representation of
$\mathfrak{A}$ on $\mathcal{H}_{\omega}$, in general by unbounded
operators, such that the vector $\Omega_{\omega}$ is cyclic for
$\pi_{\omega}(\mathfrak{A})$, i.e.\
$\overline{\pi_{\omega}(\mathfrak{A})\Omega_{\omega}}=\mathcal{H}_{\omega}$,
and $\omega(A)=\langle\Omega_{\omega},\pi_{\omega}(A)
\Omega_{\omega}\rangle$. We will denote the space of all states
on $\mathfrak{A}$ by $\mathscr{S}(\mathfrak{A})$. It is a
convex set in the (algebraic) dual space $\mathfrak{A}'$,
which is closed in the weak$^*$-topology. We will call a
state pure if for any decomposition
$\omega=\lambda\omega_1+(1-\lambda)\omega_2$ with
$\omega_1,\omega_2\in\mathscr{S}(\mathfrak{A})$ and
$0<\lambda<1$ we must have $\omega_1=\omega_2=\omega$.

For dynamical systems, the following class of states are
of special interest:
\begin{definition}\label{Def_Equilibrium}
An \emph{equilibrium state} $\omega$ for an algebraic
dynamical system $(\mathfrak{A},\alpha_t)$ is a state
$\omega$ on $\mathfrak{A}$ such that
$\alpha_t^*\omega:=\omega\circ\alpha_t=\omega$ for all
$t\in\mathbb{R}$. We denote the space of all equilibrium
states by $\mathscr{G}(\mathfrak{A})$ (suppressing the
dependence on $\alpha_t$).
\end{definition}
Note that $\mathscr{G}(\mathfrak{A})$ is a closed convex
subset of $\mathscr{S}(\mathfrak{A})$. In the
GNS-representation space of an equilibrium state $\omega$
the time evolution $\alpha_t$ is implemented by a unitary
group $U_t$ via
\[
\pi_{\omega}(\alpha_t(A))=U_t\pi_{\omega}(A)U_t^{-1},
\quad A\in\mathfrak{A}.
\]
The group $U_t$ is uniquely determined by the additional
condition that $U_t\Omega_{\omega}=\Omega_{\omega}$
(cf.\ Ref.\ \cite{Bratteli+} Cor.\ 2.3.17). If the group
$U_t$ is strongly continuous, it has a self-adjoint
generator by Stone's Theorem (Ref.\ \cite{Reed+} Thm.\ VIII.8),
so we may write $U_t=e^{ith}$, where the self-adjoint
operator $h$ is called the Hamiltonian.

\subsubsection{Ground states}

\begin{definition}\label{Def_Ground}
A \emph{ground state} $\omega$ on an algebraic dynamical
system $(\mathfrak{A},\alpha_t)$ is an equilibrium state
for which $U_t=e^{ith}$ is strongly continuous and the
Hamiltonian $h$ satisfies $h\ge0$. We denote the space of
all ground states by $\mathscr{G}^0(\mathfrak{A})$.

A ground state $\omega$ is called \emph{non-degenerate} when
the eigenspace of $h$ with eigenvalue $0$ is one-dimensional,
i.e.\ $h\psi=0$ implies $\psi=\lambda\Omega_{\omega}$ for
some $\lambda\in\mathbb{C}$.

A ground state $\omega$ is called \emph{extremal} if for any
decomposition $\omega=\lambda\omega_1+(1-\lambda)\omega_2$
with $\omega_1,\omega_2\in\mathscr{G}^0(\mathfrak{A})$
and $0<\lambda<1$ we must have $\omega_1=\omega_2=\omega$.
\end{definition}
Note that pure ground states are always extremal.
Furthermore, we have the following result, which is
essentially due to Borchers \cite{Borchers1966}:
\begin{theorem}\label{Thm_Borchers}
A non-degenerate ground state $\omega$ on an algebraic
dynamical system $(\mathfrak{A},\alpha_t)$ with
$\mathfrak{A}$ a $C^*$-algebra is pure.
\end{theorem}
\begin{proof*}
The strongly continuous unitary group $U_t$ on
$\mathcal{H}_{\omega}$ defines a group of automorphisms
on the von Neumann algebra
$\mathfrak{R}:=\pi_{\omega}(\mathfrak{A})''$. (A $'$
denotes the commutant of an algebra and $''$ the double
commutant \cite{Kadison+}.) The result of
Ref.\ \cite{Borchers1966} is that $U_t\in\mathfrak{R}$
for all $t\in\mathbb{R}$.
Now any unit vector $\psi$ of the
form $\psi=X\Omega_{\omega}$ with $X\in\mathfrak{R}'$
satisfies $h\psi=Xh\Omega_{\omega}=0$. Because
$\Omega_{\omega}$ is cyclic for $\mathfrak{R}$, it is
separating for $\mathfrak{R}'$, so
$\psi=\lambda\Omega_{\omega}$ if and only if $X=\lambda I$.
Hence if $\omega$ is nondegenerate, then
$\mathfrak{R}'=\mathbb{C}I$, which means that $\omega$ is
pure (Ref.\ \cite{Kadison+} Thm.\ 10.2.3).
\end{proof*}

% The converse is false: consider a Hilbert space $\mathcal{H}$
% with $\mathfrak{A}=\mathcal{B}(\mathcal{H})$ and $h=0$.
% Any vector in $\mathcal{H}$ defines a pure ground state,
% which is degenerate unless $\mathcal{H}$ is one-dimensional.

In the case that $\mathfrak{A}$ is commutative, ground
states have a special property which is worth singling
out. The proof involves analytic continuation arguments
which are typical for the study of ground and KMS states:
\begin{lemma}\label{Lem_ClGround}
Let $\omega$ be a state on an algebraic dynamical system
$(\mathfrak{A},\alpha_t)$ with $\mathfrak{A}$ a
commutative $^*$-algebra. Then the following statements
are equivalent:
\begin{enumerate}
\item[\textup{(i)}] $\omega$ is a ground state,
\item[\textup{(ii)}] $\omega(A\alpha_t(B))=\omega(AB)$ for all
$A,B\in\mathfrak{A}$ and $t\in\mathbb{R}$,
\item[\textup{(iii)}] $\omega$ is an equilibrium state with $U_t=I$ for
all $t\in\mathbb{R}$, in the GNS-representation of $\omega$.
\end{enumerate}
\end{lemma}
\begin{proof*}
Suppose that $\omega$ is a ground state. For arbitrarily
given $A,B\in\mathfrak{A}$ we consider the function
$f(t):=\omega(A\alpha_t(B))=\omega(\alpha_t(B)A)$. Because
$h\ge 0$ (by definition of ground states) we may use Lemma
\ref{Lem_Holo} to define a bounded, continuous function
$F_+(z)$ on the upper half plane
$\left\{z:=t+i\tau|\ \tau\ge 0\right\}$ by
\[
F_+(z):=\langle\pi_{\omega}(A^*)\Omega_{\omega},
e^{izh}\pi_{\omega}(B)\Omega_{\omega}\rangle,
\]
which is holomorphic on $\tau>0$ and satisfies
$F_+(t)=f(t)$ for $\tau=0$. Similarly we can define a
bounded continuous function $F_-(z)$ on the lower half
plane by
\[
F_-(z):=\langle\pi_{\omega}(B^*)\Omega_{\omega},
e^{-izh}\pi_{\omega}(A)\Omega_{\omega}\rangle,
\]
which is holomorphic for $\tau<0$ and which again satisfies
$F_-(t)=f(t)$ for $\tau=0$. It follows from the Edge of the Wedge
Theorem \cite{Berenstein+1991} that there is an entire
holomorphic function $F$ which extends both $F_+$ and
$F_-$. Since $F$ must be bounded as well it is constant
by Liouville's Theorem \cite{Berenstein+1991}.
Restricting to $\tau=0$ we find $f(t)=f(0)$,
i.e.\ $\omega(A\alpha_t(B))=\omega(AB)$.

Now suppose that the second item holds for $\omega$.
Then $\omega$ is an equilibrium state (taking $A=I$) and
using the group properties of $\alpha_t$ one easily shows
that $\omega(A\alpha_t(B)C)=\omega(ABC)$ for all
$t\in\mathbb{R}$ and $A,B,C\in\mathfrak{A}$. This implies
that $\pi_{\omega}(\alpha_t(B))=\pi_{\omega}(B)$ and hence
that $U_t=I$ for all $t\in\mathbb{R}$. Finally, $U_t=I$
implies $h=0$, so $\omega$ is a ground state.
\end{proof*}

Lemma \ref{Lem_ClGround} allows us to give a nice
description of all ground and equilibrium states on those
algebraic dynamical system $(\mathfrak{A},\alpha_t)$ for
which $\mathfrak{A}$ is a commutative $C^*$-algebra. For
this we make use of the classic structure theorem for
commutative $C^*$-algebras (cf.\ Ref.\ \cite{Kadison+}
Thm.\ 4.4.3), which tells us that there is a compact Hausdorff
space $X$, unique up to homeomorphism, and a $^*$-isomorphism
$\map{\alpha}{\mathfrak{A}}{C(X)}$, where $C(X)$ is the
$C^*$-algebra of continuous, complex-valued functions on
$X$ in the suppremum norm. The one-parameter group of
$^*$-isomorphisms
$\beta_t:=\alpha\circ\alpha_t\circ\alpha^{-1}$ on $C(X)$ is
then given by $\beta_t(F)=\Psi_t^*F$, where $\Psi_t$ is a
(uniquely determined) one-parameter group of homeomorphisms
of $X$. We define the set of fixed points
$X_0:=\left\{x\in X|\ \Psi_t(x)=x\mathrm{\ for\ all\ }
t\in\mathbb{R}\right\}$, which is closed in $X$ and hence
compact.
\begin{theorem}\label{Thm_ClGround}
Using the notations above, the following statements are true
for an algebraic dynamical system $(\mathfrak{A},\alpha_t)$
with $\mathfrak{A}$ a commutative $C^*$-algebra:
\begin{enumerate}
\item[\textup{(i)}] There is an affine bijection between probability measures
$\mu$ on $X$ and states on $\mathfrak{A}$ given by
$\mu\mapsto\omega_{\mu}$, where
$\omega_{\mu}(A):=\int_X d\mu\ \alpha(A)$.
\item[\textup{(ii)}] The state $\omega_{\mu}$ is pure if and only if $\mu$ is
supported at a single point.
\item[\textup{(iii)}] $\omega_{\mu}$ is an equilibrium state if and only if
$\Psi_t^*\mu=\mu$ for all $t\in\mathbb{R}$.
\item[\textup{(iv)}] $\omega_{\mu}$ is a pure equilibrium state if and only if
$\mu$ is supported at a single point in $X_0$.
\item[\textup{(v)}] $\omega_{\mu}$ is a ground state if and only if $\mu$ is
supported on $X_0$.
\item[\textup{(vi)}] $\omega$ is an extremal ground state if and only if it is
pure.
\end{enumerate}
\end{theorem}
\begin{proof*}
We only prove statement (v), as the others follow from standard
results on cummutative $C^*$-algebras and the definitions
above \cite{Kadison+}. By Lemma \ref{Lem_ClGround},
$\omega_{\mu}$ is a ground state if and only if
$\int_Xd\mu\ F(\Psi_t^*G-G)=0$ for all $F,G\in C(X)$. Because
$\Psi_t^*G-G=0$ on $X_0$ this is certainly the case when
$\mathrm{supp}(\mu)\subset X_0$ (cf.\ Ref.\ \cite{Kadison+} Remark
3.4.13). Conversely, for any $x\in X_0^c$ in the complement of
$X_0$ we can find a $t\in\mathbb{R}$ and an open set $U\subset X$
such that $x\in U$ and $\Psi_t(U)\cap U=\emptyset$. (In detail: we
may first choose a $t\in\mathbb{R}$ such that $y:=\Psi_t(x)\not=x$.
As $X$ is Hausdorff we may find an open set $V\subset X$ such that
$x\in V$ and $y\not\in\overline{V}$. Taking
$U:=V\setminus\Psi_{-t}(\overline{V})$ will do.) By Urysohn's
Lemma \cite{Kelley1955} there is a $G\in C(X)$ with $G(x)=1$ which
vanishes on $X\setminus U$. Note that $\overline{G}\Psi_t^*G=0$,
so if $\omega_{\mu}$ is a ground state we have
$\int_Xd\mu\ |G|^2=-\int_Xd\mu\ \overline{G}(\Psi_t^*G-G)=0$.
As $G(x)=1$ this entails that $x\not\in\mathrm{supp}(\mu)$, so
$\mathrm{supp}(\mu)\subset X_0$.
\end{proof*}
Note in particular that pure equilibrium states are automatically
ground states.

\subsubsection{KMS states}

In physical applications, thermal equilibrium states can be
characterised by the KMS-condition:
\begin{definition}\label{Def_KMS}
A state $\omega$ on an algebraic dynamical system
$(\mathfrak{A},\alpha_t)$is called a \emph{$\beta$-KMS state}
for $\beta>0$, when it satisfies the \emph{KMS-condition} at
\emph{inverse temperature} $\beta$, i.e.\ when for all
operators $A,B\in\mathfrak{A}$ there is a holomorphic function
$F_{AB}$ on the strip
$\mathrm{S}_{\beta}:=\mathbb{R}\times i(0,\beta)\subset\mathbb{C}$
with a bounded, continuous extension to
$\overline{\mathrm{S}_{\beta}}$ such that
\begin{equation}\label{Eqn_KMS}
F_{AB}(t)=\omega(A\alpha_t(B)),\quad
F_{AB}(t+i\beta)=\omega(\alpha_t(B)A).
\end{equation}

We will denote the space of all $\beta$-KMS states by
$\mathscr{G}^{(\beta)}(\mathfrak{A})$. A $\beta$-KMS state
$\omega$ is called \emph{extremal} if for any decomposition
$\omega=\lambda\omega_1+(1-\lambda)\omega_2$ with
$\omega_1,\omega_2\in\mathscr{G}^{(\beta)}(\mathfrak{A})$
and $0<\lambda<1$ we must have $\omega_1=\omega_2=\omega$.
\end{definition}
When $\mathfrak{A}$ is a topological $^*$-algebra and
$\omega$ is a continuous state, then it suffices to
require the existence of $F_{AB}$ for $A,B$ in a dense
sub-algebra of $\mathfrak{A}$, as we will see in
Proposition \ref{Prop_KMSEq} below. When
$(\mathfrak{A},\alpha_t)$ is a $C^*$-dynamical system
one may also drop the requirement that $F_{AB}$ is
bounded (Ref.\ \cite{Bratteli+} Prop.\ 5.3.7).

The motivations behind this condition will be discussed in
some detail in Section \ref{Sec_Thermal}, in the context of
our physical applications to the linear scalar quantum field. Note,
however, that a ground state satisfies a similar condition
with $\beta=\infty$, when we identify $\mathrm{S}_{\beta}$,
respectively $\overline{\mathrm{S}_{\beta}}$, with the open,
respectively closed, upper half plane. (This may be seen by
the same methods as used in the proof of Lemma \ref{Lem_ClGround}.)

The following general result again relies on analytic
continuation arguments:
\begin{proposition}\label{Prop_KMSEq}
Let $\omega$ be a $\beta$-KMS state on an algebraic dynamical
system $(\mathfrak{A},\alpha_t)$. Then the following hold true:
\begin{enumerate}
\item[\textup{(i)}] $\omega$ is an equilibrium state.
\item[\textup{(ii)}] For all $A,B\in\mathfrak{A}$ and
$z\in\overline{\mathrm{S}_{\beta}}$ we have
\[
|F_{AB}(z)|^2\le \mathrm{max}(\omega(AA^*)\omega(B^*B),
\omega(A^*A)\omega(BB^*)).
\]
\end{enumerate}
\end{proposition}
\begin{proof*}
For any $B$ the function $F_{IB}(z)$ satisfies
$F_{IB}(t)=F_{IB}(t+i\beta)$. Let $F(z)$ be the periodic
extension of $F_{IB}(z)$ in $\mathrm{Im}(z)$ with period
$\beta$. Then $F$ is continuous and bounded on $\mathbb{C}$
and it is holomorphic, even when
$\mathrm{Im}(z)\in\beta\mathbb{Z}$, by the Edge of the Wedge
Theorem \cite{Berenstein+1991}. $F$ must then be a constant
by Liouville's Theorem \cite{Berenstein+1991}, so
$F_{IB}(t)=F_{IB}(0)$, i.e.\ $\omega(\alpha_t(B))=\omega(B)$
and $\omega$ is in equilibrium.

For any operators $A,B\in\mathfrak{A}$ the corresponding
function $F_{AB}$ on $\overline{\mathrm{S}_{\beta}}$ satisfies
\[
|F_{AB}(z)|\le \sup_{t\in\mathbb{R}}\mathrm{max}
\left\{|F_{AB}(t)|,|F_{AB}(t+i\beta)|\right\}
\]
by the boundedness of $F_{AB}$ and Hadamard's Three Line
Theorem (Ref.\ \cite{Reed+}, Appendix to IX.4). The second
statement then follows from the first, and the Cauchy-Schwarz
inequality.
\end{proof*}

For commutative algebras a state $\omega$ is a $\beta$-KMS
state if and only if it is a ground state (cf.\ Lemma
\ref{Lem_ClGround}).
% Note that $F_{AB}$ satisfies $F_{AB}(t)=F_{AB}(t+i\beta)$
% by commutativity. As for $F_{IB}$ in the proof of
% Proposition \ref{Prop_KMSEq} we can extend $F_{AB}$
% periodically in the imaginary part of $z$ with period
% $\beta$ to a bounded entire holomorphic function on
% $\mathbb{C}$. Hence, $F_{AB}$ is constant, so item (ii)
% of Lemma \ref{Lem_ClGround} is satisfied and $\omega$ is
% a ground state.

\subsection{Weyl $C^*$-algebras}\label{SSec_Weyl}

For our physical applications to linear scalar quantum fields
we will make use of an algebraic formulation involving
Weyl $C^*$-algebras. In preparation for those applications
we will now briefly review some fundamental aspects of
these algebras \cite{Binz+2004}, especially in relation to
thermal equilibrium states.

We consider a pre-symplectic space $(L,\sigma)$, which
means that $L$ is a real linear space and $\sigma$ is an
anti-symmetric bilinear form. We call $(L,\sigma)$ a
symplectic space if $\sigma$ is non-degenerate, which means
that $\sigma(f,f')=0$ for all $f'\in L$ implies $f=0$.
For each pre-symplectic space $(L,\sigma)$ there is a
unique $C^*$-algebra generated by linearly independent
operators $W(f)$, $f\in L$, subject to the Weyl
relations \cite{Binz+2004}
\begin{equation}\label{Eqn_Weyl}
W(f)W(f')=e^{\frac{-i}{2}\sigma(f,f')}W(f+f'),\quad
W(f)^*=W(-f).
\end{equation}
This is the Weyl $C^*$-algebra,
which we will denote by $\alg{W}(L,\sigma)$. By
construction, the linear space generated by all $W(f)$,
but without taking the completion in the $C^*$-norm, is
also $^*$-algebra, which we will denote by
$\accentset{\circ}{\alg{W}}(L,\sigma)$ and which is a dense
subset of $\alg{W}(L,\sigma)$. Every state on $\alg{W}(L,\sigma)$
restricts to a state on $\accentset{\circ}{\alg{W}}(L,\sigma)$,
but we even have the following stronger result:
\begin{lemma}\label{Lem_ExtendStates}
The restriction map $\map{r}{\mathscr{S}(\alg{W}(L,\sigma))}
{\mathscr{S}(\accentset{\circ}{\alg{W}}(L,\sigma))}$ is an
affine homeomorphism for the respective weak$^*$-topologies.
\end{lemma}
This follows from Theorem 3-5 and Lemma 3-3a) of
Ref.\ \cite{Binz+2004} and the fact that the weak$^*$-topology
on a bounded set in the continuous dual space $\alg{W}(L,\sigma)'$
is already determined by the dense set
$\accentset{\circ}{\alg{W}}(L,\sigma)\subset\alg{W}(L,\sigma)$.

The Weyl $C^*$-algebra $\alg{W}(L,0)$ is commutative, so
there is a $^*$-isomorphism $\map{\alpha}{\alg{W}(L,0)}{C(X)}$,
where we may identify $X$ as the space of pure states
$\mathscr{S}(\alg{W}(L,0))$. Alternatively we may identify $X$
with the dual group $\hat{L}$ of $L$, viewed as an additive
group \cite{Binz+2004}. Elements of $\hat{L}$ are characters of
$L$, i.e.\ group homomorphisms from $L$ (as an additive group) to
the unit circle $S^1$ (as a multiplicative group). The bijection
between pure states $\rho\in X$ and characters $\chi\in\hat{L}$
is given by $\rho(W(f))=\chi(f)$ (cf.\ Ref.\ \cite{Kadison+}
Prop.\ 4.4.1).

\begin{remark}\label{Rem_Gauge2}
For any pure state $\rho\in\mathscr{S}(\alg{W}(L,0))$ we can
define a $^*$-isomorphism
$\map{\eta_{\rho}}{\alg{W}(L,\sigma)}{\alg{W}(L,\sigma)}$ by
continuous linear extension of
$\eta_{\rho}(W(f)):=\rho(W(f))W(f)$ \cite{Binz+2004}. The
$^*$-isomorphisms $\eta_{\rho}$ are sometimes known as
\emph{gauge transformations of the second kind}. We will denote
the gauge transformations on the commutative Weyl algebra
$\alg{W}(L,0)$ by $\zeta_{\rho}$.

The state space $\mathscr{S}(\alg{W}(L,0))$ contains a special
state,\footnote{Not to be confused with the tracial state
$\rho^t$, defined by $\rho^t(W(f))=0$ for all $f\not=0$, which
can be defined on any Weyl $C^*$-algebra, commutative or not.}
$\rho^0$, defined by $\rho^0(W(f))=1$ for all $f\in L$. This
state is pure, because its GNS-representation is one-dimensional.
It is easy to verify that $\rho=\zeta_{\rho}^*\rho^0$ for all
pure states $\rho\in\mathscr{S}(\alg{W}(L,0))$.
\end{remark}

The algebras $\alg{W}(L,\lambda\sigma)$, $0\le\lambda\le 1$, may
be viewed as a strict and continuous deformation \cite{Binz+2004_2}
of the commutative algebra $\alg{W}(L,0)$. It will be interesting
for us to compare the state space of the Weyl $C^*$-algebra
$\alg{W}(L,\sigma)$ with that of the commutative Weyl
$C^*$-algebra $\alg{W}(L,0)$:
\begin{lemma}\label{Lem_AffineIso}
For every $\omega'\in\mathscr{S}(\alg{W}(L,\sigma))$ there is a
unique weak$^*$-continuous, affine map $\map{\lambda_{\omega'}}
{\mathscr{S}(\alg{W}(L,0))}{\mathscr{S}(\alg{W}(L,\sigma))}$ which
is given by $\lambda_{\omega'}(\rho)=\eta_{\rho}^*\omega'$ on pure
states. For any pure state $\rho'$ on $\alg{W}(L,0)$ we have
$\lambda_{\omega'}\circ\zeta_{\rho'}^*=\eta_{\rho'}^*\circ\lambda_{\omega'}$
and $\lambda_{\omega'}$ is injective when $\omega'(W(f))\not=0$
for all $f\in L$.
\end{lemma}
\begin{proof*}
For pure states we have
\[
\lambda_{\omega'}(\rho)(W(f))=\omega'(W(f))\rho(W(f)).
\]
Because every state in $\mathscr{S}(\alg{W}(L,0))$ is a weak$^*$-limit
of finite affine combinations of pure states, $\lambda_{\omega'}$
extends uniquely to a weak$^*$-continuous, affine map from
$\mathscr{S}(\alg{W}(L,0))$ to $\mathscr{S}(\alg{W}(L,\sigma))$, which
is given by the same formula. The injectivity of $\lambda_{\omega'}$
under the stated assumptions is immediate from this formula and
Lemma \ref{Lem_ExtendStates}. The intertwining relation with the gauge
transformations of the second kind is a straightforward exercise.
\end{proof*}

\subsubsection{Quasi-free and $C^k$ states}

On any Weyl $C^*$-algebra there is a special class of states,
called quasi-free states, which are distinguished by their
algebraic form. They are obtained from the following
well known result:
\begin{theorem}\label{Thm_Twopoint}
Let $(L,\sigma)$ be a pre-symplectic space. A sesquilinear form
$\omega_2$ on the complexification $L\otimes\mathbb{C}$ defines
a state $\omega$ on $\alg{W}(L,\sigma)$ by continuous linear
extension of
\[
\omega(W(f))=e^{-\frac12\omega_2(f,f)},\quad f\in L,
\]
if and only if for all $f,f'\in L\otimes\mathbb{C}$:
\begin{enumerate}
\item[\textup{(i)}] $\omega_2(\overline{f},f)\ge 0$ (positive type),
\item[\textup{(ii)}] $2\omega_{2-}(f,f'):=\omega_2(f,f')-\omega_2(f',f)=i\sigma(f,f')$
(canonical commutator).
\end{enumerate}
\end{theorem}
We will call $\omega_2$ a two-point function, even though it is
generally not a function of two points $x,y\in M$.
The two-point function $\omega_2$ can be characterised
alternatively in terms of a one-particle structure \cite{Kay1978}:
\begin{definition}
A \emph{one-particle structure} on a pre-symplectic space
$(L,\sigma)$ is a pair $(p,\mathcal{K})$ consisting of a complex
linear map $\map{p}{L\otimes\mathbb{C}}{\mathcal{K}}$ into a
Hilbert space $\mathcal{K}$ such that
\begin{enumerate}
\item[\textup{(i)}] $p$ has dense range in $\mathcal{K}$,
\item[\textup{(ii)}] $\langle p(\overline{f}),p(f')\rangle-
\langle p(\overline{f}'),p(f)\rangle=i\sigma(f,f')$.
\end{enumerate}
\end{definition}
Given a one-particle structure, one can define an associated
two-point function by
$\omega_2(\overline{f},f'):=\langle p(f),p(f')\rangle$.
Conversely, a two-point function $\omega_2$ determines a unique
one-particle structure $(p,\mathcal{K})$ such that the above
equality holds, by similar arguments as used in the
GNS-construction. This we call the one-particle structure
associated with $\omega_2$.

A wider class of states which will be of interest is the
following:
\begin{definition}\label{Def_CkState}
A state $\omega$ on the Weyl $C^*$-algebra $\alg{W}(L,\sigma)$
is called $C^k$, $k> 0$, when the maps
\[
\omega_n(f_1,\ldots,f_n):=(-i)^n\partial_{s_1}\cdots\partial_{s_n}
\omega(W(s_1f_1)\cdots W(s_nf_n))|_{s_1=\ldots=s_n=0}
\]
are well defined on $C^{\infty}_0(M)^{\times n}$ for all
$1\le n\le k$. The $\omega_n$ are linear maps and they are called
the $n$-point functions. A state is called $C^{\infty}$, when
it is $C^k$ for all $k>0$.
\end{definition}
% This is equivalent to the existence of
% $\partial_s^n\omega(W(sf))|_{s=0}$ for all $1\le n\le k$ and $f\in L$.
When $\omega$ is a quasi-free state, it is $C^{\infty}$ and
all higher $n$-point functions can be expressed in terms of the
two-point function $\omega_2$ via Wick's Theorem. For such states
it only remains to analyze the two-point functions $\omega_2$.
% If $\omega$ is $C^2$, then for each $f\in L$ the map
% $F(t):=\omega(W(tf))$ is $C^2$ on $\mathbb{R}$. Indeed,
% $\|\pi_{\omega}(W(tf)-I)\Omega_{\omega}\|^2=2-F(t)-F(-t)$
% is continuous at $t=0$ (because it even has a derivative
% there). Hence, $\pi_{\omega}(W(tf)\Omega_{\omega}$ is continuous.
% Similarly, using the fact that $\omega$ is $C^2$,
% $\|t^{-1}\pi_{\omega}(W(tf)-I)\Omega_{\omega}\|^2
% =t^{-2}\omega((W(-tf)-I)(W(tf)-I))$ has a limit at $t=0$, so
% $t\mapsto \pi_{\omega}(W(tf))\Omega_{\omega}$ is $C^1$.
% In fact, by the Weyl relations, $t\mapsto\pi_{\omega}(W(tf))$
% is strongly $C^1$. It then follows that $F$ is $C^1$ and
% $F'(t)=i\omega(\Phi(f)W(tf))$ is $C^1$, so $F$ is $C^2$.
% For this reason the results of \cite{Kay1993} apply without
% modification.

A physical reason why quasi-free states are of interest is
the following (see also Theorems \ref{Thm_GroundExUn} and
\ref{Thm_KMSExUn} below):
\begin{theorem}\label{Thm_KayUniqueness}
Let $(L,\sigma)$ be a pre-symplectic space and let $\omega$ be
a $C^2$ state on $\alg{W}(L,\sigma)$. $\omega_2$, as defined
in Definition \ref{Def_CkState}, defines a unique quasi-free
state $\omega'$ by Theorem \ref{Thm_Twopoint} and a
one-particle structure $(p,\mathcal{K})$. Then,
\begin{enumerate}
\item[\textup{(i)}] $\omega'$ is pure if and only if $p$ has a dense range
already on $L$ (without complexification) and $p(f)=0$ for
all degenerate $f\in L$ (i.e.\ $f\in L$ for which
$\sigma(f,f')=0$ for all $f'\in L$).
\item[\textup{(ii)}] If $\omega'$ is pure, then $\omega=\omega'$.
\end{enumerate}
\end{theorem}
\begin{proof*}
The claim that $\omega_2$ satisfies the assumptions of Theorem
\ref{Thm_Twopoint} is a standard exercise. The characterization
of pure quasi-free states in terms of their one-particle
structures was established in Ref.\ \cite{Kay+1991}, Lemma A.2,
for the symplectic case. The generalization to the pre-symplectic
case is straightforward. The fact that this implies that
$\omega=\omega'$ is a theorem due to Ref.\ \cite{Kay1993}, for
the symplectic case. This result and its proof carry over to the
pre-symplectic case without modification.
\end{proof*}

A related result in the commutative case is the following
characterisation of the state $\rho^0$:
\begin{proposition}\label{Prop_rho0}
If $\rho\in\mathscr{S}(\alg{W}(L,0))$ is a $C^1$ pure state, then
$\rho(W(f))=e^{i\rho_1(f)}$ for all $f\in L$. In particular, if
$\rho_1=0$, then $\rho=\rho^0$.
\end{proposition}
\begin{proof*}
Given any $f\in L$ we consider $F(t):=\rho(W(tf))$. Because $\rho$
is pure and $\alg{W}(L,0)$ is commutative, $F(t+t')=F(t)F(t')$
(cf.\ \cite{Kadison+} Prop.\ 4.4.1) and hence
$\partial_tF(t)=F(t)\partial_t F(0)=F(t)i\rho_1(f)$. Hence,
$F(t)=e^{it\rho_1(f)}$ and the results follow.
\end{proof*}

\subsection{Quasi-free dynamics on Weyl $C^*$-algebras}\label{SSec_DynWeylSys}

A pre-symplectic isomorphism $T$ of $(L,\sigma)$ is a real-linear
isomorphism $\map{T}{L}{L}$ which preserves the pre-symplectic form,
$\sigma(Tf,Tf')=\sigma(f,f')$. Each pre-symplectic isomorphism gives
rise to a unique $^*$-isomorphism $\alpha_T$ of $\alg{W}(L,\sigma)$
such that $\alpha_T(W(f))=W(Tf)$ (see Ref.\ \cite{Binz+2004}, or also
Ref.\ \cite{Bratteli+} Thm.\ 5.2.8). Hence, a one-parameter group of
pre-symplectic isomorphisms $T_t$ gives rise to a one-parameter group
$\alpha_t$ of $^*$-isomorphisms on $\alg{W}(L,\sigma)$. Not every
one-parameter group of $^*$-isomorphisms on $\alg{W}(L,\sigma)$
arises in this way, but the time evolution that we will be
interested in for our physical applications does.

\begin{definition}\label{Def_GroundOneP}
A \emph{one-particle dynamical system} $(L,\sigma,T_t)$ is a
pre-symplectic space $(L,\sigma)$ with a one-parameter group of
pre-symplectic isomorphisms $T_t$. The associated algebraic
dynamical system $(\alg{W}(L,\sigma),\alpha_t)$ with
$\alpha_t(W(f))=W(Tf)$ is called \emph{quasi-free}.

An equilibrium one-particle structure $(p,\mathcal{K})$ on a
one-particle dynamical system $(L,\sigma,T_t)$ is a
one-particle structure on $(L,\sigma)$ for which there is a
one-parameter unitary group $\tilde{O}_t$ on $\mathcal{K}$ such
that $\tilde{O}_tp=pT_t$.

A \emph{ground one-particle structure} is an equilibrium
one-particle structure $(p,\mathcal{K})$ for which the
unitary group $\tilde{O}_t=e^{itH}$ is strongly continuous
and $H\ge 0$.

A \emph{KMS one-particle structure} at inverse temperature
$\beta>0$ is an equilibrium one-particle structure
$(p,\mathcal{K})$, with associated two-point function
$\omega_2$, such that for all $f,f'\in L$ there exists a
bounded continuous function $F_{ff'}$ on
$\overline{\mathrm{S}_{\beta}}$, holomorphic on its interior,
satisfying
\[
F_{ff'}(t)=\omega_2(f,T_tf'),\quad
F_{ff'}(t+i\beta)=\omega_2(T_tf',f).
\]

An equilibrium one-particle structure is called
\emph{non-degenerate} when $\tilde{O}_t=e^{itH}$ is
strongly continuous and 0 is not an eigenvalue for $H$.
\end{definition}
Note that a quasi-free state $\omega$ with two-point function
$\omega_2$ is in equilibrium for a quasi-free dynamical
system if and only if the associated one-particle structure
$(p,\mathcal{K})$ is in equilibrium. Furthermore, we have
\begin{proposition}\label{Prop_EqToOnePEq}
Let $\omega$ be a $C^2$ equilibrium state on a quasi-free
algebraic dynamical system $(\alg{W}(L,\sigma),\alpha_t)$.
Let $(p,\mathcal{K})$ be the one-particle structure associated
to $\omega_2$ and assume that $\omega_1=0$.
\begin{enumerate}
\item[\textup{(i)}] If $\omega$ is a (non-degenerate) ground state, then
$(p,\mathcal{K})$ is a (non-degenerate) ground one-particle
structure.
\item[\textup{(ii)}] If $\omega$ is a $\beta$-KMS state, then
$(p,\mathcal{K})$ is a $\beta$-KMS one-particle structure.
\end{enumerate}
When $\omega$ is quasi-free, the converses of these
statements are also true.
\end{proposition}
\begin{proof*}
We may identify $\mathcal{K}$ as a closed linear subspace
of the GNS-representation space $\mathcal{H}_{\omega}$,
spanned by the vectors
$p(f):=\Phi_{\omega}(f)\Omega_{\omega}:=
-i\partial_s\pi_{\omega}(W(sf))\Omega_{\omega}|_{s=0}$.
This derivative is well defined, because $\omega$ is $C^2$.
The unitary group $U_t$ on $\mathcal{H}_{\omega}$
restricts to a unitary group $\tilde{O}_t$ on $\mathcal{K}$,
because the dynamics is quasi-free, and the generator $h$ of
$U_t$ restricts to the generator $H$ of $\tilde{O}_t$. Also
note that $\mathcal{K}$ is
perpendicular to $\Omega_{\omega}$, because $\omega_1=0$.
It is then clear that when $\omega$ is a (non-degenerate)
ground state, then $H$ is (strictly) positive and
$(p,\mathcal{K})$ is a (non-degenerate) ground one-particle
structure. When $\omega$ is a $\beta$-KMS state and
$f,f'\in L$, we may take $A(s):=s^{-1}(W(sf)-I)$ and
$B(s):=s^{-1}(W(sf')-I)$ for any $s\not=0$ to find
functions $F_{A(s)B(s)}$. Because $\omega$ is $C^2$, the
functions $\omega(A^*(s)A(s))$ and $\omega(A(s)A^*(s))$
have well defined limits as $s\rightarrow 0$, and similarly
for $B$. We may then use Proposition \ref{Prop_KMSEq} to
take the uniform limit of $-F_{A(s)B(s')}$ as
$s,s'\rightarrow 0$, which yields the desired function
$F_{ff'}$. This proves both items.

If $\omega$ is quasi-free, its GNS-representation is a
Fock space, $\mathcal{H}_{\omega}=\oplus_{n=0}^{\infty}
P_{+,n}\mathcal{K}^{\otimes n}$, where $P_{+,n}$ is the
projection onto the symmetrised $n$-fold tensor product. $U_t$
is the second quantization of $\tilde{O}_t$ and $h$ is the
second quantization of $H$. For the converse of the first
statement we note that $\omega$ is a (non-degenerate) ground
state iff the restriction of $h$ to each $n$-particle space
with $n\ge 1$ is (strictly) positive. If $(p,\mathcal{K})$
is a (non-degenerate) ground one-particle structure, then
$H$ is (strictly) positive. The restriction $h_n$ of $h$ to
$P_{+,n}\mathcal{K}^{\otimes n}$ is given by
$\overline{H_n}P_{+,n}$, where $H_n$ is defined to be the
operator $H_n:=\sum_{j=1}^nI^{\otimes j-1}\otimes H
\otimes I^{\otimes n-j}$ on the algebraic tensor product
$D(H)^{\otimes n}$ of the domain $D(H)$ of $H$. By Nelson's
Analytic Vector Theorem (Ref.\ \cite{Reed+} Thm.\ X.39), $H_n$
is essentially self-adjoint (because $H$ is). The closure of
each summand in $H_n$ is a (strictly) positive operator (by
Lemma \ref{Lem_SP1}), and hence so is $\overline{H_n}$ (by
Lemma \ref{Lem_SP3}). Therefore, $h_n$ is (strictly) positive
for $n\ge 1$ and $\omega$ is a (non-degenerate) ground state.

Now we turn to the converse of the second statement. One
may use the Weyl relations and the quasi-free property to
find
\[
\omega(W(f)\alpha_t(W(f')))=\omega(W(f))\omega(W(f'))
e^{-\omega_2(f,T_tf')}.
\]
Using $F_{ff'}$ in the exponent yields the desired
$F_{W(f)W(f')}$. For finite linear combinations of Weyl
operators the desired property is now clear and for
general operators in $\alg{W}(L,\sigma)$ one appeals to
Proposition \ref{Prop_KMSEq} and a limiting argument.
\end{proof*}

One of the nice aspects of quasi-free dynamical systems is that
we may view $T_t$ also as a pre-symplectic isomorphism of $(L,0)$,
so we may compare the corresponding quasi-free dynamics on
$\alg{W}(L,\sigma)$ and on $\alg{W}(L,0)$. In this context we
prove the following result (adapted from Ref.\ \cite{Rocca+1970}):
\begin{proposition}\label{Prop_AllKMSStates}
Let $(L,\sigma,T_t)$ be a one-particle dynamical system and
consider the corresponding quasi-free dynamical systems
$(\alg{W}(L,\sigma),\alpha_t)$ and $(\alg{W}(L,0),\beta_t)$.
\begin{enumerate}
\item[\textup{(i)}] If $\omega^{(\beta)}\in\mathscr{G}^{(\beta)}(\alg{W}(L,\sigma))$
is quasi-free and $\omega^{(\beta)}_2$ defines a non-degenerate
equilibrium one-particle structure, then the map
$\lambda_{(\beta)}:=\lambda_{\omega^{(\beta)}}$ of Lemma
\ref{Lem_AffineIso} restricts to an affine homeomorphism
$\map{\lambda_{(\beta)}}{\mathscr{G}^0(\alg{W}(L,0))}
{\mathscr{G}^{(\beta)}(\alg{W}(L,\sigma))}$.
\item[\textup{(ii)}] If $\omega^0\in\mathscr{G}^0(\alg{W}(L,\sigma))$ is a quasi-free
and non-degenerate state and if the strong derivative
$\partial_t\pi_{\omega^0}(\alpha_t(W(f)))\Omega_{\omega^0}|_{t=0}$
exists for all $f\in L$, then the map $\lambda_0:=\lambda_{\omega^0}$
restricts to an affine homeomorphism
$\map{\lambda_0}{\mathscr{G}^0(\alg{W}(L,0))}
{\mathscr{G}^0(\alg{W}(L,\sigma))}$.
\end{enumerate}
\end{proposition}
\begin{proof*}
First consider the KMS case. It follows from Lemma
\ref{Lem_AffineIso} that $\lambda_{(\beta)}$ defines a
continuous affine map from $\mathscr{G}^0(\alg{W}(L,0))$ to
$\mathscr{S}(\alg{W}(L,\sigma))$, which is injective because
$\omega^{(\beta)}(W(f))=e^{-\frac12\omega^{(\beta)}_2(f,f)}\not=0$.
If $\rho\in\mathscr{G}^0(\alg{W}(L,0))$, then
$\omega:=\lambda_{(\beta)}(\rho)$ is invariant under $\alpha_t$,
because $\omega^{(\beta)}$ and $\rho$ are equilibrium states for
$\alpha_t$ and $\beta_t$, respectively, and these one-parameter
groups are quasi-free with the same underlying $T_t$. For any
$A=\sum_{i=1}^nc_iW(f_i)$ and $B=\sum_{i=1}^nd_iW(f'_i)$
in $\accentset{\circ}{\alg{W}}(L,\sigma)$ we have
\begin{equation}\label{Eq_Omega}
\omega(A\alpha_t(B))=\sum_{i,j=1}^nc_id_j
\omega^{(\beta)}(W(f_i)\alpha_t(W(f'_j)))\rho(W(f_i)W(f'_j)),
\end{equation}
by a short computation involving the Weyl relations and
the properties of $\rho$ established in Lemma
\ref{Lem_ClGround}. A similar computation for
$\omega(\alpha_t(B)A)$ and the KMS-condition for
$\omega^{(\beta)}$ now imply the existence of a function
$F_{AB}$ as needed for the KMS-condition for $\omega$. For the
operators in the $C^*$-algebraic completion $\alg{W}(L,\sigma)$
one uses Proposition \ref{Prop_KMSEq}. Hence $\omega$ is a
$\beta$-KMS state.

For ground states, Eq.\ (\ref{Eq_Omega}) (with $\omega^0$ instead
of $\omega^{(\beta)}$) implies that the
unitary group $U_t$ that implements $\alpha_t$ in the
GNS-representation of $\omega$ is weakly continuous and hence
strongly continuous. The dense domain
$\pi_{\omega}(\alg{W}(L,\sigma)))\Omega_{\omega}$ is invariant
under the action of $U_t$ and one may show that $U_t=e^{ith}$
has strong derivatives there, because the same is true for
$\omega^0$. Hence this domain is a core for the Hamiltonian $h$
(see e.g.\ Thm.\ VIII.10 of Ref.\ \cite{Reed+}). Taking the
derivative with respect to $t$ of Eq.(\ref{Eq_Omega}) and taking
$A=B$ shows that $h\ge 0$, by Schur's Product Theorem
(cf.\ Ref.\ \cite{Bellman1960} Ch.6 Sec.7 or Ref.\ \cite{Karsten1989}).
This proves that $\omega$ is a ground state.

We now turn to surjectivity. Given any
$\omega\in\mathscr{G}^{(\beta)}(\alg{W}(L,\sigma))$ we may
define the linear map $\rho$ on
$\accentset{\circ}{\alg{W}}(L,0)$ by
$\rho(W(f)):=\frac{\omega(W(f))}{\omega^{(\beta)}(W(f))}$ for
all $f\in L$. Given
any $f,f'\in L$ we now let $F^{(\beta)}_{W(-f)W(f')}(z)$ and
$F_{W(-f)W(f')}(z)$ be the functions on
$\overline{\mathrm{S}_{\beta}}$, obtained from the
KMS-condition for $\omega^{(\beta)}$ and $\omega$,
respectively. Note that
$F^{(\beta)}_{W(-f)W(f')}(z)=C\exp(-F_{-f,f'}(z))$, by the
one-particle KMS-condition for $\omega_2$ (cf.\ Proposition
\ref{Prop_EqToOnePEq}), where
$C:=\exp(-\frac12(\omega^{(\beta)}(f,f)+\omega^{(\beta)}(f',f')))$.
Hence,
\[
G(z):=(F^{(\beta)}_{W(-f)W(f')}(z))^{-1}F_{W(-f)W(f')}(z)
\]
defines a bounded and continuous function on
$\overline{\mathrm{S}_{\beta}}$ which is holomorphic in its
interior. Furthermore, $G(t)=\rho(W(-f)\beta_t(W(f')))$ and
$G(t+i\beta)=\rho(\beta_t(W(f'))W(-f))$. As $\rho$ is defined
on a commutative $C^*$-algebra it then follows that
$G(z+i\beta)=G(z)$ and we may extend $G$ periodically to
a bounded continuous function on $\mathbb{C}$, which is
entire holomorphic by the Edge of the Wedge
Theorem \cite{Berenstein+1991}. Hence, $G$ is constant (by
Liouville's Theorem \cite{Berenstein+1991}) and
$\rho(W(-f)\beta_t(W(f'))=\rho(W(-f)W(f'))$ for all
$t\in\mathbb{R}$. A similar argument holds for the case
of ground states.

For any $A=\sum_{i=1}^nc_iW(f_i)$ we have
\begin{eqnarray}
0&\le&\sum_{i,j=1}^N\overline{c_i}c_j\omega(W(-f_i)W(f_j))\nonumber\\
&=&\sum_{i,j=1}^N\overline{c_i}c_j
\exp(-\frac12\omega^{(\beta)}_2(f_j-f_i,f_j-f_i))\rho(W(-f_i)W(f_j)).
\nonumber
\end{eqnarray}
For some $t>0$ we now let $F_i^M:=\sum_{m=0}^{M-1}\frac{1}{M}T_{mt}f_i$
for any $M\in\mathbb{N}$. Using the previous paragraph one shows
that $\rho(W(-F^M_i)W(F^M_j))=\rho(W(-f_i)W(f_j))$, from which we find
\[
0\le\sum_{i,j=1}^N\overline{c_i}c_j
\exp(-\frac12\omega^{(\beta)}_2(F^M_j-F^M_i,F^M_j-F^M_i))
\rho(W(-f_i)W(f_j)).
\]
However, as the one-particle structure $(p,\mathcal{K})$ associated
to $\omega^{(\beta)}_2$ is non-degenerate, we see from von Neumann's
Mean Ergodic Theorem (Ref.\ \cite{Reed+} Thm.\ II.11) that
$\lim_{M\rightarrow\infty}p(F_i^M)=0$. The exponential term will then
converge to $1$ as $M\rightarrow\infty$, leading to the conclusion that
$\rho$ is positive. The unique extension of $\rho$ to a state on
$\alg{W}(L,0)$ is a ground state by the result of the previous
paragraph and Lemma \ref{Lem_ClGround}. The same argument works for
the case of ground states.

Finally, to see that $\lambda_{(\beta)}$
(resp.\ $\lambda_0$) is a homeomorphism it suffices to note that the
inverse map $\omega\mapsto\rho$ is weak$^*$-continuous from
$\mathscr{G}^{(\beta)}(\accentset{\circ}{\alg{W}}(L,\sigma))$
(resp.\ $\mathscr{G}^0(\accentset{\circ}{\alg{W}}(L,\sigma))$) to
$\mathscr{G}^0(\accentset{\circ}{\alg{W}}(L,0))$, by construction.
\end{proof*}

\begin{remark}\label{Rem_Gauge3}
In the setting of Proposition \ref{Prop_AllKMSStates} we note that
the space $\mathscr{G}^0(\alg{W}(L,0))$ of classical ground states
always contains the pure state $\rho^0$ and that
$\omega^{(\beta)}=\lambda_{(\beta)}(\rho^0)$. For any other pure
classical ground state $\rho\in\mathscr{G}^0(\alg{W}(L,0))$ we
consider the gauge transformations of the second kind $\eta_{\rho}$
of $\alg{W}(L,\sigma)$ and $\zeta_{\rho}$ of $\alg{W}(L,0)$
(cf.\ Remark \ref{Rem_Gauge2}). We then have
$\rho=\zeta_{\rho}^*\rho^0$ and
$\lambda^{(\beta)}\circ\zeta_{\rho}^*=\eta_{\rho}^*\circ\lambda^{(\beta)}$.
Thus every extremal $\beta$-KMS state can be obtained from
$\omega^{(\beta)}$ by a gauge transformation of the second kind.
The same holds for extremal ground states and $\omega^0$. In
particular, all extremal ground states are pure.
\end{remark}

\section{Review of geometric results}\label{Sec_Geometry}

Before we consider the details of the linear scalar quantum field it is in
order to study the spacetime in which it propagates. In the paragraphs
below we will describe the class of stationary, globally hyperbolic
spacetimes and the subclass of standard static spacetimes. For the latter
case we also introduce the complexification and Euclideanization that are
necessary in order to perform a Wick rotation. Most of our exposition here
is a brief review of recent results of Refs.\ \cite{Caponio+2011} and
\cite{Sanchez2005}.

We assume that the reader is already familiar with the following standard
terminology, which will be used throughout (cf.\ the reference book
\cite{Wald}):
\begin{definition}\label{Def_(GH)Spac}
A \emph{spacetime} $M=(\mathcal{M},g)$ is a smooth, connected, oriented
manifold $\mathcal{M}$ of dimension $d\ge 2$ with a smooth Lorentzian
metric $g$ of signature $(-+\ldots+)$.

A \emph{Cauchy surface} $\Sigma$ in $M$ is a subset $\Sigma\subset M$
that is intersected exactly once by every inextendible time-like curve
in $M$. A spacetime is said to be \emph{globally hyperbolic} when it
has a Cauchy surface.
\end{definition}
For a spacetime $M$ we note that the manifold $\mathcal{M}$ is
automatically paracompact \cite{Geroch1968}.
We are mainly interested in spacetimes that are globally hyperbolic,
because they allow us to formulate the linear field equation as an initial
value (or Cauchy) problem. We will only consider Cauchy surfaces that
are space-like, smooth hypersurfaces \cite{Bernal+2003}. A
globally hyperbolic spacetime is automatically time-orientable and we
will assume that a choice of time-orientation has been fixed. It follows
that any Cauchy surface is also oriented. Our notions and notations for
causal relations, the Levi-Civita connection, etc.\ follow standard
usage \cite{Wald}. We will let $h$ denote the Riemannian
metric on a Cauchy surface $\Sigma$ induced by the Lorentzian metric $g$
on $M$, and we let $\nabla^{(h)}$ denote the corresponding Levi-Civita
connection on $\Sigma$. Spacetime indices $a,b,\ldots$ are chosen from
the beginning of the alphabet and run from $0$ to $d-1$, whereas spatial
indices are denoted by $i,j,\ldots$ and run from $1$ to $d-1$.

\subsection{Stationary spacetimes}\label{SSec_Stationary}

Stationary spacetimes come equipped with a preferred notion of
time-flow, which is mathematically encoded in the presence of a
time-like vector field. To be precise:
\begin{definition}\label{Def_StatnSpac}
A \emph{stationary spacetime} $(M,\xi)$ is a spacetime $M$ together
with a smooth, complete, future-pointing, time-like Killing vector
field $\xi$ on $M$.
\end{definition}
Here completeness means that the corresponding flow
$\map{\Xi}{\mathbb{R}\times\mathcal{M}}{\mathcal{M}}$, defined by
$\Xi(0,x)=x$ and $d\, \Xi(t,x;\partial_t,0)=\xi(\Xi(t,x))$, is well
defined for all $t\in\mathbb{R}$. This flow is interpreted
physically as the flow of time and following standard usage we write
$\map{\Xi_t}{\mathcal{M}}{\mathcal{M}}$ for the map
$\Xi_t(x):=\Xi(t,x)$.

$\xi$ is a Killing vector field if it satisfies Killing's equation,
$\nabla_{(a}\xi_{b)}=0$, where the round brackets in the subscript
denote symmetrization as an idempotent operation. Equivalently, it
means that the metric is invariant under the time flow of $\xi$,
$\Xi_t^*g=g$ for all $t\in\mathbb{R}$.

\begin{example}\label{Ex_SStatn}
{\bf Standard stationary spacetimes:}
Examples of stationary spacetimes are easily obtained by the
following construction. Let $S$ be a manifold of dimension $d-1$,
let $h$ be a Riemannian metric on $S$, let $v>0$ be a smooth,
strictly positive function on $S$ and let $w$ be a smooth
one-form on $S$ such that $h^{ij}w_iw_j<v^2$. One now defines
$\mathcal{M}:=\mathbb{R}\times S$ with canonical projection map
$\map{\pi}{\mathcal{M}}{S}$ and the canonical time coordinate
$\map{t}{\mathcal{M}}{\mathbb{R}}$ is the canonical projection
onto the first factor. A stationary spacetime
$M=(\mathcal{M},g)$ is then obtained by defining
\[
g:=-(\pi^*v)^2dt^{\otimes 2}+ 2\pi^*(w)\otimes_s dt+\pi^*h,
\]
where $\otimes_s$ is the symmetrised tensor product. We will
always choose adapted local coordinates on $M$, i.e.\ coordinates
$(t,x^i)$ such that the $x^i$ are local coordinates on $S$,
unless stated otherwise.

Note that $g$ indeed has a Lorentz signature and that the canonical
vector field $\partial_t$ on $\mathbb{R}$ gives rise to a Killing
vector field $\xi$ on $M$. On $S_0:=\left\{0\right\}\times S$ we
can write $\xi^a=Nn^a+N^a$, where $n^a$ is the future pointing unit
normal vector field to $S_0\subset M$ and $n_aN^a=0$. The function
$N$ is known as the lapse function and $N^a$ as the shift vector
field. They are related to $v$ and $w$ by
\[
N=(v^2+h^{ij}w_iw_j)^{\frac12},\quad
N^i=h^{ij}w_j,
\]
where we use the fact that $N^a$ is tangent to $\Sigma$, so the
component for $a=0$ vanishes (in adapted local coordinates).
The inverse of the metric takes the form
\[
g^{-1}=-N^{-2}\partial_t^{\otimes 2}+
2N^{-2}N^j\partial_j\otimes_s\partial_t
+(h^{ij}-N^{-2}N^iN^j)\partial_i\otimes\partial_j,
\]
where $h^{ij}$ is the inverse of the Riemannian metric $h$.
\end{example}

\begin{definition}\label{Def_SStatnSpac}
A stationary spacetime of the form of Example \ref{Ex_SStatn} is
called a \emph{standard stationary spacetime}.
\end{definition}
Note that a standard stationary spacetime $M$ is uniquely determined
by the data $(S,h,v,w)$. However, different data may give rise
to the same spacetime, because there is a lot of freedom in the
choice of the surface $S\subset M$. This is another way of saying
that a stationary spacetime has a preferred time-flow, given by
the Killing vector field, but it does not have a preferred time
coordinate, because we can choose different canonical time
coordinates which vanish on different spatial hypersurfaces.

Although not all stationary spacetimes are standard,\footnote{
Consider e.g.\ Minkowski spacetime and compactify an inertial
time coordinate to a circle.} they are the only ones of interest
to us because of the following result:
\begin{proposition}\label{Prop_GHSStatn}
Let $M$ be a stationary spacetime which is globally hyperbolic.
Then $M$ is isometrically diffeomorphic to a standard stationary
spacetime.
\end{proposition}
This is Proposition 3.3 of Ref.\ \cite{Sanchez2005}. The proof
is elegant and short and we include it here for completeness:
\begin{proof*}
Fix a Cauchy surface $\Sigma\subset M$ and use the flow
$\Xi$ of the Killing vector field to define a local
diffeomorphism $\map{\psi}{\mathbb{R}\times\Sigma}{M}$ by
$\psi(t,x)=\Xi(t,x)$, The curves $t\mapsto\psi(t,x)$ are
time-like and inextendible, because $\xi$ is assumed to be
complete. This means that they intersect $\Sigma$ exactly
once, proving that $\psi$ is both injective and surjective and
hence a diffeomorphism. We define
$M':=(\mathbb{R}\times\Sigma,\psi^*g)$ and it remains to show
that $M'$ is standard stationary. This follows easily from the
fact that $\psi^*\xi=\partial_t$, where $t$ is the canonical
time-coordinate on $M'$, together with the fact that
$\partial_t\psi^*g=0$, which is Killing's equation.
\end{proof*}

A more complicated issue is the converse question, whether a
standard stationary spacetime is globally hyperbolic. A full
characterization of those data $(S,h,v,w)$ that give rise to
a standard stationary spacetime $M$ which is globally hyperbolic
was recently given by Ref.\ \cite{Caponio+2011}. It should be noted
that $S$ need not be a Cauchy surface, even if $M$ is globally
hyperbolic. A full characterization of those data for which
$S$ is a Cauchy surface was also given in Ref.\ \cite{Caponio+2011}.
To close this section we will sketch the main ingredients of
this analysis and state the main results, although they will
not be needed in the remainder of this paper.

Let $s\mapsto \gamma(s):=(t(s),x(s))$ be a smooth, time-like
curve in a standard stationary spacetime $M$ with data
$(S,h,v,w)$. The fact that $\gamma$ is time-like can be
stated as the quadratic inequality
\[
h_{ij}\dot{x}^i\dot{x}^j+2w_i\dot{x}^i\dot{t}-v^2\dot{t}^2\le 0,
\]
where $\dot{\ }$ denotes a derivative w.r.t.\ $s$. If $\gamma$
is future pointing this leads to
\[
\dot{t}\ge v^{-2}w_i\dot{x}^i+\left(v^{-4}(w_i\dot{x}^i)^2
+v^{-2}h_{ij}\dot{x}^i\dot{x}^j\right)^{\frac12}=:F(\dot{x}),
\]
whereas for past-pointing $\gamma$ we find
\[
\dot{t}\le v^{-2}w_i\dot{x}^i-\left(v^{-4}(w_i\dot{x}^i)^2
+v^{-2}h_{ij}\dot{x}^i\dot{x}^j\right)^{\frac12}
=:-\tilde{F}(\dot{x}).
\]
$F$ and $\tilde{F}$ are smooth, strictly positive functions on
$TS\setminus 0$, where $0$ denotes the zero section. (In fact,
$F$ and $\tilde{F}$ define Finsler metrics on $S$ of Randers
type. We refer the interested reader to
Ref.\ \cite{Caponio+2011} for a brief introduction or to
Ref.\ \cite{Bao+2000} for a full exposition on Finsler geometry.)

It turns out that the questions concerning the causality of the
standard stationary spacetime with data $(S,h,w,v)$ can be
determined entirely from the properties of $S$ with respect to
$F$ and $\tilde{F}$. As for a Riemannian metric, we can use $F$
to define the length of a smooth curve $\map{\gamma}{[0,1]}{S}$
as $l_F(\gamma):=\int_0^1F(\dot{\gamma}(s))ds$ and from that we
can define a generalised distance function
\[
d(p,q):=\inf_{\gamma\in C(p,q)}l_F(\gamma),
\]
where $C(p,q)$ is the set of all piecewise smooth curves from
$p$ to $q$. $d$ satisfies all properties of a distance function,
except symmetry. Indeed, if $\tilde{\gamma}(s):=\gamma(1-s)$ we
have $l_F(\tilde{\gamma})=l_{\tilde{F}}(\gamma)$, which in
general differs from $l_F(\gamma)$. However, taking the ordering
into account one can still define notions of forward and
backward Cauchy sequences and corresponding notions of forward
and backward completeness for the pair
$(S,F)$ \cite{Bao+2000,Caponio+2011}.

We now state without proof the results on the causality of
standard stationary spacetimes (Thm.\ 4.3b, Thm.\ 4.4 and
Cor.\ 5.6 of Ref.\ \cite{Caponio+2011}).
\begin{theorem}\label{Thm_SStatnGH}
Let $M$ be a standard stationary spacetime with data $(S,h,v,w)$.
\begin{enumerate}
\item[\textup{(i)}] $M$ is globally hyperbolic if and only if for all $x\in S$
and all $r>0$ the symmetrised closed ball
$B_s(p,r):=\left\{x|\ d(p,x)+d(x,p)\le r\right\}$
is compact.
\item[\textup{(ii)}] $S\subset M$ is a Cauchy surface if and only if $(S,F)$
is both forward and backward complete. In this case all
hypersurfaces $S_t:=\left\{t\right\}\times S$ are Cauchy.
\item[\textup{(iii)}] If $M$ is globally hyperbolic, then $(S,\tilde{h})$ is a
complete Riemannian manifold with
\[
\tilde{h}:=v^{-2}h+v^{-4}w\otimes w.
\]
\end{enumerate}
\end{theorem}
We record for completeness that the inverse metric of
$\tilde{h}$ is given by
$\tilde{h}^{ij}=v^2h^{ij}-v^2N^{-2}N^iN^j=v^2g^{ij}$,
where $g^{ij}$ is expressed in adapted coordinates.

\subsection{Standard static spacetimes}\label{SSec_Static}

We have seen that stationary spacetimes have a preferred time
flow, but no preferred time
coordinate. This is different for standard static spacetimes,
which we will describe now. For a fuller discussion of static
spacetimes we refer the reader to Ref.\ \cite{Sanchez2005} and
references therein.

\begin{definition}\label{Def_StatSpac}
A \emph{static spacetime} $M=(\mathcal{M},g,\xi)$ is a
stationary spacetime with a Killing vector field $\xi$ that
is irrotational.
\end{definition}
The property that $\xi$ is irrotational means that the
distribution of vectors orthonogal to $\xi$ is involutive,
i.e.\ $[X,Y]^a\xi_a=0$ when $X^a\xi_a=Y^a\xi_a=0$. This can
be expressed equivalently as
\[
\xi_{[a}\nabla_b\xi_{c]}=0,
\]
where the square brackets in the subscript denote
anti-symmetrization as an idempotent operation.
By Frobenius' Theorem (Ref.\ \cite{Wald} Thm.\ B.3.2) $\xi$ is
irrotational if and only if $M$ can be foliated by
hypersurfaces orthogonal to $\xi$.

If $x^i$, $i=1,\ldots d-1$, are local coordinates on a
$(d-1)$-dimensional hypersurface $H\subset M$ orthogonal to
$\xi$ we can (locally) supplement them by the parameter $t$
appearing in the flow $\Xi_t$ to define coordinates on a
portion of $M$. When used like this, we call $t$ a Killing
time coordinate. Note that the surfaces of constant $t$
remain orthogonal to $\xi=\partial_t$, because they are the
image of $H$ under $\Xi_t$.
\begin{remark}\label{Rem_KillingTime}
Although the definition of a (local) Killing time coordinate
depends on the choice of the hypersurface $H$, any two Killing
time coordinates on the same open set differ at most by a
constant, because both are constant on the hypersurfaces
orthogonal to $\xi$. In this sense, static spacetimes have a
preferred time coordinate up to a constant, which we will
often call \emph{the} Killing time coordinate, with some
slight abuse of language.
\end{remark}

In the local coordinates $(t,x^i)$ the metric can be expressed
as
\[
g=-v^2dt^{\otimes 2}+g_{ij}dx^i\otimes dx^j,
\]
with $1\le i,j\le d-1$ and the smooth coefficient functions
$v,g_{ij}$ are independent of $t$. We introduce a special
name for the class of static spacetimes for which this form
of the metric can be obtained globally:
\begin{definition}\label{Def_StStatSpac}
A \emph{standard static spacetime} $M=(\mathcal{M},g,\xi)$
is a standard stationary spacetime with a vanishing shift
vector field, i.e.\ $\mathcal{M}\simeq\mathbb{R}\times S$,
$\xi=\partial_t$ and
\[
g=-(\pi^*N)^2dt^{\otimes 2}+\pi^*h,
\]
where the Killing time coordinate $t$ is the projection on
the first factor of $\mathbb{R}\times S$, $\pi$ is the
projection on the second factor, $h$ is a Riemannian metric
on $S$ and $N$ is a smooth, strictly positive function on
$S$.
\end{definition}
The data $(S,h,N)$ determine a unique standard static
spacetime, which is the standard stationary spacetime with
data $(S,h,v=N,w=0)$. The canonical time coordinate of the
latter coincides with the Killing time coordinate.

Unlike the stationary case, there is only a limited freedom
in the choice of data that describe a fixed standard static
spacetime $M$. Indeed, suppose that $(S,h,v)$ and $(S',h',v')$
determine the same standard static spacetime $M$ and consider
the hypersurfaces $S_0=\left\{0\right\}\times S$ and
$S'_0=\left\{0\right\}\times S'$ in $M$. By Remark
\ref{Rem_KillingTime} there is a $T\in\mathbb{R}$ such that
the diffeomorphism $\Xi_T$ of $M$ has $S'_0=\Xi_T(S_0)$,
$\Xi_T^*h'=h$ and $\Xi_T^*v'=v$.

For our applications to Wick rotations we are particularly
interested in spacetimes which are both standard static and
globally hyperbolic. To determine whether a standard static
spacetime is globally hyperbolic we quote from Theorem 3.1
in Ref.\ \cite{Sanchez2005}:
\begin{theorem}\label{Thm_TestGH}
For a standard static spacetime $M$ with data $(S,h,v)$
the following are equivalent:
\begin{enumerate}
\item[\textup{(i)}] $M$ is globally hyperbolic.
\item[\textup{(ii)}] $S$ is complete in the conformal metric
$\tilde{h}_{ij}=v^{-2}h_{ij}$.
\item[\textup{(iii)}] Each constant Killing time hypersurface is Cauchy.
\end{enumerate}
\end{theorem}
This is in fact a special case of Theorem \ref{Thm_SStatnGH},
when $w=0$. In the ultra-static case $v\equiv 1$, it
essentially reduces to Proposition 5.2 in Ref.\ \cite{Kay1978}.
Note, however, that $(S,h)$ itself need not be a complete
Riemannian manifold in general.

\begin{remark}\label{Rem_OpticalMetric}
The metric $\tilde{h}$ is also called the optical
metric \cite{Abramowicz+1988}, because geodesics of $\tilde{h}$
are the projections onto $\Sigma$ of light-like geodesics in
$M$. To see this we first note that the light-like geodesics of
$M=(\mathcal{M},g)$ coincide with those of
$\tilde{M}:=(\mathcal{M},v^{-2}g)$ after a reparametrization
(cf.\ Ref.\ \cite{Wald} Appendix D). Because $\tilde{M}$ is
ultra-static, the geodesic equation for a curve
$\gamma(s)=(t(s),x(s))$ decouples into the geodesic equation
for $x$ in $(S,\tilde{h})$ and $\partial_s^2t=0$.
(Ref.\ \cite{Abramowicz+1988} also uses the term optical
metric in the stationary case for the metric $N^{-2}h$,
although the motivation is less convincing in that case. It
might be more appropriate to refer to the Finsler metrics
$F,\tilde{F}$ of Section \ref{SSec_Stationary} as optical
metrics.)
\end{remark}

When the spacetime $M$ is both globally hyperbolic and
static, it is automatically a standard \emph{stationary} spacetime
by Proposition \ref{Prop_GHSStatn}. However, it may yet
fail to be a standard static spacetime. A simple
counter-example, taken from Ref.\ \cite{Candela+2008} (see
also Ref.\ \cite{Sanchez2005}), is the cylinder spacetime
$M=(\mathbb{R}\times \mathbb{S}^1,g)$ with the metric
$g:=-dt^{\otimes 2}+d\theta^{\otimes 2}+2dt\otimes_s d\theta$.
This is a globally hyperbolic spacetime with Cauchy
surfaces diffeomorphic to the circle $\mathbb{S}^1$. The vector
field $\xi=\partial_t$ is a time-like Killing field, which
is irrotational on dimensional grounds. However,
hypersurfaces orthogonal to $\xi$ must be diffeomorphic
to $\mathbb{R}$, as they wind around the cylinder.

A complete characterization of which static, globally
hyperbolic spacetimes are standard static is given by
\begin{proposition}\label{Prop_GHSStat}
Let $(M,\xi)$ be a static, globally hyperbolic spacetime.
Then $M$ is isometrically diffeomorphic to a standard
static spacetime if and only if it admits a Cauchy
surface that is Killing field orthogonal.
\end{proposition}
\begin{proof*}
If $M$ is isometrically diffeomorphic to a standard
static spacetime, the existence of a Killing field
orthogonal Cauchy surface follows from Theorem
\ref{Thm_TestGH}. Conversely, if such a Cauchy surface
exists we may choose this surface in the proof of
Proposition \ref{Prop_GHSStatn}, which simultaneously
shows that $M$ is isometrically diffeomorphic to a
standard stationary spacetime $M'$ and that the metric
$g'$ has no cross terms involving $w$. Hence, $M'$ is
standard static.
\end{proof*}

\subsection{Spacetime complexification}\label{SSec_Complexify}

To conclude our geometric considerations we now define
complexifications and Riemannian manifolds associated
to any given standard static spacetime. With a view to
our applications to thermal states it is necessary to
consider the case where the domain of the imaginary
time variable is compactified. For this purpose we let
$R>0$ and we define the cylinder
\[
\mathcal{C}_R:=\mathbb{C}/\sim,\quad z\sim z'
\Leftrightarrow z-z'\in 2\pi iR\mathbb{Z}.
\]
This equivalence relation compactifies the imaginary
axis of $\mathbb{C}$ to a circle $\mathbb{S}^1_R$ of
circumference $2\pi R$.
$\mathcal{C}_{\infty}:=\mathbb{C}$ can be taken as a
degenerate case with $R=\infty$ and
$\mathbb{S}^1_{\infty}:=\mathbb{R}$.

Let $M$ be a standard static spacetime with data
$(S,h,N)$. For any $R>0$ we define the complexification
$M^c_R$ as the real manifold $M^c_R=\mathcal{C}_R\times S$
endowed with the symmetric, complex-valued, tensor field
\[
g^c_R(z,x)=-N^2(x)(dt+id\tau)^{\otimes 2}+h(x),
\]
where $z=t+i\tau$ is the coordinate on $\mathcal{C}_R$.
$M$ can be embedded into $M^c_R$ as the $\tau=0$ surface
and $g^c_R$ is the analytic continuation of $g$ in
$z$. Furthermore, we define the Riemannian manifold
$M_R:=\left\{(z,x)\in M^c_R|\ t=0\right\}$ endowed with
the pull-back metric of $g^c_R$
\[
g_R(\tau,x)=N^2(x)d\tau^{\otimes 2}+h(x).
\]
Note that $M_R\simeq \mathbb{S}^1_R\times S$ as a manifold
and since $S=M\cap M_R$ in $M^c_R$, we can identify $S$
also as the $\left\{\tau=0\right\}$ surface in $M_R$.
$M_R$ has a Killing field $\xi_R=\partial_{\tau}$, which
can be viewed as the analytic continuation of
$\xi=\partial_t$.

The constructions above do not depend on any freedom in the
choice of $S$, because this freedom boils down to a Killing
time translation (see Remark \ref{Rem_KillingTime}) which
has a unique analytic continuation to $M^c_R$. It is also
unnecessary for $S$ to be a Cauchy surface at this stage. Note
that in the standard stationary case there is more freedom
to choose canonical time coordinates, so it would be unclear
whether an analogous construction can be made independent
of the choice of such a coordinate. Besides, any cross terms
$w\otimes dt$ in the metric would spoil the real-valuedness
of the restriction $g_R$ of the analytically continued metric,
so it would not be Riemannian.

Whereas the Killing time coordinate on $M$ is used to define
the complexifications $M^c_R$ and the Riemannian manifolds
$M_R$, it may be a bad choice of coordinate to analyze the
behaviour near the edge of $S$. This will be the case e.g.\ if
$M$ is the right wedge of a static black hole spacetime
with a bifurcate Killing horizon and we wish to study the
behaviour near the bifurcation surface.\footnote{This setting
will be studied in detail in a forthcoming
publication \cite{SandersIP}.} Anticipating these problems we
now consider Gaussian normal coordinates near $S$, instead of
the Killing time coordinate, and we study the properties of the
complexification procedure above with respect to these new
coordinates.

\begin{proposition}\label{Prop_complexGnormal}
Let $M$ be a standard static spacetime, let $R>0$ and let
$x^i$ denote local coordinates on a portion $U$ of $S$. Let
$x=(x^0,x^i)$ be the corresponding Gaussian normal
coordinates on a portion of $M$, containing $U$, and let
$x'=((x')^0,x^i)$ be Gaussian normal coordinates on a
portion of $M_R$, containing $U$. We may express the
metrics $g$ and $g_R$ in these coordinates as
\[
g=-(dx^0)^{\otimes 2}+h_{ij}dx^idx^j,\quad
g_R=(d(x')^0)^{\otimes 2}+h'_{ij}dx^idx^j,
\]
and we then have for all $n\ge 0$:
\begin{eqnarray}\label{Eqn_Infholo}
\partial_0^{2n}h_{ij}|_U&=&
(-1)^n(\partial'_0)^{2n}h'_{ij}|_U,\nonumber\\
\partial_0^{2n+1}h_{ij}|_U\ =&0&
=\ (\partial'_0)^{2n+1}h'_{ij}|_U.
\end{eqnarray}
\end{proposition}
In the ultra-static case we have $x^0=t$, which means that
the metric $g$ is real-analytic in $x^0$ and its analytic
continuation satisfies $g_{ab}(ix^0,x^i)=(g_R)_{ab}(x^0,x^i)$,
This immediately implies Eq.\ (\ref{Eqn_Infholo}),
by the Cauchy-Riemann equations and the reality of $g$ and
$g_R$. In the general case, the Proposition can be interpreted
as saying that $g$ is ``infinitesimally holomorphic'' in
$z:=x^0+i(x')^0$.
\begin{proof*}
The form of the metrics follows from the construction of
Gaussian normal coordinates, as is well known \cite{Wald}.
The idea is now to use the fact that the geometries of $M$ and
$M_R$ are entirely determined by $(S,h,N)$. The number of
coefficients in $(h_{ij},\xi^a)$ equals $\frac{d(d+1)}{2}$,
which is exactly the number of components
of Killing's equation. We may write out Killing's equation in
the chosen local coordinates, for which the Christoffel symbol
vanishes when two or more indices are $0$. The $(00)$-component
of Killing's equation is then $\partial_0\xi^0=0$, which means
that $\xi^0(x)=N(x^i)$. Substituting this back in the remaining
equations yields\footnote{In these coordinates it is less clear
that the Cauchy problem is well posed, unless the initial data
are analytic, in which case the Cauchy-Kowalewsky Theorem
applies \cite{Wald}. However, we know that the data
$(\Sigma,h,N)$ determine a unique, smooth solution, which is
easily written down in adapted coordinates.}
\begin{eqnarray}
h_{ij}\partial_0\xi^j&=&\partial_iN\nonumber\\
N\partial_0h_{ij}&=&-2h_{k(i}\partial_{j)}\xi^k
-\xi^k\partial_kh_{ij}.\nonumber
\end{eqnarray}
All normal derivatives of $\xi^i$ and $h_{ij}$ are uniquely
determined by the initial data, as can be shown by induction,
taking successive normal derivatives of the equations
above. In the Riemannian case we find similarly
$\xi_R^0(x')=N(x^i)$ and
\begin{eqnarray}
h'_{ij}\partial'_0\xi_R^j&=&-\partial'_iN\nonumber\\
N\partial'_0h'_{ij}&=&-2h'_{k(i}\partial'_{j)}\xi_R^k
-\xi_R^k\partial'_kh'_{ij}.\nonumber
\end{eqnarray}
Note the change of sign in the first equation when compared to
the Lorentzian case.

One now proves by induction on $n\ge 0$ that\footnote{The
vanishing of the odd normal derivatives on $\Sigma$ can also
be seen by a symmetry argument involving a reflection in the
Killing time around the Cauchy surface.}
\[
\partial_0^nh_{ij}|_U=i^n(\partial'_0)^nh'_{ij}|_U,\quad
\partial_0^n\xi^i|_U=i^{n+1}(\partial'_0)^n\xi_R^i|_U.
\]
For $n=0$ these equalities are true, because they just
express the equality of the initial data. (Note in
particular that $\xi^i|_U=0=\xi_R^i|_U$.) Now suppose they
hold true for all $0\le l\le n$. We use Killing's equation
and $\partial_0N=\partial'_0N=0$ to compute
\begin{eqnarray}
\partial_0^{n+1}h_{ij}|_U&=&-N^{-1}\partial_0^n(
2h_{k(i}\partial_{j)}\xi^k+\xi^k\partial_kh_{ij})|_U\nonumber\\
&=&-i^{n+1}N^{-1}(\partial'_0)^n(
2h'_{k(i}\partial'_{j)}\xi_R^k+\xi_R^k\partial'_kh'_{ik})|_U
=i^{n+1}(\partial'_0)^{n+1}h'_{ij}|_U,\nonumber
\end{eqnarray}
where the induction hypothesis was used in the second
equality. Similarly, by the binomial formula,
\[
h_{ij}\partial_0^{n+1}\xi^j|_U=
-\sum_{l=0}^{n-1}\left(\begin{array}{c}n\\ l\end{array}\right)
\partial_0^{n-l}h_{ij}\cdot \partial_0^{l+1}\xi^j|_U
=i^{n+2}h_{ij}(\partial'_0)^{n+1}\xi_R^j|_U,
\]
where we used that fact that .
As $h_{ij}$ is invertible, the result for $n+1$ follows,
completing the proof by induction. The statement of the
proposition is then immediately clear, because both $h_{ij}$
and $h'_{ij}$ are real-valued.
\end{proof*}

\begin{corollary}\label{Cor_CauchyGeodesics}
For a smooth curve $\map{\gamma}{[0,1]}{S}$ the following
are equivalent:
\begin{enumerate}
\item[\textup{(i)}] $\gamma$ is a geodesic in $(S,h)$,
\item[\textup{(ii)}] $\gamma$ is a geodesic in $M$,
\item[\textup{(iii)}] $\gamma$ is a geodesic in $M_R$.
\end{enumerate}
\end{corollary}
\begin{proof*}
We express the geodesic equation in $M$ in terms of local
coordinates $x^i$ on $S$ and a Gaussian normal coordinate
$x^0$ near $S\subset M$. Using the notation
$\gamma^a:=x^a\circ\gamma$, with $\gamma^0=0$, the
components
\[
\partial_s^2\gamma^i=-\Gamma^i_{\phantom{i}ab}
\partial_s\gamma^a\partial_s\gamma^b
=-\Gamma^i_{\phantom{i}jk}\partial_s\gamma^j\partial_s\gamma^k
\]
form exactly the geodesic equation in $(S,h)$. The remaining
equation is
\[
0=\partial_s^2\gamma^0=-\Gamma^0_{\phantom{0}ij}\partial_s\gamma^i
\partial_s\gamma^j=\frac{-1}{2} \partial_0h_{ij}|_U
\partial_s\gamma^i\partial_s\gamma^j,
\]
which is true by Proposition \ref{Prop_complexGnormal}. This
proves the equivalence of the first and second statements.
The equivalence of the first and third statement is shown in
a similar manner.
\end{proof*}

\section{The linear scalar quantum field}\label{Sec_Field}

It is well understood how to quantize a linear real scalar
field on any globally hyperbolic
spacetime \cite{Dimock1980,Baer+2009,Wald1994,Brunetti+2003}. In
this section we will present this quantization, with a special
focus on the case where the spacetime is stationary \cite{Kay1978}.
This extra structure allows one to obtain additional results
concerning e.g.\ ground states for the Killing flow.

As a matter of convention we will identify distributions
on $M,\ M_R$ and $\Sigma$ with distribution densities,
using the natural volume forms determined by the metrics.
To unburden our notation we will often leave the volume form
implicit, which should not lead to any confusion. However, we
point out that the volume form is important when restricting to
submanifolds, because in that case a change in volume form is
involved. We will also make use of the natural Hilbert spaces
of square-integrable functions on the various spacetimes and
Riemannian manifolds, where integration is performed with
respect to the volume forms determined by the metrics.
This understood we may leave the volume forms implicit in
our notation, writing e.g. $L^2(M)$, $L^2(\Sigma)$ instead of
$L^2(M,d\mathrm{vol}_g)$ and $L^2(\Sigma,d\mathrm{vol}_h)$.

\subsection{The classical scalar field in stationary spacetimes}\label{SSec_FFStat}

The classical theory of a linear scalar field on a spacetime
$M$ is described by the (modified) Klein-Gordon equation
for $\phi\in C^{\infty}(M)$,
\begin{equation}\label{Eqn_KG}
K\phi:=(-\Box+V)\phi=0,
\end{equation}
where $\Box:=\nabla^a\nabla_a$ denotes the Laplace-Beltrami
operator and the potential $V$ is a smooth, real-valued
function. $V$ is often chosen to be of the form
\[
V=cR+m^2,\quad m\ge 0,\ c\in\mathbb{R}
\]
with mass $m$ and scalar curvature coupling $c$. In any
globally hyperbolic spacetime, the Klein-Gordon equation has
a well posed initial value formulation (see
e.g.\ Ref.\ \cite{Baer+2009} Ch.3\ Thm.\ 3.). To formulate it we
introduce the space of initial data
\[
\mathcal{D}(\Sigma):=C_0^{\infty}(\Sigma)\oplus C_0^{\infty}(\Sigma),
\]
as a topological direct sum, where each summand carries
the test-function topology.
\begin{theorem}\label{Thm_Initval}
Let $\Sigma\subset M$ be a Cauchy surface in a globally
hyperbolic spacetime $M$ with future pointing normal vector
field $n^a$. For each $(\phi_0,\phi_1)\in\mathcal{D}(\Sigma)$
there is a unique $\phi\in C^{\infty}(M)$ such that
\begin{equation}\label{Eqn_CauchyProblem}
K\phi=0,\quad \phi|_{\Sigma}=\phi_0,\ n^a\nabla_a\phi|_{\Sigma}
=\phi_1.
\end{equation}
Moreover, $\mathrm{supp}(\phi)\subset J(\mathrm{supp}(\phi_0)
\cup\mathrm{supp}(\phi_1))$ and the linear map
$\map{S}{\mathcal{D}(\Sigma)}{C^{\infty}(M)}$ which sends
$(\phi_0,\phi_1)$ to the corresponding solution $\phi$ of
Eq.\ (\ref{Eqn_CauchyProblem}) is continuous, if
$C^{\infty}(\Sigma)$ is endowed with the usual Fr\'echet
topology.
\end{theorem}
It follows from Theorem \ref{Thm_Initval} that the Klein-Gordon
operator $K$ has unique advanced ($-$) and retarded ($+$)
fundamental solutions $E^{\pm}$ and we define $E:=E^--E^+$.

The solution map $S$ and the operator $E$ will be used
frequently to translate between the spacetime and the initial
data formulations of the theory and we note that
\begin{eqnarray}\label{Eqn_CCRInit}
E(f,f')&:=&\int_MfEf':=\int_{M^{\times 2}}d\mathrm{vol}_g(x)\
d\mathrm{vol}_g(x')f(x)E(x,x')f'(x')\nonumber\\
&=&\int_{\Sigma}Ef\cdot n^a\nabla_aEf'-n^a\nabla_aEf\cdot Ef',
\end{eqnarray}
where $\Sigma\subset M$ is any Cauchy surface and
$f,f'\in C_0^{\infty}(M)$. The kernel of $E$, acting on
$C_0^{\infty}(M)$, is exactly $KC_0^{\infty}(M)$ \cite{Dimock1980}
and for later use we introduce the real-linear space
\[
L:=C_0^{\infty}(M,\mathbb{R})/KC_0^{\infty}(M,\mathbb{R}).
\]

In a stationary, globally hyperbolic spacetime $(M,\xi)$, the
Killing vector field determines a natural time evolution. We
fix a Cauchy surface $\Sigma\subset M$ and use it to write $M$
as a standard stationary spacetime
(cf.\ Sec.\ \ref{SSec_Stationary}). We will work throughout in
adapted coordinates $x^a=(t,x^i)$ and assume that the potential
$V$ is stationary,
\[
\xi^a\nabla_aV=\partial_0V=0.
\]

As the potential $V$ is real-valued we may view $K$ as a
symmetric operator on the dense domain $C_0^{\infty}(M)$ in
$L^2(M)$. We will now separate off the canonical time
dependence of this operator and write the spatial dependence
in terms of $h_{ij}$, $N$, $N^i$ and $V$. The cleanest way
to do so is by ensuring that we obtain symmetric
operators in $L^2(\Sigma)$ for the spatial parts. For this
reason it is convenient to consider the unitary isomorphism
\[
\map{U}{L^2(M)}{L^2(\mathbb{R})\otimes L^2(\Sigma)}:f\mapsto\sqrt{N}f
\]
onto the Hilbert tensor product, where $\mathbb{R}$ is
viewed as a Riemannian manifold with the standard metric
$dt$. To see that $U$ is indeed an isomorphism we use
Schwartz Kernels Theorem, the diffeomorphism
$M\simeq\mathbb{R}\times\Sigma$ and the fact that
$\det g=-N^2\det h$ and $d\mathrm{vol}_g=Ndt\ d\mathrm{vol}_h$,
which may be seen by choosing local coordinates on $\Sigma$ that
diagonalize $h_{ij}$ at a point. The symmetric operator
$UNKNU^{-1}$ can now be written as
\begin{eqnarray}\label{Eqn_Ksplit}
N^{\frac32}KN^{\frac12}&=&N^{\frac32}(-\Box+V)N^{\frac12}\nonumber\\
&=&\partial_0^2-(\nabla^{(h)}_iN^i+N^i\nabla^{(h)}_i)\partial_0\nonumber\\
&&-N^{\frac12}\nabla^{(h)}_i(Nh^{ij}-N^{-1}N^iN^j)\nabla^{(h)}_jN^{\frac12}
+VN^2.
\end{eqnarray}
The computation that leads to this expression has been omitted,
because it is straightforward.\footnote{Instead of the Riemannian
manifold $(\Sigma,h)$ one may also consider $(\Sigma,{\tilde{h}})$,
cf.\ Theorem \ref{Thm_SStatnGH}. In this case the unitary map takes
the form $U:f\mapsto v^{\frac{d}{2}}f$ and
\begin{eqnarray}
v^{\frac{d}{2}}NKNv^{-\frac{d}{2}}&=&\partial_0^2-
(\nabla^{(\tilde{h})}_iN^i+N^i\nabla^{(\tilde{h})}_i)\partial_0
-N\nabla^{(\tilde{h})}_iv^{-2}\tilde{h}^{ij}\nabla^{(\tilde{h})}_jN
\nonumber\\
&&+N^2v^{-4}\frac{d}{2}(v(\Box_{\tilde{h}}v)+\frac{d-6}{2}\tilde{h}^{ij}
(\nabla^{(\tilde{h})}_iv)(\nabla^{(\tilde{h})}_jv))+VN^2.\nonumber
\end{eqnarray}
Although the metric $\tilde{h}$ has the advantage of being
complete, it may be a less natural choice than $h$, especially
when the spacetime $M$ is isometrically embedded into a larger
spacetime.}\label{Ft_KsplitConf}

Because $\xi$ is a Killing field, the flow $\Xi_t$ preserves
the Klein-Gordon equation: $K\Xi_t^*\phi=\Xi_t^*(K\phi)$ for
all $t\in\mathbb{R}$. Moreover, if $K\phi=0$ and $\phi$ has
compactly supported initial data on some Cauchy surface, then
the same is true for $\Xi_t^*\phi$. This means that the time
flow determines a time evolution on the initial data in
$\mathcal{D}(\Sigma)$. Indeed, let $S$ be the solution
operator of Theorem \ref{Thm_Initval} and let $S^{-1}$ be its
inverse, i.e.\
$S^{-1}(\phi)=(\phi|_{\Sigma},n^a\nabla_a\phi|_{\Sigma})$.
We may define the time evolution maps $T_t$ on
$\mathcal{D}(\Sigma)$ by $T_t:=S^{-1}\Xi^*_tS$. The
maps $T_t$ form a continuous (even smooth) one-parameter
group for $t\in\mathbb{R}$, by Theorem \ref{Thm_Initval}.
The infinitesimal generator $H_{cl}$ of the group $T_t$ is
the classical Hamiltonian:
\begin{lemma}\label{Lem_Hamiltonian}
The (classical) Hamiltonian operator $H_{cl}$ is given (in
matrix notation on $\mathcal{D}(\Sigma)$) by
\[
H_{cl}\left(\begin{array}{c}\phi_0\\ \phi_1\end{array}\right)
:=-i\partial_tT_t\left.\left(\begin{array}{c}
\phi_0\\ \phi_1\end{array}\right)\right|_{t=0}
=-i\left(\begin{array}{cc}
N^i\nabla^{(h)}_i&N\\
\nabla^{(h)}_iNh^{ij}\nabla^{(h)}_j-VN\ &\ \nabla^{(h)}_iN^i
\end{array}\right)
\left(\begin{array}{c}
\phi_0\\ \phi_1\end{array}\right).
\]
\end{lemma}
\begin{proof*}
The computation is simplified somewhat by defining
$X:=\left(\begin{array}{cc}I&0\\ N^i\nabla^{(h)}_i&
N\end{array}\right)$, with inverse
$X^{-1}:=N^{-1}\left(\begin{array}{cc}
N&0\\ -N^i\nabla^{(h)}_i&I\end{array}\right)$. Note that
$X\left(\begin{array}{c}\phi_0\\ \phi_1\end{array}\right)
=\left(\begin{array}{c}\phi_0\\ \partial_0\phi|_{\Sigma}
\end{array}\right)$, where $\phi:=S(\phi_0,\phi_1)$. Now
the first row of $XH_{cl}X^{-1}$ is simply $(0\ -iI)$ and
the second row can be be found by writing
$\partial_0^2=N^{\frac12}\partial_0^2N^{-\frac12}$ and
by eliminating the second order time derivative using
Eq.\ (\ref{Eqn_Ksplit}) and $K\phi=0$. $H_{cl}$ is then
obtained from a straightforward matrix multiplication.
The details are omitted.
\end{proof*}

For any solution $\phi\in C^{\infty}(M)$ of the
Klein-Gordon equation one defines the
stress-energy-momentum tensor
\[
T_{ab}(\phi):=\nabla_{(a}\overline{\phi}\nabla_{b)}\phi-\frac12 g_{ab}
\left(\nabla^c\overline{\phi}\nabla_c\phi+V|\phi|^2\right),
\]
which is symmetric and
\[
\nabla^aT_{ab}(\phi)=\frac{-1}{2}(\nabla_bV)|\phi|^2,
\]
because $K\phi=0$. By Killing's equation $\nabla^a\xi^b$ is
anti-symmetric, so the energy-momentum one-form
\[
P_a(\phi):=\xi^bT_{ab}(\phi)
\]
satisfies
\[
\nabla^aP_a(\phi)=\xi^b\nabla^a(T_{ab}(\phi))
=\frac{-1}{2}(\partial_0V)|\phi|^2=0,
\]
where we used the assumption that $V$ is stationary. Note
that energy-momentum is conserved, even though the
stress-tensor may not be divergence free. On a Cauchy
surface $\Sigma$ with future pointing normal $n^a$, the
energy density is defined by
\[
\varepsilon_{\Sigma}(\phi):=n^aP_a(\phi)|_{\Sigma}
=n^a\xi^bT_{ab}(\phi)|_{\Sigma}.
\]
If $\phi=S(\phi_0,\phi_1)$ for some
$(\phi_0,\phi_1)\in\mathcal{D}(\Sigma)$, then we can also define
the total energy on $\Sigma$ by
\[
\mathcal{E}(\phi):=\int_{\Sigma}\varepsilon_{\Sigma}(\phi).
\]
The conservation of $P_a(\phi)$ implies that $\mathcal{E}(\phi)$
is independent of the choice of Cauchy surface, by Stokes'
Theorem. In particular,
$\mathcal{E}(\Xi_t^*\phi)=\mathcal{E}(\phi)$ for all $t$,
because the left-hand side is the integral of
$\varepsilon_{\Sigma'}(\phi)$ over the Cauchy surface
$\Sigma':=\Xi_t(\Sigma)$.
\begin{lemma}\label{Lem_Eoperator}
Viewing $\mathcal{D}(\Sigma)$ as a dense domain in
$L^2(\Sigma)^{\oplus 2}$ we have
\[
\mathcal{E}(S(\phi_0,\phi_1))=\langle(\phi_0,\phi_1),
A(\phi_0,\phi_1)\rangle,
\]
where the operator $A$ is given by
\[
A:=\frac12\left(\begin{array}{cc}
-\nabla^{(h)}_iNh^{ij}\nabla^{(h)}_j+VN\ &\ -\nabla^{(h)}_iN^i\\
N^i\nabla^{(h)}_i&N\end{array}\right).
\]
In particular, $A=\frac{i}{2}\sigma H_{cl}$ with
$\sigma:=\left(\begin{array}{cc}0&-1\\ 1&0\end{array}\right)$
and $A\ge\frac12\left(\begin{array}{cc}VN&0\\
0&N^{-1}v^2\end{array}\right)$.
\end{lemma}
\begin{proof*}
The form of $\mathcal{E}$ can be computed by expressing the
energy density on $\Sigma$ in terms of the initial
data. The computation is straightforward, so the details are
omitted. The final equality is then obvious from Lemma
\ref{Lem_Hamiltonian}, whereas the final inequality follows
from
\begin{eqnarray}
\langle(\phi_0,\phi_1),A(\phi_0,\phi_1)\rangle&=&\frac12
\int_{\Sigma}Nh^{ij}(\nabla^{(h)}_i\overline{\phi_0}+N^{-1}
N_i\overline{\phi_1})(\nabla^{(h)}_j\phi_0+N^{-1}N_j\phi_1)
\nonumber\\
&&+VN|\phi_0|^2+(N-N^{-1}N^iN_i)|\phi_1|^2,\nonumber
\end{eqnarray}
where the first term in the integrand is non-negative and
may be dropped.
\end{proof*}

When $V>0$ everywhere, we may define an energetic inner
product on $L\otimes\mathbb{C}$ by setting
\[
\langle f,f'\rangle_e:=\langle S^{-1}Ef,AS^{-1}Ef'\rangle,
\]
where the inner product on the right-hand side is in
$L^2(\Sigma)^{\oplus 2}$. Note that $\langle,\rangle_e$
is indeed positive and non-degenerate, by the properties
of $A$ established in Lemma \ref{Lem_Eoperator} and the
positivity of $VN$ and $N^{-1}v^2$. Since $V$ is
stationary, the energetic inner product is independent of
the choice of Cauchy surface, like the energy, because
\[
\|f\|_e^2=\mathcal{E}(Ef).
\]
\begin{definition}
When $V$ is stationary and $V>0$, the \emph{energetic
Hilbert space} $\mathcal{H}_e$ is the Hilbert space
completion of $L\otimes\mathbb{C}$ in the energetic
norm.
\end{definition}
$\mathcal{H}_e$ can be interpreted as the space of all
(complex) finite energy solutions of the Klein-Gordon
equation (\ref{Eqn_KG}).

The following detailed description of the energetic
Hilbert space is the main result of this section. The
proof makes use of strictly positive operators and we
have collected some basic results on such operators in
\ref{App_SP} (see also Ref.\ \cite{Kay1985}).
\begin{theorem}\label{Thm_Hinvertable}
Let $M$ be a stationary, globally hyperbolic spacetime
with a Cauchy surface $\Sigma$ and assume that $V$ is
stationary and $V>0$. Let $\hat{A}$ denote the
Friedrichs extension of the operator $A$ of Lemma
\ref{Lem_Eoperator}. The linear map
$\map{q_{cl}}{\mathcal{D}(\Sigma)}{L^2(\Sigma)^{\oplus 2}}$
defined by $q_{cl}(\phi_0,\phi_1):=\sqrt{\hat{A}}
\left(\begin{array}{c}\phi_0\\ \phi_1\end{array}\right)$
is continuous, injective, has dense range, commutes with
complex conjugation and satisfies
$\|q_{cl}(\phi_0,\phi_1)\|^2=\mathcal{E}(S(\phi_0,\phi_1))$.
Hence, $\mathcal{H}_e\simeq L^2(\Sigma)^{\oplus 2}$.

There is a unique, strongly continuous unitary group
$O_t=e^{itH_e}$ on $L^2(\Sigma)^{\oplus 2}$ such that
$O_tq_{cl}=q_{cl}T_t$. Its infinitesimal generator
is given by $H_e:=2i\sqrt{\hat{A}}\sigma\sqrt{\hat{A}}$.
$iH_e$ commutes with complex conjugation, $H_e$ and all
its powers $H_e^n$, $n\in\mathbb{N}$, are essentially
self-adjoint on the range of $q_{cl}$, $H_e$ is
invertible and the range of $q_{cl}$ is a core for
$|H_e|^{-1}$.
\end{theorem}
The explicit characterization of $\mathcal{H}_e$ in terms
of $L^2(\Sigma)^{\oplus 2}$ is often very useful, although
it is less aesthetically appealing, because it requires the
choice of an arbitrary Cauchy surface $\Sigma$.
\begin{proof*}
We first consider the Friedrichs extension $\hat{A}$
of $A$, which is a positive, self-adjoint operator. By
Lemma \ref{Lem_Friedrichs}, $\mathcal{D}(\Sigma)$ is a
core for $\hat{A}^{\frac12}$. Furthermore,
$\hat{A}\ge B$, where the operator
$B:=\frac12\left(\begin{array}{cc}VN&0\\
0&N^{-1}v^2\end{array}\right)$ is defined on
$\mathcal{D}(\Sigma)$ (cf.\ Lemma \ref{Lem_Eoperator}).
Note that $B$ is essentially self-adjoint with a strictly
positive closure, by Proposition \ref{Prop_mul}. Hence,
$\hat{A}$ is also strictly positive, by Lemma
\ref{Lem_SP3}, and $\mathcal{D}(\Sigma)$ is in the domain
of $\hat{A}^{-\frac12}$. Moreover, as
$\mathcal{D}(\Sigma)$ is a core for $\hat{A}^{\frac12}$,
the latter has a dense range on $\mathcal{D}(\Sigma)$.
Therefore, $q_{cl}$ is a well defined, injective linear
map with dense range $\mathcal{R}$ and by Lemma
\ref{Lem_Eoperator}, $\|q_{cl}(\phi_0,\phi_1)\|^2=
\mathcal{E}(S(\phi_0,\phi_1))$. As $S$ is continuous,
the last equation also entails the continuity of
$q_{cl}$. (Alternatively one may use Theorem
\ref{Thm_Kernel} of \ref{App_SP}.) Also note
that $A$ commutes with complex conjugation in
$L^2(\Sigma)^{\oplus 2}$, hence the same is true for
$\hat{A}^{\frac12}$ and for $q_{cl}$.

Because $q_{cl}$ is invertible we may define $O_t$ by
$O_t=q_{cl}T_tq_{cl}^{-1}$ on $\mathcal{R}$. Note that
the total energy
$\|O_tq_{cl}(\phi_0,\phi_1)\|^2=\mathcal{E}(\Xi_t^*S(\phi_0,\phi_1))$
is independent of $t$, so each $O_t$ is a densely defined
isometry, which extends uniquely to a unitary isomorphism
on the entire Hilbert space, again denoted by $O_t$.
$O_t^{-1}=O_{-t}$ and the continuity of $f\mapsto T_tf$
in the test-function topology entails the strong
continuity of $O_t$.

Because the time-derivative of $T_t(\phi_0,\phi_1)$
converges in the test-function topology of
$\mathcal{D}(\Sigma)$ and $q_{cl}$ is continuous, the
infinitesimal generator of $O_t$ is well defined on
the range $\mathcal{R}$ of $q_{cl}$, where it is given
by
\[
H_e=q_{cl}H_{cl}q_{cl}^{-1}=
2i\sqrt{\hat{A}}\sigma\sqrt{\hat{A}},
\]
because of the Lemmas \ref{Lem_Hamiltonian},
\ref{Lem_Eoperator}. Both $H_e$ and $O_t$ preserve
$\mathcal{R}$, so $H_e$ and all its powers are
essentially self-adjoint on $\mathcal{R}$ by
Lemma 2.1 in Ref.\ \cite{Chernoff1973}.

$\hat{A}$ commutes with complex conjugation, so it is
clear that $iH_e$ also commutes with it. Furthermore,
the map
$M:=\frac{i}{2}\hat{A}^{-\frac12}\sigma\hat{A}^{-\frac12}$
is well defined on $\mathcal{R}$ and it satisfies
$MH_e=I$ there. Note that $M$ is closable, because it
is symmetric and densely defined. By Lemma
\ref{Lem_Cl1}, $H_e$ must be invertible. Lemma
\ref{Lem_SP2} implies that $H_e^{-1}$ is self-adjoint
and invertible and a core is given by
$H_e\mathcal{R}\subset\mathcal{R}$. As $M$ is a
symmetric extension of $H_e^{-1}$ on this domain,
we must have $\overline{M}=H_e^{-1}$ and the domain
$\mathcal{R}$ of $M$ is a core for $H_e^{-1}$ and
hence also for $|H_e|^{-1}$, by the Spectral
Calculus Theorem.
\end{proof*}

\subsection{The scalar quantum field in stationary spacetimes
and equilibrium one-particle structures}\label{SSec_Ground}

We now study the quantised scalar field in a stationary spacetime,
where the ground states play a similarly important role for the
theory as the vacuum state in Minkowski spacetime. Because of the
importance of quasi-free equilibrium states (cf.\ Sec.\ \ref{Sec_Alg})
we first focus on equilibrium one-particle structures, whereas
the ground and equilibrium states (beyond their two-point
distributions) will be discussed in Section \ref{Sec_GroundProp}
below.

The well posedness of the Cauchy problem established in
Theorem \ref{Thm_Initval} remains true if we specify arbitrary
distributional initial data, allowing distributional solutions
and using distributional topologies \cite{Sanders+2012}. In this
setting it is natural to introduce local observables, associated to
arbitrary $f\in C_0^{\infty}(M)$, which measure the distributional
field $\phi$ by the formula $\phi(f):=\int\phi f$. These observables
$\phi\mapsto\phi(f)$ can be regarded as functions on the space of
classical solutions $\phi$ and we may use them to generate an algebra
of observables. We choose to work with the Weyl $C^*$-algebra
$\alg{W}^{cl}:=\alg{W}(L,0)$, whose elements we interpret as
$e^{i\phi(f)}$, which remains bounded when $\phi$ and $f$ are
real-valued.

Interpreting the right-hand side of Eq.\ (\ref{Eqn_CCRInit})
in terms of initial values and momenta motivates the
introduction of the symplectic space $(L,E)$, so that the
corresponding quantum theory is described by
$\alg{W}:=\alg{W}(L,E)$. For each open subset $O\subset M$
we will denote by $\alg{W}(O)$ the $C^*$-subalgebra generated
by those $W(f)$ with $f$ supported in $O$ (and similarly for
$\alg{W}^{cl}(O)$). In this way one obtains a net of local
$C^*$-algebras \cite{Haag,Brunetti+2003}.

When $(M,\xi)$ is a stationary, globally hyperbolic
spacetime and $V$ is stationary, $(L,0,T_{-t})$ and
$(L,E,T_{-t})$ become one-particle dynamical systems.
This follows from the fact that $\Xi_{-t}^*$ preserves the
metric and that the $E^{\pm}$ are unique, so the symplectic
form $E(f,f'):=\int_MfEf'$ is preserved. We may consider
the associated quasi-free dynamical systems
$(\alg{W}^{cl},\alpha_t^{cl})$ and
$(\alg{W},\alpha_t)$, so that
\[
\alpha^{cl}_t(W(f))=W(\Xi_{-t}^*f),\quad
\alpha_t(W(f))=W(\Xi_{-t}^*f)
\]
for all $f\in L$.\footnote{The sign in $T_{-t}$ is explained
by the desire to have $\alpha^{cl}_t\phi=\Xi_t^*\phi$ for the
field $\phi$, so that
$\alpha^{cl}_t(\phi(f))=(\Xi_t^*\phi)(f)=\phi(\Xi_{-t}^*f)$
in the distributional perspective. The same argument
applies to the quantum case.}\label{Ft_Sign}
$\alpha^{cl}_t$ and $\alpha_t$ describe
the Killing time flow at an algebraic level and we note
that $\alpha_t(\alg{W}(O))=\alg{W}(\Xi_t(O))$ and similarly
in the classical case. However, neither $\alpha^{cl}_t$ nor
$\alpha_t$ is norm-continuous in $t$, as $\|w(f)-w(g)\|=2$
for all $f\not=g\in L$ (Ref.\ \cite{Binz+2004} Prop.\ 3-10).
For this reason, general results on $C^*$-dynamical
systems \cite{Bratteli+,Sakai1991} do not
apply directly to our situation. (Nor can we view
$(\alg{W},\alpha_t)$ as a $W^*$-dynamical system, because
$\alg{W}$ is not a $W^*$-algebra or von Neumann algebra.)

In order to take advantage of the smoothness of the
time evolution maps $T_t$ on $\mathcal{D}(\Sigma)$ we need
the following definition.
\begin{definition}\label{Def_DkState}
We call a state $\omega$ on the Weyl $C^*$-algebra $\alg{W}$
(or $\alg{W}^{cl}$) $D^k$, $k> 0$, when it is $C^k$
(cf.\ Def.\ \ref{Def_CkState}) and the maps
\[
\omega_n(f_1,\ldots,f_n):=(-i)^n\partial_{s_1}\cdots\partial_{s_n}
\omega(W(s_1f_1)\cdots W(s_nf_n))|_{s_1=\ldots =s_n=0}
\]
are \emph{distributions} on $M^{\times n}$ for all
$1\le n\le k$. The $\omega_n$ are called the $n$-point
distributions. A state is called \emph{regular}, or
$D^{\infty}$, when it is $D^k$ for all $k>0$.
\end{definition}
In our setting the distributional character of the $\omega_n$
is natural and useful.

\begin{remark}
An alternative description of the scalar quantum field
uses the $^*$-algebra $\alg{A}$, generated by the identity
$I$ and the smeared field operators $\Phi(f)$,
$f\in C_0^{\infty}(M)$, satisfying
\begin{enumerate}
\item[\textup{(i)}] $f\mapsto\Phi(f)$ is $\mathbb{C}$-linear,
\item[\textup{(ii)}] $\Phi(f)^*=\Phi(\overline{f})$,
\item[\textup{(iii)}] $K\Phi(f):=\Phi(Kf)=0$,
\item[\textup{(iv)}] $[\Phi(f),\Phi(f')]=iE(f,f')I$.
\end{enumerate}
Although the algebras $\alg{A}$ and $\alg{W}$ are
technically different, their relation can be
understood from a physical point of view by formally
setting $W(f)=e^{i\Phi(f)}$. In suitable representations
this can be made rigorous. This applies in particular
to regular states $\omega$ on $\alg{W}$, which give
rise to a corresponding state on $\alg{A}$.
\end{remark}

\subsubsection{Two-point distributions}

When $\omega$ is a $D^2$ state on $\alg{W}$, we may
identify the one-particle structure $(p,\mathcal{K})$ of
$\omega_2$ as a map into a subspace of the
GNS-representation space $\mathcal{H}_{\omega}$, as in
the proof of Proposition \ref{Prop_EqToOnePEq}. A similar
construction applies to the so-called truncated two-point
distribution,
$\omega^T_2(x,x'):=\omega_2(x,x')-\omega_1(x)\omega_1(x')$,
where we now take
$p(f):=\pi_{\omega}(\Phi(f)-\omega_1(f)I)\Omega_{\omega}$.
Note that $\omega^T_2$ is indeed a two-point distribution,
(cf.\ Theorem \ref{Thm_Twopoint}) and that $\omega_2=\omega^T_2$
when $\omega_1=0$, so in that case the two constructions
coincide.

When $\omega_2$ is a distribution, the associated
one-particle structure can be viewed as a
$\mathcal{K}$-valued distribution $p$ which satisfies the
Klein-Gordon equation \cite{Strohmaier+2002}.
(Conversely, when $p$ is a distribution, the associated
$\omega_2$ is also a distribution.) For any Cauchy surface
$\Sigma$, $p$ is uniquely determined by its initial data,
which form a continuous linear map
$\map{q_{\Sigma}}{\mathcal{D}(\Sigma)}{\mathcal{K}}$
with dense range and such that
\[
\langle q_{\Sigma}(\overline{\phi_0},\overline{\phi_1}),
q_{\Sigma}(\phi'_0,\phi'_1)\rangle-
\langle q_{\Sigma}(\overline{\phi'_0},\overline{\phi'_1}),
q_{\Sigma}(\phi_0,\phi_1)\rangle
=i\int_{\Sigma}\phi_0\phi'_1-\phi_1\phi'_0
\]
(cf.\ Eq.\ (\ref{Eqn_CCRInit})). Conversely, any such
linear map $q_{\Sigma}$ determines a unique one-particle
structure. Indeed, just like smooth solutions to the
Klein-Gordon equation, two-point distributions are
uniquely determined by their initial data on a Cauchy
surface:
\begin{proposition}\label{Prop_StateInitval}
Let $\Sigma\subset M$ be a Cauchy surface in a globally
hyperbolic spacetime with future pointing normal $n^a$ and
let $\omega$ be a distribution density in $M^{\times 2}$.
If $K_x\omega(x,y)=K_y\omega(x,y)=0$, then the restrictions
\[
\omega_{ij}:=(n^a\nabla_a)_x^i(n^b\nabla_b)_y^j
\omega|_{\Sigma^{\times 2}}
\]
are well defined distribution densities in
$\Sigma^{\times 2}$ for all $i,j\in\mathbb{N}$.

Conversely, for any four distribution densities
$\omega_{ij}$, $0\le i,j\le 1$, on $\Sigma^{\times 2}$,
there is a unique distribution density $\omega$ on
$M^{\times 2}$ such that
\begin{equation}\label{Eqn_CauchyProblem2}
K_x\omega=K_y\omega=0,\quad
(n^a\nabla_a)_x^i(n^b\nabla_b)_y^j\omega|_{\Sigma^{\times 2}}
=\omega_{ij}.
\end{equation}
\end{proposition}
Support and continuity properties analogous to Theorem
\ref{Thm_Initval} also hold, but we will not need them.
We omit the proof of this basic result.

There is a preferred class of $D^2$ states, called
Hadamard states, which are characterised by the fact that
their two-point distribution has a singularity structure
that is of the same form as for the Minkowski vacuum state.
These states are important, because the renormalised Wick
powers and stress tensor of the quantum field have finite
expectation values in them. To put it more precisely,
$\omega_2$ is of Hadamard form if and only
if \cite{Radzikowski1996}
\begin{eqnarray}\label{Eqn_DefHad}
WF(\omega_2)&=&\left\{(x,k;y,l)\in T^*M^{\times 2}|\ l\not=0 \mathrm{\ is\
future\ pointing\ and\ light-like\ and\ }\right.\nonumber\\
&&(y,l) \mathrm{\ generates\ a\ geodesic\ } \gamma
\mathrm{\ which\ goes\ through\ } x \mathrm{\ with\ tangent}\nonumber\\
&&\left.\mathrm{vector\ } -k\right\}.
\end{eqnarray}
This condition is already implied by one of the following
apparently weaker, and often more convenient, estimates on
$\omega_2$ or its associated one-particle structure
$(p,\mathcal{K})$:
\[
WF(\omega_2)\subset V^-M\times V^+M,\quad WF(p)\subset V^+M,
\]
where $V^{\pm}M\subset T^*M$ is the space of future ($+$)
or past ($-$) pointing causal co-vectors on $M$
(cf.\ Ref.\ \cite{Strohmaier+2002}, Prop.\ 6.1). For any
regular state (even if it is not quasi-free) the Hadamard
condition allows one to estimate the singularity structure of
all higher $n$-point distributions too \cite{Sanders2010}, so
that the state satisfies the microlocal spectrum condition of
Ref.\ \cite{Brunetti+1995}.

By the Propagation of Singularities Theorem and the fact
that $\omega_2$ solves the Klein-Gordon equation in both
variables it suffices to check the condition in
Eq.\ (\ref{Eqn_DefHad}) on a Cauchy surface $\Sigma$:
\[
WF(\omega_2)|_{\Sigma}\subset\left\{(x,-k;x,k)|\ (x,k)\in V^+M|_{\Sigma}\right\}.
\]
Unfortunately it is somewhat complicated to see whether a
state $\omega_2$ is Hadamard by inspecting its initial data
on a Cauchy surface $\Sigma$. The initial data of $\omega_2$
should be smooth away from the diagonal in $\Sigma^{\times 2}$,
so it suffices to characterize the singularities on the
diagonal. However, for the singularities on the diagonal we
are not aware of any argument that avoids the use of the
Hadamard parametrix construction, which involves the Hadamard
series for which Hadamard states were originally named.

\subsubsection{Equilibrium two-point distributions}

An equilibrium one-particle structure $(p,\mathcal{K})$
has some nice additional structure when $p$ is a
distribution:
\begin{lemma}\label{Lem_EqStruc}
If $(p,\mathcal{K})$ is an equilibrium one-particle
structure such that $p$ is a distribution, then the
unitary group $\tilde{O}_t$ on $\mathcal{K}$ defined
by $\tilde{O}_tp=p\Xi_{-t}^*$ (on $C_0^{\infty}(M)$)
is strongly continuous, $\tilde{O}_t=e^{itH}$. Its
strong derivative is well defined on the range of $p$,
$H$ is essentially self-adjoint on this range and
$Hp(f)=ip(\partial_0f)$ for all $f\in C_0^{\infty}(M)$.
\end{lemma}
\begin{proof*}
The strong continuity of $\tilde{O}_t$ follows from the
continuity of $t\mapsto\Xi_{-t}^*f$ in the test-function
topology and the fact that $p$ is a distribution. The
formula for $H$ on the range of $p$ can be deduced from
the continuity of $p$ by a direct calculation:
\[
Hp(f):=-i\partial_t\tilde{O}_tp(f)|_{t=0}
=-i\partial_t p(\Xi_{-t}^*(f))|_{t=0}=ip(\partial_0f).
\]
The essential self-adjointness of $H$ on the range of
$p$ then follows from Chernoff's Lemma \cite{Chernoff1973}.
\end{proof*}

The next two results are the main results of this section.
They are existence and uniqueness results for non-degenerate
ground and $\beta$-KMS one-particle structures. For the
existence of a non-degenerate ground we adapt a result of
Ref.\ \cite{Kay1978}, which imposed additional restrictions
on the potential $V$ and on the Killing field in order to
obtain such a ground one-particle structure with, in
addition, a mass gap. For the existence of a non-degenerate
$\beta$-KMS one-particle structure see
Refs.\ \cite{Kay1985_1,Rocca+1970}.
\begin{theorem}\label{Thm_GroundEx}
Let $M$ be a globally hyperbolic, stationary spacetime
and consider a linear scalar field with a stationary
potential $V$ such that $V>0$.
\begin{enumerate}
\item[\textup{(i)}] There exists a non-degenerate ground one-particle
structure $(p_0,\mathcal{K}_0)$, with
$\mathcal{K}_0\subset\mathcal{H}_e$ the closed range of
\[
p_0(f):=\sqrt{2}|H_e|^{-\frac12}P_-p_{cl}(f),
\]
where $P_-$ is the spectral projection onto the negative
part of the spectrum of $H_e$ and
$p_{cl}(f):=q_{cl}S^{-1}E(f)$.
\item[\textup{(ii)}] For every $\beta>0$ there exists a non-degenerate
$\beta$-KMS one-particle structure
$(p_{(\beta)},\mathcal{K}_{(\beta)})$, with
$\mathcal{K}_{(\beta)}\subset\mathcal{H}_e^{\oplus 2}$
the closed range of
\begin{eqnarray}
p_{(\beta)}(f)&=&
\sqrt{2}P_-|H_e|^{-\frac12}(I-e^{-\beta |H_e|})^{-\frac12}p_{cl}(f)
\oplus\nonumber\\
&&\sqrt{2}P_+|H_e|^{-\frac12}e^{-\frac{\beta}{2}|H_e|}
(I-e^{-\beta |H_e|})^{-\frac12}p_{cl}(f).\nonumber
\end{eqnarray}
\end{enumerate}
\end{theorem}
The occurrence of $P_-$, rather than $P_+$, is in line
with the footnote on page \pageref{Ft_Sign}.
\begin{proof*}
We start with the $\mathcal{H}_e$-valued distribution
$p_{cl}(f):=q_{cl}S^{-1}E(f)$ and the unitary group $O_t$
determined by Theorem \ref{Thm_Hinvertable}.
Define $p_0(f):=\sqrt{2}|H_e|^{-\frac12}P_-p_{cl}(f)$ and
let the closed range of $p_0$ be denoted by
$\mathcal{K}_0$. It is not hard to see that
$O_tp_0(f)=p_0(\Xi^*_tf)$, so $O_t$ preserves $\mathcal{K}_0$
and we may let $\tilde{O}_t:=O_{-t}|_{\mathcal{K}}$. The
generator $H$ of this strongly continuous unitary group is
the restriction of $-H_e$, which is strictly positive there.
The range of $p_0$ is in the domain of $H$ and $H^{-\frac12}$,
by Theorem \ref{Thm_Hinvertable}. If we let $C$ denote the
complex conjugation on $L^2(\Sigma)^{\oplus 2}$, then
$CH_eC=-H_e$, so $CP_-C=P_+$, the spectral projection onto
the positive part of the spectrum of $H_e$. Thus,
\begin{eqnarray}
\langle p_0(\overline{f}),p_0(f')\rangle
&=&2\langle p_{cl}(\overline{f}),|H_e|^{-1}P_-p_{cl}(f')\rangle
=-2\langle CH_e^{-1}P_-p_{cl}(f'),Cp_{cl}(\overline{f})\rangle\nonumber\\
&=&2\langle H_e^{-1}P_+p_{cl}(\overline{f'}),p_{cl}(f)\rangle\nonumber\\
&=&2\langle p_{cl}(\overline{f'}),|H_e|^{-1}P_-p_{cl}(f)\rangle
+2\langle p_{cl}(\overline{f'}),H_e^{-1}p_{cl}(f)\rangle\nonumber\\
&=&\langle p_0(\overline{f'}),p_0(f)\rangle
+i\langle S^{-1}E\overline{f'},\sigma S^{-1}Ef\rangle
=\langle p_0(\overline{f'}),p_0(f)\rangle-iE(f,f').\nonumber
\end{eqnarray}
This proves that $(p_0,\mathcal{K}_0)$ is a non-degenerate
ground one-particle structure.

The formula for $p_{(\beta)}$ is well defined, because the
range of $p_{cl}$ is in the domain of $|H_e|^{-1}$ by Theorem
\ref{Thm_Hinvertable}. It defines a $\mathcal{K}_{(\beta)}$-valued
distribution with dense range, which solves the Klein-Gordon
equation. Just like for the ground one-particle structure one
may check that
$\langle p_{(\beta)}(\overline{f}),p_{(\beta)}(f')\rangle
-\langle p_{(\beta)}(\overline{f'}),p_{(\beta)}(f)\rangle=iE(f,f')$,
so $(p_{(\beta)},\mathcal{K}_{(\beta)})$ does indeed define a
one-particle structure.

Viewing $\mathcal{K}_{(\beta)}$ as a subspace of
$\mathcal{H}_e^{\oplus 2}$ we note that $O_t^{\oplus 2}$ preserves
the range of $p_{(\beta)}$, because
$O_t^{\oplus 2}p_{(\beta)}(f)=p_{(\beta)}(\Xi_t^*f)$. We can
therefore define a strongly continuous unitary group $\tilde{O}_t$
on $\mathcal{K}^{(\beta)}$ as the restriction of
$O_{-t}^{\oplus 2}$. The generator $H$ of $\tilde{O}_t$ is given
by the restriction of $|H_e|\oplus -|H_e|$ and the range of
$p_{(\beta)}$ is contained in $D(e^{-\frac{\beta}{2}H})$. One may
then compute
\begin{eqnarray}\label{Eqn_StrongOnePKMS}
&&\langle e^{-\frac{\beta}{2}H}p_{(\beta)}(\overline{f}),
e^{-\frac{\beta}{2}H}p_{(\beta)}(f')\rangle\nonumber\\
&=&\langle p_{cl}(\overline{f}),|H_e|^{-1}(I-e^{-\beta|H_e|})^{-1}
(P_++e^{-\beta|H_e|}P_-)p_{cl}(f')\rangle\nonumber\\
&=&\langle p_{(\beta)}(\overline{f'}),p_{(\beta)}(f)\rangle.
\end{eqnarray}
This implies the one-particle KMS-condition, because for any
$f,f'\in C_0^{\infty}(M,\mathbb{R})$ the function
\[
F_{ff'}(z):=\langle e^{-\frac{i}{2}\overline{z}H}p_{(\beta)}(\overline{f}),
e^{\frac{i}{2}zH}p_{(\beta)}(f')\rangle
\]
is bounded and continuous on $\overline{\mathrm{S}_{\beta}}$ and
holomorphic in its interior. The correct boundary conditions
follow from Eq.\ (\ref{Eqn_StrongOnePKMS}).
\end{proof*}
As $(p_0,\mathcal{K}_0)$ is non-degenerate, the associated
quasi-free state is non-degenerate too (Proposition \ref{Prop_EqToOnePEq})
and hence it is pure (by Borchers' Theorem \ref{Thm_Borchers}).
We then see from Theorem \ref{Thm_KayUniqueness} that $p_0$
already has dense range on the real subspace. (Of course a
direct proof of this fact is also possible.)

% Eq.\ (\ref{Eqn_StrongOnePKMS}) implies the one-particle KMS-condition,
% but the converse is less obvious.

% In the light of Section \ref{SSec_Gibbs} the quasi-free $\beta$-KMS
% state is probably degenerate: on compact Cauchy surfaces $H$ has a
% discrete spectrum and if $\psi_{\pm}$ are eigenvectors of $H$ with
% opposite eigenvalues, then
% $(H\otimes I+I\otimes H)(\psi_+\otimes\psi_-+\psi_-\otimes\psi_+)=0$.

\begin{remark}
Note that there is a connection between the classical energy
and the Hamiltonian operator $H_0$ in the ground one-particle
structure, which is given by
\[
\langle p_0(f),H_0p_0(f)\rangle+\langle p_0(\overline{f}),
H_0p_0(\overline{f})\rangle=2\mathcal{E}(Ef),
\]
as may be shown by the same techniques employed in the proof of
Theorem \ref{Thm_GroundEx}.
\end{remark}

% If $f$ is real-valued one verifies by a direct computation that
% \[
% \omega^{(\beta)}_2(f,f)=\langle p_0(f),\coth(\frac{\beta}{2}H_0)
% p_0(f)\rangle,
% \]
% where the complex conjugation of $\mathcal{H}_e\simeq L^2(\Sigma)$
% is used to eliminate the term involving $P_+$ in favor of a term
% involving $P_-$. Hence,
% \[
% \omega^{(\beta)}_2(f,f)-\omega^{0}_2(f,f)=
% 2\langle p_0(f),e^{-\beta H_0}(I-e^{-\beta H_0})^{-1}p_0(f)\rangle,
% \]
% exhibiting the characteristic behaviour of Bose-Einstein
% statistics (cf.\ Ref.\ \cite{Bratteli+} Sec.\ 5.2.5 and 5.3.)). An
% alternative formula\footnote{We thank Christoph Solveen for bringing
% this derivation to our attention.} which exhibits the characteristic
% Bose-Einstein statistics can be obtained by Fourier transformation
% of $J(t):=E(f,\Xi_t^*f')$, which leads to
% \[
% i\widehat{J}(-k)=\widehat{F}_{ff'}(k)-e^{-\beta k}\widehat{F}_{ff'}(k)
% \]
% by the KMS-condition. Solving this for $\widehat{F}$, performing an
% inverse Fourier transform and a change of variables ($k\rightarrow -k$)
% yields
% \[
% \omega^{(\beta)}_2(f,f')=F(0)=\frac{1}{2\pi i}\int dk\
% \widehat{J}(k)\frac{e^{-\beta k}}{1-e^{-\beta k}}.
% \]

Next we establish a uniqueness result for non-degenerate ground
and $\beta$-KMS one-particle
structures \cite{Kay1979,Kay1985_1}.\footnote{Our uniqueness
result is a slight strengthening of the results of
Refs.\ \cite{Kay1979,Kay1985_1}, in our setting, because our
definition of $\beta$-KMS one-particle structures is
slightly less stringent and we provide a bit more detail on
degenerate one-particle structures. Note that
Ref.\ \cite{Kay1979} formulates and proves uniqueness in the
class of non-degenerate ground one-particle structures, for which
$p$ already has a dense range on the real linear space $L$
(which entails that the associated quasi-free ground state
is pure by Theorem \ref{Thm_KayUniqueness}). That this
extra condition is not needed for the proof was pointed
out by the same author in Ref.\ \cite{Kay1985_1}, which
also proves uniqueness of non-degenerate $\beta$-KMS
one-particle structures.}
\begin{proposition}\label{Prop_OnePuniqueness}
Let $(p_2,\mathcal{K}_2,\tilde{O}^{(2)}_t)$ be a ground,
resp.\ $\beta$-KMS, one-particle structure (with $\beta>0$) and
let $P_2$ be the orthogonal projection onto the space of
$\tilde{O}_t^{(2)}$-invariant vectors. Let
$(p_1,\mathcal{K}_1,\tilde{O}^{(1)}_t)$ be the non-degenerate
ground, resp.\ $\beta$-KMS, one-particle structure of
Theorem \ref{Thm_GroundEx}. Then there is a unique isometry
$\map{U}{\mathcal{K}_1}{\mathcal{K}_2}$ such that
$\tilde{O}^{(2)}_tU=U\tilde{O}^{(1)}_t$ and $Up_1=(I-P_2)p_2$.
In particular, if $P_2=0$, then $U$ is an isomorphism.

Let $w:=\omega_2^{(2)}-\omega_2^{(1)}$ denote the difference
of the associated two-point functions $\omega_2^{(i)}$. Then
$w$ is a real-valued, symmetric (weak) bi-solution to the
Klein-Gordon equation which is of positive type and independent
of the Killing time (in both entries). If $p_2$ is a distribution
on $M$, then $w\in C^{\infty}(M^{\times 2})$.
\end{proposition}
\begin{proof*}
The proof follows Ref.\ \cite{Kay1979} (see also Ref.\ \cite{Weinless1969}).
For arbitrary $f,f'\in C_0^{\infty}(M,\mathbb{R})$ the function
\begin{eqnarray}
F(t)&:=&\langle p_2(f),\tilde{O}^{(2)}_tp_2(f')\rangle_{\mathcal{K}_2}
-\langle p_1(f),\tilde{O}^{(1)}_tp_1(f')\rangle_{\mathcal{K}_1}\nonumber\\
&=&\langle p_2(f),e^{itH_2}p_2(f')\rangle_{\mathcal{K}_2}
-\langle p_1(f),e^{itH_1}p_1(f')\rangle_{\mathcal{K}_1}\nonumber
\end{eqnarray}
is continuous (by the Definition \ref{Def_GroundOneP} of ground
and $\beta$-KMS one-particle structures) and real-valued on
$\mathbb{R}$. Suppose both one-particle structures satisfy the
one-particle $\beta$-KMS condition at the same $\beta>0$. There
is then a bounded continuous extension $\tilde{F}$ of $F$ to
$\overline{\mathrm{S}_{\beta}}$, holomorphic in the interior.
By repeatedly applying Schwarz' reflection
principle \cite{Berenstein+1991}, $\tilde{F}$ extends to a bounded
holomorphic function on all of $\mathbb{C}$, which means that
$\tilde{F}$ and $F$ are constant, by Liouville's Theorem \cite{Berenstein+1991}.
Similarly, if both are ground one-particle structures, the positivity of
the infinitesimal generators $H_i$ implies that there is a bounded,
holomorphic function $F_+$ in the upper half plane, which has
$F$ as its boundary value. By Schwarz' reflection principle,
$F_+$ can be extended to a bounded holomorphic function on the
entire plane, which again means that $F$ is constant.

Note that the range of $p_1$ is in the domain of $H_1$, because
the strong derivative $\partial_t\tilde{O}^{(1)}_tp_1(f)|_{t=0}$
exists (cf.\ Theorem \ref{Thm_GroundEx}). The same is true for
$p_2$ and $H_2$, because
$\|(\tilde{O}^{(2)}_t-I)p_2(f)\|^2-\|(\tilde{O}^{(1)}_t-I)p_1(f)\|^2\equiv 0$,
by the previous paragraph. The constancy of $F$ implies
$\partial_t^2F|_{t=0}=0$, i.e.
\[
\langle p_1(f),H_1^2p_1(f')\rangle_{\mathcal{K}_1}
=\langle p_2(f),H_2^2p_2(f')\rangle_{\mathcal{K}_2}.
\]
This equality must hold for all $f,f'\in C_0^{\infty}(M)$, by
complex (anti-)linearity. We may therefore define linear maps
$X_i:=H_ip_i$ and we let $V_i:=\mathrm{ker}(X_i)$ denote
their kernels. By the previous equation, $V_1=V_2=:V$, so
the $X_i$ descend to linear injections
$\map{\tilde{X}_i}{C_0^{\infty}(M)/V}{\mathcal{K}_i}$.
We set $U:=\tilde{X}_2\tilde{X}_1^{-1}$ between the ranges
of the $X_i$. It is obvious from the previous paragraph
that $U$ is an isometry, because $UH_1p_1=H_2p_2$. The
non-degeneracy of the first one-particle structure
implies that $H_1$ is injective, while the range of $p_1$ is
a core for it. It follows that the map $\tilde{X}_1$ has a
dense range, so $U$ extends by continuity to an isometry from
$\mathcal{K}_1$ into $\mathcal{K}_2$. Note that $U$
intertwines between the unitary groups, because
$\tilde{O}_t^{(i)}H_ip_i(f)=H_ip_i(\Xi_{-t}^*f)$. Hence
$UH_1=H_2U$ and $P_2UH_1=(P_2H_2)U=0$, which means that
$P_2U=0$, because $H_1$ has a dense range. Let $R$ be the
unique linear map such that $RP_2=0$ and $RH_2=I-P_2$. Then
$U=RH_2U=RUH_1$ and $Up_1=RUH_1p_1=RH_2p_2=(I-P_2)p_2$. The
uniqueness of $U$ is then obvious, as $p_1$ has a dense
range.

By construction, $w:=\omega_2^{(2)}-\omega_2^{(1)}$ is a
real-valued, symmetric bi-solution to the Klein-Gordon
equation (in a weak sense). Moreover, as $U$ is isometric
and $Up_1=(I-P_2)p_2$,
\[
w(\overline{f},f)=\|p_2(f)\|^2-\|Up_1(f)\|^2=\|P_2p_2(f)\|^2
\ge 0,
\]
so $w$ is of positive type. For fixed $f,f'\in C_0^{\infty}(M)$,
$w(\overline{f},\Xi_{-t}^*f')=F(t)=w(\overline{\Xi_t^*f},f')$
is constant, as we saw in the first paragraph of this proof.
If $p_2$ is a distribution on $M$, then $w$ is a distribution on
$M^{\times 2}$ and, in adapted coordinates,
$\partial_0w=\partial'_0w=0$. The equation $K_xK_{x'}w=0$ then
reduces to an elliptic equation on $\Sigma^{\times 2}$, which
implies that $w$ is smooth (see e.g.\ Ref.\ \cite{Hoermander}
Thm.\ 8.3.1).
\end{proof*}

\begin{remark}\label{Rem_AllHad}
Proposition \ref{Prop_OnePuniqueness} shows in particular that
there is at most one non-degenerate ground one-particle
structure and at most one non-degenerate $\beta$-KMS one-particle
structure at any fixed $\beta>0$, up to unitary equivalence.
These are the ones of Theorem \ref{Thm_GroundEx}. The degenerate
ones may be classified in terms of $w$. In spacetimes with a
compact Cauchy surface $\Sigma$ we note that the only smooth
function $w$ with the stated properties is $w=0$. Indeed,
for any fixed $y\in\Sigma$, $v_y(x):=w(x,y)$ solves $Cv_y=0$ for
$C:=-\nabla^{(h)}_i(Nh^{ij}-N^{-1}N^iN^j)\nabla^{(h)}_i+VN$.
(This is because $w$ solves the Klein-Gordon equation and is
Killing time independent.) As
$0=\langle v_y,Cv_y\rangle\ge \|\sqrt{VN}v_y\|^2$ in
$L^2(\Sigma)$ this implies $v_y=0$ and hence $w=0$.
\end{remark}

\subsubsection{Simplifications in the standard static case}

On a standard static spacetime $M$, the construction of
the non-degenerate ground and $\beta$-KMS one-particle
structures in the proof of Theorem \ref{Thm_GroundEx}
simplifies. For later convenience we formulate these
results as a proposition \cite{Kay1978}:
\begin{proposition}\label{Prop_StStatGround}
Let $\Sigma\subset M$ be a Cauchy surface orthogonal to
the Killing field of the standard static, globally
hyperbolic spacetime $M$. Under the assumptions of
Theorem \ref{Thm_GroundEx} we have:
\begin{enumerate}
\item[\textup{(i)}] The unique non-degenerate ground one-particle
structure is given, up to equivalence, by
$\mathcal{K}_0=L^2(\Sigma)$, and
$p_0=q_{0,\Sigma}S^{-1}E$ with
\[
q_{0,\Sigma}(f_0,f_1):=\frac{1}{\sqrt{2}}\left(
C^{\frac14}N^{-\frac12}f_0-iC^{-\frac14}N^{\frac12}f_1\right).
\]
Furthermore, the unitary group $\tilde{O}_t$ of Lemma
\ref{Lem_EqStruc} is given by $\tilde{O}_t=e^{it\sqrt{C}}$.
\item[\textup{(ii)}] For any $\beta>0$ the unique non-degenerate
$\beta$-KMS one-particle structure is given, up to equivalence,
by $\mathcal{K}_{(\beta)}=L^2(\Sigma)^{\oplus 2}$, and
$p_{(\beta)}=q_{(\beta),\Sigma}S^{-1}E$ with
\begin{eqnarray}
q_{(\beta),\Sigma}(f_0,f_1)&:=&\frac{1}{\sqrt{2}}\left(
(I-e^{-\beta\sqrt{C}})^{-\frac12}
(C^{\frac14}N^{-\frac12}f_0-iC^{-\frac14}N^{\frac12}f_1)
\right.\nonumber\\
&&\oplus\left. e^{-\frac{\beta}{2}\sqrt{C}}
(I-e^{-\beta\sqrt{C}})^{-\frac12}
(C^{\frac14}N^{-\frac12}f_0+iC^{-\frac14}N^{\frac12}f_1)
\right).\nonumber
\end{eqnarray}
Furthermore, the unitary group $\tilde{O}_t$ of Lemma
\ref{Lem_EqStruc} is given by
$\tilde{O}_t=e^{it\sqrt{C}}\oplus e^{-it\sqrt{C}}$.
\end{enumerate}
Here $C$ is the closure of the partial differential operator
\[
C_0:=-\sqrt{N}\nabla^{(h),i}N\nabla^{(h)}_i\sqrt{N}+VN^2
\]
defined on $C_0^{\infty}(\Sigma)$. $C_0$ and all integer powers
of it are essentially self-adjoint on the invariant domain
$C_0^{\infty}(\Sigma)$. Furthermore, $C$ is strictly positive
with $C\ge VN^2$ and $C_0^{\infty}(\Sigma)$ is contained in
the domain of $C^{\pm\frac12}$ for both signs.
\end{proposition}
% In the comment below, can we show directly that the
% additional potential term does not spoil the result?
One may also write $C$ in terms of the conformal metric
$\tilde{h}$ as
\[
C=\Box_{\tilde{h}}+VN^2+\frac{d-2}{2}N^{-2}
\left(N(\Box_{\tilde{h}}N)+\frac{d-4}{2}\tilde{h}^{ij}
(\partial_iN)(\partial_jN)\right),
\]
on $L^2(\Sigma,d\mathrm{vol}_{\tilde{h}})$, where we used
the footnote on page \pageref{Ft_KsplitConf} and the fact
that $v=N$ in the static case. The completeness of $\tilde{h}$
(Theorem \ref{Thm_TestGH}) implies that all powers of
$-\Box_{\tilde{h}}$ are essentially self-adjoint on the
test-functions. Proposition \ref{Prop_StStatGround} shows,
among other things, that the additional terms do not spoil
this result.
\begin{proof*}
In the standard static case $N^i\equiv 0$, so the operator $A$
of Lemma \ref{Lem_Eoperator} can be written as a diagonal matrix
$A=\frac12\left(\begin{array}{cc}\alpha&0\\ 0&N\end{array}\right)$,
where $\alpha:=-\nabla^{(h),i}N\nabla^{(h)}_i+VN$. Let
$\hat{\alpha}$ denote the Friedrichs extension of $\alpha$, which
is strictly positive by Lemmas \ref{Lem_Friedrichs}, \ref{Lem_SP3}.
We may then compute $\sqrt{\hat{A}}$ and hence, on the range of
$\sqrt{\hat{A}}$,
\begin{eqnarray}
H_e&=&2i\sqrt{\hat{A}}\sigma\sqrt{\hat{A}}=
\left(\begin{array}{cc}0&-i\sqrt{\hat{\alpha}}\sqrt{N}\\
i\sqrt{N}\sqrt{\hat{\alpha}}&0\end{array}\right).\nonumber
\end{eqnarray}
Both $\sqrt{\hat{\alpha}}\sqrt{N}$ and $\sqrt{N}\sqrt{\hat{\alpha}}$
are closable operators, because $H_e$ is closeable. Furthermore,
their closures are each others adjoints, because $H_e$ is
self-adjoint. By the Polar Decomposition Theorem (Ref.\ \cite{Kadison+}
Thm.\ 6.1.11) there is then a partial isometry $U$ such that
$\overline{\sqrt{\hat{\alpha}}\sqrt{N}}=UC^{\frac12}$ and
$\overline{\sqrt{N}\sqrt{\hat{\alpha}}}=C^{\frac12}U^*$, where
$C=\overline{\sqrt{N}\hat{\alpha}\sqrt{N}}=\overline{C_0}$. Now
$H_e^2=\left(\begin{array}{cc}\sqrt{\hat{\alpha}}N\sqrt{\hat{\alpha}}&0\\
0&C_0\end{array}\right)$ on the range of $\sqrt{\hat{A}}$, which
is invariant. The essential self-adjointness of all even powers
of $H_e$ on this range (Theorem \ref{Thm_Hinvertable}),
restricted to the second summand of $L^2(\Sigma)^{\oplus 2}$,
implies that all integer powers of $C_0$ are essentially
self-adjoint on the range of $\sqrt{N}$, which is just
$C_0^{\infty}(\Sigma)$. The estimate $C\ge VN^2$ follows from a
partial integration, whereas strict positivity follows from
Lemma \ref{Lem_SP3}. That $C_0^{\infty}(\Sigma)$ is in the
domain of $C^{\frac12}$ is clear, because it is in the domain of
$C$, and that it is in the domain of $C^{-\frac12}$ follows again
from Lemma \ref{Lem_SP3}. Finally, the domain and range of $U$ are
the entire $L^2(\Sigma)$, because $C^{\frac12}$ and
$\hat{\alpha}^{\frac12}$ have dense ranges. This establishes all
the claims concerning $C$.

Returning to one-particle structures, we may write, after some
short computations:
\[
V^*|H_e|V=\left(\begin{array}{cc}C^{\frac12}&0\\ 0&C^{\frac12}
\end{array}\right)\quad
V^*P_{\pm}V=\frac12 I\pm\frac12
\left(\begin{array}{cc}0&i\\ -i&0\end{array}\right)
\]
\[
q_{cl}(f_0,f_1)=\frac{1}{\sqrt{2}}V\left(\begin{array}{cc}
C^{\frac12}N^{-\frac12}&0\\ 0&N^{\frac12}\end{array}\right)
\left(\begin{array}{c}f_0\\ f_1\end{array}\right),
\]
where we introduced the unitary operator $V:=\left(\begin{array}{cc}
U&0\\ 0&I\end{array}\right)$. A comparison with the proof of Theorem
\ref{Thm_GroundEx} yields
\begin{eqnarray}
q(f_0,f_1)&=&\frac12
\left(\begin{array}{c}U\\ iI\end{array}\right)
(I-e^{-\beta\sqrt{C}})^{-\frac12}
(C^{\frac14}N^{-\frac12}f_0-iC^{-\frac14}N^{\frac12}f_1)
\nonumber\\
&&\oplus\frac12
\left(\begin{array}{c}U\\ -iI\end{array}\right)
e^{-\frac{\beta}{2}\sqrt{C}}(I-e^{-\beta\sqrt{C}})^{-\frac12}
(C^{\frac14}N^{-\frac12}f_0\ +iC^{-\frac14}N^{\frac12}f_1),
\end{eqnarray}
where we made use of the fact that
$P_{\pm}V=\frac12\left(\begin{array}{c}U\\ \mp iI\end{array}\right)
(I\ \pm iI)$. As $\|U\psi\oplus \pm i\psi\|^2=\|\sqrt{2}\psi\|^2$,
the first factors in each summand can safely be replaced by
$\sqrt{2}$, leading to a unitary equivalent formulation,
$q_{(\beta),\Sigma}$. Note that the range of $q_{(\beta),\Sigma}$
is dense in $L^2(\Sigma)^{\oplus 2}$, because if $\psi\oplus\chi$
is orthogonal to this range, then we may use the strict positivity
of the operators $(I-e^{-\beta\sqrt{C}})^{-\frac12}C^{\pm\frac14}$
to show that $\psi\pm e^{-\frac{\beta}{2}\sqrt{C}}\chi=0$ for both
signs and hence $\psi=\chi=0$. The proof of the fact that
$H=\sqrt{C}\oplus -\sqrt{C}$ is an easy exercise which we omit.
The case of the ground one-particle structure is similar, but
simpler.
% The fact that $C$ commutes with complex conjugation
% implies that $q_{0,\Sigma}$ has dense range in $L^2(\Sigma)$
% already on either $C_0^{\infty}(\Sigma,\mathbb{R})^{\oplus 2}$
% or on $C_0^{\infty}(\Sigma)\oplus\left\{0\right\}$. That
% $H=\sqrt{C}$ follows from
% \[
% \left(\begin{array}{cc}UC^{\frac12}U^*&0\\ 0&C^{\frac12}
% \end{array}\right)\left(\begin{array}{c}U\\ iI\end{array}\right)
% =\left(\begin{array}{c}U\\ iI\end{array}\right)C^{\frac12}.
% \]
\end{proof*}

The result of Proposition \ref{Prop_StStatGround} can be
interpreted in terms of positive and negative frequency
solutions \cite{Wald1979}. Indeed, any solution
$\phi=Ef\in\mathcal{S}$ with initial data $(f_0,f_1)$ can be
decomposed into positive and negative frequency parts
\begin{equation}\label{Eqn_FrequencySplit}
(N^{-\frac12}\phi)(t,.)=e^{it\sqrt{C}}N^{-\frac12}f_+
+e^{-it\sqrt{C}}N^{-\frac12}f_-,
\end{equation}
where $f_{\pm}=\frac12(f_0\mp i N^{\frac12}C^{-\frac12}N^{\frac12}f_1)$.
In the ground state we have
\begin{equation}\label{Eqn_GroundStruc}
\omega^0_2(\overline{f},f)=\frac{1}{2}\|C^{\frac14}N^{-\frac12}f_0
-iC^{-\frac14}N^{\frac12}f_1\|^2,
\end{equation}
which vanishes when $f_0=iN^{\frac12}C^{-\frac12}N^{\frac12}f_1$,
which is the case precisely when $f_+=0$, i.e.\ when $\phi$ is a
negative frequency solution. (The occurrence of negative, rather
than positive, frequency solutions here is explained by the
footnote on page \pageref{Ft_Sign}.)

\section{Ground states and their properties}\label{Sec_GroundProp}

We are now ready to study the space $\mathscr{G}^0(\alg{W})$
of ground states, under the assumptions of Theorem
\ref{Thm_GroundEx}, and to consider some of their
properties. These properties often generalize the special
properties of the Minkowski vacuum. Note that a
characterization of all classical equilibrium and ground
states on the commutative Weyl $C^*$-algebra $\alg{W}^{cl}$
can be given, in principle, using the results of Section
\ref{Sec_Alg}.

\subsection{The space of ground states}

The following theorem gives a full description of the space
$\mathscr{G}^0(\alg{W})$ of all ground states. (This result
may be compared to Theorem \ref{Thm_ClGround}.)
\begin{theorem}\label{Thm_GroundExUn}
Let $M$ be a globally hyperbolic, stationary spacetime and
consider a linear scalar field with a stationary potential
$V$ such that $V>0$.
\begin{enumerate}
\item[\textup{(i)}] There exists a unique $C^2$ ground state $\omega^0$
with vanishing one-point function. It is also the unique
extremal $C^1$ ground state with vanishing one-point function.
We denote its GNS-triple by $(\mathcal{H}_0,\pi_0,\Omega_0)$
and the one-particle structure of its two-point function is
$(p_0,\mathcal{K}_0)$ (cf.\ Theorem \ref{Thm_GroundEx}).
\item[\textup{(ii)}] $\omega^0$ is quasi-free and regular ($D^{\infty}$)
and $\pi_0$ is faithful and irreducible.
\item[\textup{(iii)}] The map $\lambda_0:=\lambda_{\omega^0}$ of Lemma
\ref{Lem_AffineIso} restricts to an affine homeomorphism
$\map{\lambda_0}{\mathscr{G}^0(\alg{W}^{cl})}{\mathscr{G}^0(\alg{W})}$.
\item[\textup{(iv)}] Any $D^2$ ground state is Hadamard and any regular ground
state satisfies the microlocal spectrum condition. A ground
state $\omega=\lambda_0(\rho)$ is $C^k$, resp.\ $D^k$,
$k=1,2,\ldots$, if and only if $\rho$ is $C^k$, resp.\ $D^k$.
\item[\textup{(v)}] Any extremal ground state $\omega$ on $\alg{W}$ is of the
form $\omega=\eta_{\rho}^*\omega^0$ for some gauge transformation
of the second kind $\eta_{\rho}$. Hence it is pure and it is
regular (resp.\ $C^{\infty}$) if and only if it is $D^1$
(resp.\ $C^1$). Furthermore, it has the Reeh-Schlieder property,
i.e.\ for any open set $O\subset M$ the linear space
$\pi_{\omega}(\alg{W}(O))\Omega_{\omega}$ is dense in
$\mathcal{H}_{\omega}$.
\item[\textup{(vi)}] If there exists an $\epsilon>0$ such that $VN\ge\epsilon$
and $N^{-1}v^2\ge\epsilon$ everywhere, then $(p_0,\mathcal{K}_0)$
has a mass gap,\footnote{Our condition is weaker than that of
Ref.\ \cite{Kay1978}, which requires $N^{-1}v^2\ge\epsilon$,
$V\ge\epsilon$ and $N\ge\epsilon$.}
namely $\|H^{-1}\|\le\epsilon^{-1}$.
\item[\textup{(vii)}] For $d=4$, Haag duality holds: if $\Sigma\subset M$ is a
Cauchy surface, $U\subset\Sigma$ an open, relatively compact
subset whose boundary $\partial U$ is a smooth submanifold of
$\Sigma$, and $O:=D(U)$, then
\[
\pi_0(\alg{W}(O))'=\pi_0(\alg{W}(O^{\perp}))'',
\]
where $O^{\perp}:=\mathrm{int}(M\setminus J(O))$ denotes the
causal complement for any subset $O\subset M$.
\end{enumerate}
\end{theorem}
Recall that the Reeh-Schlieder property means that the ground
state has many non-local correlations \cite{Clifton+2001,Haag}.
In fact, the Reeh-Schlieder property is known for all quasi-free
$D^{\infty}$ equilibrium states \cite{Strohmaier2000}.
\begin{proof*}
Let $\omega^0$ be the quasi-free state whose two-point distribution
is associated to the non-degenerate ground one-particle structure
$(p_0,\mathcal{K}_0)$ of Theorem \ref{Thm_GroundEx}. Then $\omega^0$
is a non-degenerate and pure (and hence extremal) ground state, by
Theorems \ref{Prop_EqToOnePEq} and \ref{Thm_Borchers}. As $\omega^0$
is quasi-free and $\omega^0_2$ is a distribution (density), $\omega^0$
is a regular state. Furthermore, the representation $\pi_0$ is
irreducible, because $\omega^0$ is pure, and it is faithful, because
the space $(L,E)$ is symplectic (by construction) and hence
$\alg{W}$ is simple (Ref.\ \cite{Bratteli+} Thm.\ 5.2.8).

Using Lemma \ref{Lem_EqStruc} and the fact that $\omega^0$ is
quasi-free one may show that the strong derivatives of
$t\mapsto \pi_0(\alpha_t(W(f)))\Omega_0$ are well defined for all
$f\in L$. The map $\lambda_0:=\lambda_{\omega^0}$ of Lemma
\ref{Lem_AffineIso} then restricts to the stated affine
homeomorphism by Proposition \ref{Prop_AllKMSStates}.

For regular ground states, the Hadamard property is known to
hold \cite{Sahlmann+2000} and the microlocal spectrum then
follows \cite{Sanders2010,Brunetti+1995}. The Hadamard property
for $D^2$ ground states then follows from the last statement of
Proposition \ref{Prop_OnePuniqueness}. From the definition of
$\lambda_0$ we have
\[
(\lambda_0\rho)(W(f_1)\cdots W(f_n))=\omega^0(W(f_1)\cdots W(f_n))
\rho(W(f_1)\cdots W(f_n)).
\]
As $\omega^0$ is regular and quasi-free it follows that
$\lambda_0(\rho)$ is $C^k$ (resp.\ $D^k$) if and only if $\rho$
is $C^k$ (resp.\ $D^k$).

Extremal ground states $\omega$ on $\alg{W}$ are of the form
$\lambda_0(\rho)$ for an extremal ground state $\rho$ on
$\alg{W}^{cl}$. Such $\rho$ are pure by Theorem \ref{Thm_ClGround},
so by Lemma \ref{Lem_AffineIso} this entails
$\omega=\eta_{\rho}^*\omega^0$. Because $\eta_{\rho}^*$ preserves
pure states it follows that every extremal ground state on
$\alg{W}$ is pure (cf.\ Remark \ref{Rem_Gauge3}). Furthermore,
$\eta_{\rho}^*$ preserves the local algebras $\alg{W}(O)$, so the
extremal ground states have the Reeh-Schlieder property, because
$\omega^0$ does \cite{Strohmaier2000}. The statement on the
regularity of extremal ground states follows directly from
Proposition \ref{Prop_rho0}. This also proves the second
uniqueness clause for $\omega^0$. The first uniqueness clause
follows from Theorem \ref{Thm_KayUniqueness}.

To prove the existence of the mass gap we note that,
under the stated assumptions, $\hat{A}\ge \frac{\epsilon}{2}I$
by Lemma \ref{Lem_Eoperator}. In the energetic Hilbert space we
then use $(i\sigma)^*=i\sigma$ to estimate
\[
H_e^2=4\hat{A}^{\frac12}i\sigma\hat{A}i\sigma\hat{A}^{\frac12}
\ge 2\epsilon \hat{A}^{\frac12}(i\sigma)^2\hat{A}^{\frac12}
=2\epsilon\hat{A}\ge \epsilon^2I.
\]
Hence, $|H_e|\ge\epsilon I$, $H\ge\epsilon I$ and
$\|H^{-1}\|<\epsilon^{-1}$.

Finally, the fact that $\omega$ is pure entails Haag duality,
at least when $d=4$ (Ref.\ \cite{Verch1997}, Thm.\ 3.6), even
for slightly more general regions $O$ than used here.
\end{proof*}

A few remarks concerning the interpretation of the results
of this section and their implications are in order:
\begin{remark}\label{Rem_Gauge4}
The gauge transformations of the second kind, which appeared
in the proof of Theorem \ref{Thm_GroundExUn}, can be
physically interpreted as field redefinitions. If
$\omega_1$ is a linear map on $L$, then $\chi:=e^{-i\omega_1}$
is a character and $\rho(W(f)):=e^{-i\omega_1(f)}$ defines a
pure state on $\alg{W}^{cl}$. If we write (formally)
$W(f)=e^{i\Phi(f)}$ we have
\[
\eta_{\rho}(W(f))=e^{i(\Phi(f)-\omega_1(f)I)}.
\]
In particular, if $\omega$ is any pure $C^2$ ground state
with one-point distribution $\omega_1$ and $\rho$ is defined
as above, then we must have $\eta_{\rho}^*\omega=\omega^0$ by
Theorem \ref{Thm_GroundExUn}. Hence,
$\omega(W(f))=e^{i\omega_1(f)}\omega^0(W(f))$. Because pure
states $\rho$ of this exponential form are dense
(Ref.\ \cite{Binz+2004} Lemma 4-2) we may argue on physical
grounds that we may as well restrict attention to the pure
ground state with vanishing one-point distribution, $\omega^0$.
\end{remark}

\begin{remark}\label{Rem_vonNeumann}
Because $\omega^0$ is a uniquely distinguished ground
state and $\pi_0$ is faithful we may perform the
following standard modification of the original theory.
For each bounded region $O\subset M$ we define the
von Neumann algebra $\alg{R}(O):=\pi_0(\alg{W}(O))''$.
This gives rise to a local net of von Neumann algebras
in the spacetime $M$ and we let the $C^*$-algebra
$\alg{R}$ be their inductive limit. Each
$\alg{R}(O)$ contains the corresponding $\alg{W}(O)$,
so that $\alg{R}\supset\alg{W}$. We may then consider
the class of states on $\alg{R}$ which are locally
normal, i.e.\ they restrict to normal states on each
von Neumann algebra $\alg{R}(O)$. Such states clearly
restrict to a state on $\alg{W}$ and a state $\omega$
on $\alg{W}$ has at most one extension
to $\alg{R}$. This extension exists if and only if
$\omega$ is locally normal w.r.t.\ $\omega^0$ (by
definition). This includes at least all quasi-free
Hadamard states \cite{Verch1994}.
% (Using this fact one could argue that we could have
% constructed the algebra $\alg{R}$ a priori, but this
% would be a reversal of the actual physical motivation.
% Hadamard states are of interest exactly because their
% two-point distributions differ from that of the
% ground state by a smooth function.)

There are good physical reasons to consider only
states on $\alg{W}$ that are locally normal with
respect to $\omega^0$. For any self-adjoint operator
$A\in\alg{W}(O)$ for any bounded region $O$, the
algebra $\alg{R}(O)$ contains all the spectral
projectors of $A$, so the operational question
whether the measured value of $A$ attains a value
in some Borel set $I\subset\mathbb{R}$ corresponds
to the same projection operator for all locally
normal states. Another reason to restrict only to
locally normal states is of a more technical nature.
The action of the one-parameter group $\alpha_t$ on
$\alg{W}$ is not norm continuous, but the larger
algebra $\alg{R}$ contains a $C^*$-algebra
$\alg{R}_0$ which is dense in $\alg{R}$ in the strong
operator topology and on which $\alpha_t$ is norm
continuous (cf.\ Ref.\ \cite{Fewster+2003} Sec.\ 4,
or also Ref.\ \cite{Sakai1991} Thm.\ 1.18 for a closely
related result). This means that a large number of
results on $C^*$-dynamical systems can be brought to
bear on $(\alg{R}_0,\alpha_t)$, and hence indirectly
also on $\alg{W}$, if one considers states that are
locally normal \cite{Bratteli+,Sakai1991} with respect to
$\omega^0$.

Let us briefly describe the constructions of
Ref.\ \cite{Fewster+2003} (adapted to a stationary,
globally hyperbolic spacetime and with a possibly
non-compact Cauchy surface). The $C^*$-algebra
$\alg{R}_0$ may be generated by operators of the form
\[
A_f:=\int dt\ f(t)\alpha_t(A),
\]
where $A\in\alg{W}(O)$ for some bounded region $O$
and $f\in C_0^{\infty}(\mathbb{R})$. Then
$A_f\in\alg{R}(O')$, where $O'$ is another bounded
region that depends on $O$ and on the support of
$f$. Such operators form a $^*$-algebra which is
invariant under the action of $\alpha_t$ and on
which $\alpha_t$ is norm continuous. $\alg{R}_0$
is the norm closure of this $^*$-algebra.
\end{remark}

\subsection{The ground state representation and the quantum
stress-energy-momentum tensor}\label{SSec_T}

As $\omega^0$ is quasi-free, $\mathcal{H}_0$ is a Fock space
(cf.\ Sec.\ 3.2 of Ref.\ \cite{Kay+1991}) and we may introduce a
particle interpretation for the field, based on creation and
annihilation operators. Note that such an interpretation
fails in general spacetimes, because there are many unitarily
inequivalent Fock space representations and there is no
generally covariant prescription to single out a preferred
one \cite{Fewster+2011,Wald1994}.

Following standard notations \cite{Bratteli+} we will write
$\mathcal{H}_0=\bigoplus_{n=0}^{\infty}\mathcal{H}_0^{(n)}$,
where the $n$-particle Hilbert space is
$\mathcal{H}_0^{(n)}:=P_+(\mathcal{K}_0)^{\otimes n}$, in
which $(p_0,\mathcal{K}_0)$ is the one-particle structure
associated to $\omega^0_2$ and $P_+$ denotes the projection
onto the symmetric tensor product. We write $N$ for the
number operator, so that $N|_{\mathcal{H}_0^{(n)}}=nI$. We
will use the notation $a^*(\psi)$ and $a(\psi)$ for creation
and annihilation operators, respectively, where
$\psi\in\mathcal{K}_0$. As $a^*(\psi)^*=a(\psi)$ we see that
$a$ is complex anti-linear in $\psi$, whereas $a^*$ is linear.
The field $\Phi$ is given by
\[
\Phi(f)=\frac{1}{\sqrt{2}}(a^*(p_0(f))+a(p_0(\overline{f})))
\]
and is complex linear, as desired. We may introduce the initial
value and normal derivative of the quantum field as
\begin{eqnarray}
\Phi_0(f_1)&:=&\frac{-1}{\sqrt{2}}(a^*(q_0(0,f_1))
+a(q_0(0,\overline{f_1}))),\nonumber\\
\Phi_1(f_0)&:=&\frac{1}{\sqrt{2}}(a^*(q_0(f_0,0))
+a(q_0(\overline{f_0},0)))\nonumber
\end{eqnarray}
so that $\Phi(f)=\Phi_1(f_0)-\Phi_0(f_1)$, where
$(f_0,f_1)=S^{-1}Ef$. This is in line with what one would get
if $\Phi$ were a classical solution to the Klein-Gordon
equation (cf.\ Eq.\ (\ref{Eqn_CCRInit})). It will also be
convenient to introduce the operators
\[
\Pi(f):=\frac{i}{\sqrt{2}}(a^*(p_0(f))-a(p_0(\overline{f}))).
\]

Because the classical stress-energy-momentum tensor played a
significant role in the classical and quantum descriptions of
the linear scalar field in a stationary spacetime, it seems
fitting to also spend a few words on the quantum
stress-energy-momentum tensor. If the field theory on $M$ can
be extended to all globally hyperbolic spacetimes in a locally
covariant way \cite{Brunetti+2003}, e.g.\ if $V=cR+m^2$, then
there is a generally covariant way to define the renormalised
stress-energy-momentum tensor \cite{Hollands+2001}. However,
in our setting it will be advantageous not to renormalize the
stress tensor in a generally covariant way, but instead to
exploit the extra structure of the stationary spacetime.
(Nevertheless, our presentation of the classical and quantum
stress tensor is based on existing treatments that fit in a
generally covariant framework, e.g.\ Ref.\ \cite{Fewster+2008}.)

We may define a tensor field $G_{ab}$ on a sufficiently small
neighborhood $U\subset M^{\times 2}$ of the diagonal
$\Delta:=\left\{(x,x)|\ x\in M\right\}$ by the property that
for any vector $v^b\in T_{x'}M$, the vector
$g^{ac}(x)G_{cb}(x,x')v^b(x')\in T_xM$ is the parallel
transport of $v$ along a unique geodesic connecting $x$ to
$x'$. (The uniqueness of the geodesic can be ensured by
choosing $U$ sufficiently small.)
% Refer to B. O'Neill for the following:
% Let $V_i$ be a set of convex normal neighborhoods covering
% $M$ and recall that each $V_i$ is starshaped around each of
% its points. Let $U=\left\{(x,x')|\ \exists i\ x,x'\in V_i\right\}$.
% For any $(x,x')\in U$ there is a unique geodesic in each $V_i$
% which contains both points, and the geodesics for all
% different $V_i$ must coincide (using the starshape around $x$).
Using $G_{ab}$ and
$G^{ab}(x,x'):=g^{ac}(x)g^{bd}(x')G_{cd}(x,x')$ we may write
the classical stress-energy momentum tensor in terms of a
differential operator as
\begin{eqnarray}
T_{ab}(\phi)&=&(T_{ab}^{\mathrm{split}}\phi^{\otimes 2})(x,x)
\nonumber\\
T_{ab}^{\mathrm{split}}&=&
\nabla_a\otimes\nabla_b-\frac12G_{ab}G^{cd}
\nabla_c\otimes\nabla_d-\frac12G_{ab}\sqrt{V}\otimes\sqrt{V}.
\end{eqnarray}
Instead of letting the operator $T_{ab}^{\mathrm{split}}$ act
on the classical fields $\phi^{\otimes 2}$, we can let it act
on the normal ordered quantum field,
\[
:\Phi^{\otimes 2}:(x,x')=\Phi(x)\Phi(x')-\omega^0_2(x,x').
\]
For any vector $\psi\in\pi_0(\alg{A})\Omega_0$ we may define
the $\mathcal{H}_0$-valued distribution (density)
\[
T_{ab}^{\mathrm{ren}}(f^{ab})\psi:=\lim_{n\rightarrow\infty}
T_{ab}^{\mathrm{split}}:\Phi^{\otimes 2}:(f^{ab}\delta_n)\psi,
\]
where $\delta_n\in C^{\infty}(M^{\times 2})$ is a sequence
of functions that approximates the delta distribution
$\delta(x,x')$ and $f^{ab}$ is a compactly supported, smooth
test-tensor \cite{Brunetti+1995}. The operator
$T_{ab}^{\mathrm{ren}}(f^{ab})$ is densely defined and it is
a symmetric operator when $f^{ab}$ is real-valued. Moreover,
if $V>0$ everywhere one can show that
$T_{ab}^{\mathrm{ren}}(\chi^a\chi^b)$ is semi-bounded from
below for real-valued test-vector fields
$\chi^a$ \cite{Fewster+2008}. (Note that the method of proof
in Ref.\ \cite{Fewster+2008} is not affected by the presence
of the non-negative potential energy term $V$ in the equation
of motion.)
% $V$ only comes in through the Hadamard coefficients
% and therefore it does not affect the regularity of the
% terms in the Hadamard series.

In analogy with the classical case we define the quantum
energy-momentum one-form and the energy density by
\[
P^{\mathrm{ren}}_a(f^a):=T_{ab}^{\mathrm{ren}}(f^a\xi^b),\quad
\epsilon^{\mathrm{ren}}(f):=T_{ab}^{\mathrm{ren}}(n^a\xi^bf)
\]
in the sense of $\mathcal{H}_0$-valued distributions, when
acting on $\pi_0(\alg{A})\Omega_0$. One may check that
$T^{\mathrm{ren}}_{ab}$ is symmetric in its indices $a,b$
and that
\[
\nabla^aT^{\mathrm{ren}}_{ab}=-(\nabla_bV):\Phi^2:,
\]
where the Wick square $:\Phi^2:$ is the restriction of
$:\Phi^{\otimes 2}:$ to the diagonal
$\Delta\subset M^{\times 2}$. It follows from
$\partial_0V=0$ that $\nabla^aP^{\mathrm{ren}}_a=0$, just
like in the classical case.
% Proof:
% For any unit vector $\psi\in\pi_0(\alg{A})\Omega_0$
% let $\omega_2(x,x'):=\langle\psi,\Phi(x)\Phi(x')\psi\rangle$.
% Then $\langle\psi,T^{\mathrm{ren}}_{ab}(x)\psi\rangle
% =T^{\mathrm{split}}_{ab}w(x,x)$, where
% $w(x,x'):=(\omega_2-\omega^0_2)(x,x')$ is smooth,
% real-valued and symmetric. It follows that
% $\langle\psi,T^{\mathrm{ren}}_{ab}(x)\psi\rangle$
% is symmetric in $a,b$. Furthermore, using the fact that
% $0=\nabla_a(G_{bc}|_{\Delta})=(\nabla_aG_{bc}+
% G_{ad}(\nabla')^dG_{bc}$ one computes that
% $\nabla^a\langle\psi,T^{\mathrm{ren}}_{ab}(x)\psi\rangle
% =-(\nabla_bV)\langle\psi,:\Phi^2:(x)\psi\rangle$.
% Polarisation allows us to draw the same conclusions
% for the matrix elements with different $\psi,\psi'$,
% so the statement for the operators follows.

\begin{remark}
From a physical point of view it seems reasonable to
expect that for real-valued $f$ the operator
$\epsilon^{\mathrm{ren}}(f^2)$ is semi-bounded from
below, using the same motivation as for existing quantum
inequalities \cite{Fewster+2008}. However, the details
of the argument require that we can write
$\xi^an^b+n^a\xi^b=\sum_{j=1}^k\chi_j^a\chi_j^b$ for
some finite number of (real) vectors $\chi_j^a$. An
easy exercise shows that this is possible if and only
if we are in the static case, where $\xi^a=Nn^a$, in
which case the single vector $\chi^a=N^{-\frac12}\xi^a$
will suffice. Thus, in the static case, the results of
Ref.\ \cite{Fewster+2008} apply and
$\epsilon^{\mathrm{ren}}(f^2)$ is semi-bounded from
below.
\end{remark}

There is another result, however, which does work very
nicely in the general stationary setting:
\begin{theorem}\label{Thm_TSA}
Under the assumptions of Theorem \ref{Thm_GroundExUn},
let $\omega^0$ be the unique ground state. For any
real-valued test-tensor $f^{ab}$, the operator
$T_{ab}^{\mathrm{ren}}(f^{ab})$ is essentially
self-adjoint on $\pi_0(\alg{A})\Omega_0$.
\end{theorem}
A similar essential self-adjointness result for the
smeared stress-energy-momentum tensor in general globally
hyperbolic spacetimes is much harder to obtain by a direct
proof (cf.\ Ref.\ \cite{Sanders2012} for partial results).
\begin{proof*}
It follows from Lemma \ref{Lem_EqStruc} (and second
quantization) that the Hamiltonian operator $h$ is
essentially self-adjoint on the dense, invariant domain
$\pi_0(\alg{A})\Omega_0$ and that
\[
\langle\psi,[h+I,T_{ab}^{\mathrm{ren}}(f^{ab})]\psi'\rangle
=\langle\psi,iT_{ab}^{\mathrm{ren}}(\partial_0f^{ab})\psi'
\rangle
\]
for all $\psi,\psi'$ in that domain. (Here we have used
the fact that $\omega^0$ is an equilibrium state.) The
idea is now to use the Commutator Theorem X.36' of
Ref.\ \cite{Reed+} to prove essential self-adjointness of
$T_{ab}^{\mathrm{ren}}(f^{ab})$. This means we need to
prove that for any test-tensor $f^{ab}$ there is a $C>0$
such that
\begin{equation}\label{Eqn_CommutatorEst}
|\langle\psi,T_{ab}^{\mathrm{ren}}(f^{ab})\psi'\rangle|\le
C\|(h+I)^{\frac12}\psi\|\cdot\|(h+I)^{\frac12}\psi'\|
\end{equation}
for all $\psi,\psi'\in\pi_0(\alg{A})\Omega_0$. By
polarization it suffices to take $\psi=\psi'$. It also
suffices to consider $f^{ab}$ to be supported in a convex
normal neighborhood, by a partition of unity argument.
Moreover, the antisymmetric part of $f^{ab}$ does not
contribute and the symmetric part can be written as a
finite sum of terms of the form $\chi^a\chi^b$, so it
suffices to consider $f^{ab}=\chi^a\chi^b$.

Now consider the operators $\Pi(f)$ for
$f\in C_0^{\infty}(M)$.
$[\Pi(f),\Pi(f')]=[\Phi(f),\Phi(f')]=iE(f,f')$,
so for any $\psi\in\pi_0(\alg{A})\Omega_0$ the
distribution
\[
\omega^{\psi}_2(f,f'):=\|\psi\|^{-2}\langle\psi,
\Pi(f)\Pi(f')\psi\rangle
\]
is a Hadamard two-point distribution. As for the field
$\Phi(f)$ one may introduce the normal-ordered product
$:\Pi(f)\Pi(f'):\ :=\Pi(f)\Pi(f')-\omega^0_2(f,f')$
and following Ref.\ \cite{Fewster+2008} one proves that the
operator
\[
\tilde{T}^{\mathrm{ren}}_{ab}(\chi^a\chi^b):=
(T^{\mathrm{split}}_{ab}:\Pi^{\otimes 2}:)(\chi^a\chi^b\delta)
\]
is semi-bounded from below. Hence, for some $c>0$,
\begin{equation}
T^{\mathrm{ren}}_{ab}(\chi^a\chi^b)\le
T^{\mathrm{ren}}_{ab}(\chi^a\chi^b)
+\tilde{T}^{\mathrm{ren}}_{ab}(\chi^a\chi^b)+cI
=2(T^{\mathrm{split}}_{ab}a^*\otimes a)(\chi^a\chi^b\delta)+cI.
\label{Eqn_SAproof1}
\end{equation}
The first term on the right-hand side is the second
quantization of an operator $T$ on $\mathcal{H}_0^{(1)}$,
for which we have
\begin{eqnarray}\label{Eqn_SAproof2}
\langle\Phi(f)\Omega_0,T\Phi(f)\Omega_0\rangle&=&
2(T^{\mathrm{split}}_{ab}\overline{\phi}\otimes\phi)
(\chi^a\chi^b\delta)\nonumber\\
&=&\int_M|\chi^a\nabla_a\phi|^2
-\chi^a\chi_a g^{bc}\overline{\nabla_b\phi}\nabla_c\phi
-\chi^a\chi_aV|\phi|^2\nonumber\\
&\le&c'\int_{\mathrm{supp}(\chi^a)}
|\partial_0\phi|^2+h^{ij}
\overline{\nabla^{(h)}_i\phi}\nabla^{(h)}_j\phi+|\phi|^2
\end{eqnarray}
for some $c'>0$, where we defined $\phi:=\omega^0_2(.,f)$.
On the other hand, because the classical energy is independent
of the Cauchy surface, $h$ satisfies (cf.\ Lemma
\ref{Lem_Eoperator})
\begin{eqnarray}\label{Eqn_SAproof3}
\mathcal{E}(\phi)&=&
\langle\Phi(f)\Omega_0,h\Phi(f)\Omega_0\rangle\nonumber\\
&=&\int_M \frac{\tau(t)}{2N^2}\left(
|\partial_0\phi|^2+(N^2h^{ij}-N^iN^j)
\overline{\nabla^{(h)}_i\phi}\nabla^{(h)}_j\phi
+VN^2|\phi|^2\right),
\end{eqnarray}
where $\tau\in C_0^{\infty}(\mathbb{R})$ satisfies
$\int\tau=1$. Choosing $\tau\ge0$ and $\tau>0$ on
the compact support of $\chi^a$ and using the fact that
$Nh^{ij}-N^{-1}N^iN^j$ is positive definite, the desired
estimate Eq.\ (\ref{Eqn_CommutatorEst}) easily follows
from Eq.'s\ (\ref{Eqn_SAproof1}, \ref{Eqn_SAproof2},
\ref{Eqn_SAproof3}).
\end{proof*}

Note that $[T_{ab}^{\mathrm{ren}}(f^{ab}),\pi_0(W(f'))]=0$
whenever $\mathrm{supp}(f')\cap J(\mathrm{supp}(f^{ab}))=\emptyset$.
It follows from Haag duality that $T_{ab}^{\mathrm{ren}}(f^{ab})$
is affiliated to the local von Neumann algebra
$\alg{R}(D(\mathrm{supp}(f^{ab})))$.

\begin{lemma}\label{Lem_ApproxH}
Let $\Sigma$ be Cauchy surface in a stationary, globally
hyperbolic spacetime $M$. Let $f\in C_0^{\infty}(M)$,
$\tau\in C_0^{\infty}(\mathbb{R})$ with $\int\tau=1$ and
$\chi\in C_0^{\infty}(\Sigma)$ such that $\chi\equiv 1$ on
$\mathrm{supp}(\tau)\cap J(\mathrm{supp}(f))$, where we view
$\tau,\chi$ as functions on $M$ in adapted coordinates. Then
\[
[\epsilon^{\mathrm{ren}}(\tau\otimes N^{-1}\chi),\Phi(f)]=
\Phi(i\partial_0f)
\]
on $\pi_0(\alg{A})\Omega_0$.
\end{lemma}
\begin{proof*}
We follow the computations in Ref.\ \cite{Fewster+2003}, Appendix
A.2. Fix a vector $\psi\in\pi_0(\alg{A})\Omega_0$, so that
$\phi':=\langle\psi,\Phi(.)\psi\rangle$ is a smooth function.
Let $\phi:=E(.,f)$ and note that $\partial_0\phi=E(.,\partial_0f)$,
by the uniqueness of $E^{\pm}$. Using
$\omega([:\Phi^{\otimes 2}:(x,x'),\Phi(f)])=i\phi(x)\phi'(x')
+i\phi'(x)\phi(x')$ we find after some algebra
\[
\omega([\epsilon^{\mathrm{ren}}(.),\Phi(f)])=
i(Nh^{ij}-N^{-1}N^iN^j)\partial_i\phi\partial_j\phi'
+iVN\phi\phi'+iN^{-1}\partial_0\phi\partial_0\phi'.
\]
Using the Klein-Gordon equation and Eq.\ (\ref{Eqn_CCRInit})
we may then compute for any Cauchy surface $\Sigma'$
\begin{eqnarray}
\omega(\Phi(i\partial_0f))&=&i\int_M(\partial_0f)\phi'
=-i\int_{\Sigma'}(n^a\nabla_a\partial_0\phi)\phi'-(\partial_0\phi)
n^a\nabla_a\phi'\nonumber\\
&=&i\int_{\Sigma'}(Nh^{ij}-N^{-1}N^iN^j)\partial_i\phi\partial_j\phi'
+VN\phi\phi'+N^{-1}\partial_0\phi\partial_0\phi'\nonumber\\
&=&\int_{\Sigma'}\omega([\epsilon^{\mathrm{ren}}(.),\Phi(f)])
=\int_M\tau(t)N^{-1}\chi\omega([\epsilon^{\mathrm{ren}}(.),\Phi(f)]).
\nonumber
\end{eqnarray}
By polarization the desired operator equality now holds on the
indicated dense domain.
\end{proof*}

\section{KMS states in stationary spacetimes}\label{Sec_Thermal}

We now come to the thermal equilibrium states at non-zero
temperature. We still consider a linear scalar field in a
stationary, globally hyperbolic spacetime and we assume that
the theory has a unique $C^2$ ground state $\omega^0$ as in
Section \ref{Sec_GroundProp} and a Hamiltonian operator $h$.
In Section \ref{Sec_KMSstates} below we will review the
states satisfying the KMS-condition, which exist for every
inverse temperature $\beta>0$. Afterwards, in Section
\ref{SSec_Wick}, we show that their two-point distributions
can be obtained from a Wick rotation, in case $M$ is standard
static (see also Ref.\ \cite{Fulling+1987}).

Before we come to this, however, we study the motivation to
use the KMS-condition as a characterization of thermal
equilibrium in Section \ref{SSec_Gibbs}. In particular we
show that for a standard static spacetime $M$ with compact
Cauchy surfaces we may also define Gibbs states to describe
thermal equilibrium and these Gibbs states satisfy the
KMS-condition.

\subsection{Gibbs states and the KMS-condition}\label{SSec_Gibbs}

Consider, then, a stationary, globally hyperbolic spacetime
$M$ and a linear scalar field satisfying the assumptions of
Theorem \ref{Thm_GroundExUn}. If, for some inverse
temperature $\beta>0$, the operator $e^{-\beta h}$ is of
trace-class in the ground state representation $\pi_0$,
i.e.\ if it has a finite trace, one may define the thermal
equilibrium state to be the Gibbs state
\begin{equation}\label{Eqn_Gibbs}
\omega^{(\beta)}(A):=\frac{\mathrm{Tr}(e^{-\beta h}A)}
{\mathrm{Tr}\ e^{-\beta h}}.
\end{equation}
Here we use the fact that the set of bounded trace-class
operators on a Hilbert space forms a $^*$-ideal in the algebra
of all bounded operators (Ref.\ \cite{Kadison+} Rem.\ 8.5.6 or
Ref.\ \cite{Reed+} Thm.\ VI.19).
% Ref.\ \cite{Reed+} assumes that the Hilbert space is separable
% (see Thm.\ VI.18 loc.cit.), but according to Ref.\ \cite{Kadison+}
% this is not necessary.

We now show that these Gibbs states are well defined whenever
$M$ is standard static and has compact Cauchy surfaces. Moreover,
we explain that these Gibbs states satisfy the KMS-condition.
\begin{theorem}\label{Thm_Gibbs}
We make the assumptions of Theorem \ref{Thm_GroundExUn} with
the additional assumptions that $M$ is a standard static
spacetime with compact Cauchy surfaces, so that the theory has
a mass gap. For any $\beta>0$
\begin{enumerate}
\item[\textup{(i)}] $e^{-\beta h}$ is of trace-class and in particular the
Gibbs state $\omega^{(\beta)}$ of Eq.\ (\ref{Eqn_Gibbs}) is
well defined and normal w.r.t.\ the ground state $\omega^0$;
\item[\textup{(ii)}] the Gibbs state $\omega^{(\beta)}$ is quasi-free and
satisfies the KMS-condition at inverse temperature $\beta>0$.
\end{enumerate}
\end{theorem}
\begin{proof*}
By Ref.\ \cite{Bratteli+} Proposition 5.2.27, the operator
$e^{-\beta h}$ has a finite trace on $\mathcal{H}_0$ if and only
if $e^{-\beta H}$ has a finite trace on
$\mathcal{H}_0^{(1)}\simeq\mathcal{K}$ and $\beta H$ is strictly
positive. The latter is satisfied by our assumptions, so we only
need to show that $e^{-\beta H}$ has a finite trace. Our proof
of this fact is adapted from the proof of nuclearity in
Ref.\ \cite{Verch1993}.

We refer to Proposition \ref{Prop_StStatGround} for a convenient
formulation of the ground one-particle structure, with
$\mathcal{K}\simeq L^2(\Sigma)$ and $H=\sqrt{C}$. By assumption,
the theory has a mass gap, so $\sqrt{C}\ge \epsilon I>0$. The
exponential $e^{-\beta\sqrt{C}}$ is bounded and may be written
as $C^{-n}(C^ne^{-\beta\sqrt{C}})$ for any $n\ge 1$, where both
$C^{-n}$ and the product in brackets are bounded. Because
trace-class operators form an ideal in the algebra of bounded
operators, it suffices to prove that $C^{-n}$ is trace-class.
The operator $C$ is a partial differential operator, while
$C^{-2n}$ defines a distribution density $u$ on
$\Sigma^{\times 2}$ by Theorem \ref{Thm_Kernel}. We then have
$(C^nuC^n)(x,y)=\delta(x,y)$. Note that $C\otimes C$ is an
elliptic operator on $\Sigma^{\times 2}$. Choosing $n$ large
enough, we can make $u$ continuous.
% We use elliptic regularity here, in terms of Sobolev wave-front
% sets. Verch claims we need $n>d-1$, the dimension of $\Sigma$.
Because $\Sigma$ is compact it follows that
$u\in L^2(\Sigma^{\times 2})$, which implies that it is
Hilbert-Schmidt (Ref.\ \cite{Reed+} Thm.\ VI.23) and, by
definition of Hilbert-Schmidt operators, $C^{-n}$ is
trace-class. $\omega^{(\beta)}$ is normal with respect to
the ground state by definition. This completes the proof
of the first item.

The quasi-free property follows
from Proposition 5.2.28 of Ref.\ \cite{Bratteli+}. For the
KMS-condition we follow Ref.\ \cite{Haag+1967} and note that the
function
\[
f(z):=\pi_0(A)e^{izh}\pi_0(B)e^{-izh}e^{-\beta h}
=\pi_0(A)e^{-\tau h}e^{ith}\pi_0(B)e^{-ith}e^{(\tau-\beta) h}
\]
takes values in the bounded operators on $\mathcal{H}_0$ for
$z=t+i\tau\in\overline{\mathrm{S}_{\beta}}$, as
$0\le \tau\le \beta$. By Lemma \ref{Lem_Holo} it is continuous
on $\overline{\mathrm{S}_{\beta}}$ and holomorphic on the
interior $\mathrm{S}_{\beta}$. Moreover, $f(z)$ is
trace-class, because either $e^{(\tau-\beta) h}$ or
$e^{-\tau h}$ is trace-class. Using the fact that
$|\mathrm{Tr}(CD)|\le \|C\|\mathrm{Tr}|D|$ for all bounded
operators $C$ and trace-class operators $D$,\footnote{Proof:
if $D$ is trace-class, we may choose an orthonormal
eigenbasis $\psi_n$ of $|D|$ and use the Polar Decomposition
Theorem (Ref.\ \cite{Kadison+} Thm.\ 6.1.11) to write $D=U|D|$
for some partial isometry $U$. Then,
\[
|\mathrm{Tr}(CD)|=\left|\sum_n\langle U^*C^*\psi_n,|D|\psi_n\rangle
\right|\le \|U^*C^*\|\sum_n\||D|\psi_n\|=\|C\|\ \mathrm{Tr}|D|.
\]}
we see that $\mathrm{Tr}f(z)$ is a bounded, continuous
function on $\overline{\mathrm{S}_{\beta}}$, which is
holomorphic in the interior. Dividing by
$\mathrm{Tr}e^{-\beta h}$ proves the second item.
\end{proof*}

We see that, under suitable physical (and technical) conditions,
Gibbs states are well defined for systems in a finite spatial
volume. In fact, we will see in Theorem \ref{Thm_KMSExUn} below
that for given $\beta>0$ it is the only $\beta$-KMS state on
$\alg{W}$ satisfying some natural additional conditions. In
general, however, the given exponential operator is not of
trace-class and the definition of the Gibbs state does not make
sense. In such cases one takes the KMS-condition to be the
defining property of thermal equilibrium states.
Theorem \ref{Thm_Gibbs}, together with the uniqueness result
of Theorem \ref{Thm_KMSExUn} below, is a good indication that
such a definition is justified. Further evidence comes from
the analysis of Ref.\ \cite{Pusz+1978}, who investigated the second
law of thermodynamics for general $C^*$-dynamical systems.
They call a state $\omega$ of such a system completely
passive, if it is impossible to extract any work from any
finite set of identical copies of this system, all in the
same state, by a cyclic process. They then showed, among
other things, that a state is completely passive if and only
if it is a ground state or a KMS state at an inverse
temperature $\beta\ge 0$.\footnote{If it is impossible to
extract any work from only one copy of this system in the
given state, the state is called passive. The set of passive
states also contains convex combinations of the ground and
KMS states.} This analysis applies to our situation, if we
restrict attention to states which are locally normal with
respect to the ground state (cf.\ Remark \ref{Rem_vonNeumann}).
We will see in Section \ref{Sec_KMSstates} that quasi-free,
$D^2$ KMS states do indeed satisfy this local normality
condition, because they are Hadamard. A more general and
detailed study of the relations between passivity, the
Hadamard condition and quantum energy inequalities was made
by Ref.\ \cite{Fewster+2003}.

Probably the most direct motivation in favor of the
KMS-condition is an analysis of Ref.\ \cite{Haag+1967} (see also
Ref.\ \cite{Bratteli+}) which shows, in the context of quantum
statistical mechanics, that a thermodynamic (infinite volume)
limit of Gibbs states satisfies the KMS-condition.
Reformulated to our geometric setting, the idea is to
approximate $h$ by operators $h_O$, where $O\subset\Sigma$
has finite volume, such that
$e^{ith_O}\in\alg{R}(D(O))=\pi_0(\alg{W}(D(O)))''$ for all
$t\in\mathbb{R}$, where $D(O)\subset M$ denotes the domain of
dependence. If $e^{-\beta h_O}$ is a trace-class operator
on $\mathcal{H}_0(O):=\overline{\pi_0(\alg{W}(D(O)))\Omega_0}$
for some $\beta>0$, then it gives rise to a Gibbs state
$\omega^{(\beta,O)}$. The argument of Ref.\ \cite{Haag+1967} shows
that, under some additional assumptions on the $h_O$, one may
show that the thermodynamic limit
$\omega^{(\beta)}:=\lim_{O\rightarrow\Sigma}\omega^{(\beta,O)}$
exists and is a $\beta$-KMS state. In the case of
non-relativistic point-particles in Minkowski spacetime, an
explicit construction of the approximate Hamiltonians $h_O$
and the corresponding limiting procedure is described in detail
in Ref.\ \cite{Bratteli+} (see also the classic paper
Ref.\ \cite{Araki+1963}, where the thermodynamic limit of a
non-relativistic free Bose gas was investigated in detail).

For a quantum field it is tempting to choose $h_O$ to be of
the form $h_O=\epsilon^{\mathrm{ren}}(f)$ for some suitable
$f\in C_0^{\infty}(D(O))$, in view of Theorem \ref{Thm_TSA}
and Lemma \ref{Lem_ApproxH}. However, the argument becomes
more problematic for two reasons. Firstly, the restriction
to a bounded open region $O$ does not entail the desired
reduction in the degrees of freedom, due to the Reeh-Schlieder
property: if $O$ is non-empty, the subalgebra $\alg{R}(D(O))$
already generates the entire Hilbert space $\mathcal{H}_0$
when acting on the ground state vector $\Omega_0$. Secondly,
and more to the point, the operators $e^{-\beta h_O}$ cannot
be trace-class. In fact, $\alg{R}(D(O))$ is a type
$\mathrm{III}_1$ factor (Thm.\ 3.6g) of Ref.\ \cite{Verch1997}),
so the only trace-class operator $X\in\alg{R}(D(O))$ is
$X=0$.\footnote{For a proof, consider a trace-class
operator $X\in\alg{R}(D(O))$, so that $|X|$ has a discrete
spectrum. Suppose that $P\in\alg{R}(D(O))$ is a spectral
projection operator onto an eigenspace with an eigenvalue
$c\not=0$. As $|X|$ is trace-class, $P$ must project onto a
finite-dimensional subspace, so it is a finite projection in the
von Neumann algebra $\alg{R}(D(O))$. However, since
$\alg{R}(D(O))$ is a type $\mathrm{III}_1$ factor, it does not
have any non-trivial finite projections \cite{Kadison+}. Thus,
$P=0$ and the only possible
eigenvalue of $|X|$ is $0$, which entails $X=0$.} This
means that no $h_O$ can possibly satisfy the assumptions
made in Ref.\ \cite{Haag+1967}. Even in a spacetime with a
compact Cauchy surface $\Sigma$, the Reeh-Schlieder
property of the ground state and the type of the local
von Neumann algebras prevent us from finding appropriate
Gibbs states to define thermal equilibrium states in any
bounded region $V\subset\Sigma$ which is strictly
smaller than $\Sigma$. All this in spite of naive physical
intuition and the positive results for quantum statistical
mechanics.

It is possible that other techniques, such as local entropy
arguments \cite{Ohya+2004}, can be employed to elucidate
the local aspects of thermal equilibrium for quantum fields,
but we are not aware of a detailed treatment of this issue.
We must therefore conclude that, even though it is still
perfectly satisfactory to use the KMS-condition as the
defining property of global thermal equilibrium, the local
aspects of thermal equilibrium and temperature of a quantum
field are presently not well understood.

% The reason why the choice of $h_O$ is harder in our
% quantum field theoretic case than for the quantum
% mechanics of free bosons can be seen by considering
% one-particle Hilbert spaces. In both cases one deals with
% a Fock space over $\mathcal{K}=L^2(\Sigma)$. For the
% quantum field the Hamiltonian is $C^{\frac12}$, whereas
% it is given by $C$ for the case of particles. However,
% the main issue is that the local con Neumann algebras are
% generated by $e^{i(A^*(\psi)+A(\overline{\psi}))}$
% where $\psi\in L^2(\Sigma)$ satisfies
% $\mathrm{supp}(\psi)\subset O$ in the quantum mechanical
% case, whereas in the quantum field theoretic case we have
% $\psi=q_0(f_0,f_1)$ for some $f_0,f_1\in C_0^{\infty}(O)$.

\subsection{The space of KMS states}\label{Sec_KMSstates}

We now give a full description of the space
$\mathscr{G}^{(\beta)}(\alg{W})$ of all $\beta$-KMS states
in general stationary, globally hyperbolic spacetimes.
(This result may be compared to Theorem \ref{Thm_ClGround}
and \ref{Thm_GroundExUn}.)
\begin{theorem}\label{Thm_KMSExUn}
Let $M$ be a globally hyperbolic, stationary spacetime and
consider a linear scalar field with a stationary potential
$V$ such that $V>0$. Let $\beta>0$.
\begin{enumerate}
\item[\textup{(i)}] There exists a unique extremal $C^1$ $\beta$-KMS
state $\omega^{(\beta)}$ with vanishing one-point function.
We denote its GNS-triple by $(\mathcal{H}_{(\beta)},\pi_{(\beta)},
\Omega_{(\beta)})$ and we let $h$ be the self-adjoint generator
of the unitary group that implements $\alpha_t$ in this
GNS-representation. The one-particle structure of its two-point
function is $(p_{(\beta)},\mathcal{K}_{(\beta)})$ (cf.\ Theorem
\ref{Thm_GroundEx}).
\item[\textup{(ii)}] $\omega^{(\beta)}$ is quasi-free, regular ($D^{\infty}$),
locally quasi-equivalent to $\omega^0$ and $\pi_{(\beta)}$ is
faithful.
\item[\textup{(iii)}] The map $\lambda_{(\beta)}:=\lambda_{\omega^{(\beta)}}$ of
Lemma \ref{Lem_AffineIso} restricts to an affine homeomorphism
$\map{\lambda_{(\beta)}}{\mathscr{G}^0(\alg{W}^{cl})}
{\mathscr{G}^{(\beta)}(\alg{W})}$.
\item[\textup{(iv)}] Any $D^2$ $\beta$-KMS state is Hadamard and any regular
$\beta$-KMS state satisfies the microlocal spectrum condition. A
$\beta$-KMS state $\omega=\lambda_{(\beta)}(\rho)$ is $C^k$,
resp.\ $D^k$, $k=1,2,\ldots$, if and only if $\rho$ is $C^k$,
resp.\ $D^k$.
\item[\textup{(v)}] Any extremal $\beta$-KMS state $\omega$ on $\alg{W}$ is of
the form $\omega=\eta_{\rho}^*\omega^0$ for some gauge
transformation of the second kind $\eta_{\rho}$. It is regular
(resp.\ $C^{\infty}$) if and only if it is $D^1$ (resp.\ $C^1$).
Furthermore, it has the Reeh-Schlieder property.
\item[\textup{(vi)}] $\lim_{\beta\rightarrow\infty}\omega^{(\beta)}=\omega^0$
in the weak$^*$-topology.
\item[\textup{(vii)}] $\pi_{\omega}(\alg{W})\in D(e^{-\frac{\beta}{2}h})$ and
for all $A,B\in\alg{W}$
\[
\langle\pi_{\omega}(A^*)\Omega_{\omega},\pi_{\omega}(B^*)\Omega_{\omega}\rangle=
\langle e^{-\frac{\beta}{2}h}\pi_{\omega}(B)\Omega_{\omega},
e^{-\frac{\beta}{2}h}\pi_{\omega}(A)\Omega_{\omega}\rangle.
\]
\end{enumerate}
\end{theorem}
\begin{proof*}
Let $\omega^{(\beta)}$ be the quasi-free state whose two-point
distribution is associated to the non-degenerate $\beta$-KMS
one-particle structure $(p_{(\beta)},\mathcal{K}_{(\beta)})$ of
Theorem \ref{Thm_GroundEx}. Then $\omega^{(\beta)}$ is a $\beta$-KMS
state, by Theorem \ref{Prop_EqToOnePEq}. As $\omega^{(\beta)}$ is
quasi-free and $\omega^{(\beta)}_2$ is a distribution (density),
$\omega^{(\beta)}$ is a regular state. The representation
$\pi_{(\beta)}$ is faithful, as in the proof of Theorem
\ref{Thm_GroundExUn}.

The map $\lambda_{(\beta)}:=\lambda_{\omega^{(\beta)}}$ of Lemma
\ref{Lem_AffineIso} restricts to the stated affine homeomorphism
by Proposition \ref{Prop_AllKMSStates}. For regular $\beta$-KMS
states the Hadamard property is known to hold \cite{Sahlmann+2000}
and the microlocal spectrum then
follows \cite{Sanders2010,Brunetti+1995}. The Hadamard property
for $D^2$ $\beta$-KMS states then follows from the last statement
of Proposition \ref{Prop_OnePuniqueness}. The fact that
$\lambda_{(\beta)}(\rho)$ is $C^k$ (resp.\ $D^k$) if and only if
$\rho$ is, is shown as in Theorem \ref{Thm_GroundExUn}.

Local quasi-equivalence of all quasi-free Hadamard states was
proved in Ref.\ \cite{Verch1994}, which applies in particular to
$\omega^{(\beta)}$ and $\omega^0$.

Extremal $\beta$-KMS states $\omega$ on $\alg{W}$ are of the form
$\omega=\eta_{\rho}^*\omega^0$, as in Theorem \ref{Thm_GroundExUn},
and the Reeh-Schlieder property for $\omega$ follows from that of
$\omega^{(\beta)}$ \cite{Strohmaier2000}. The statement on the
regularity of extremal $\beta$-KMS states follows directly from
Proposition \ref{Prop_rho0}. This also proves the uniqueness
clause for $\omega^{(\beta)}$.

Using Theorem \ref{Thm_GroundEx} one may show that
$\lim_{\beta\rightarrow\infty}\omega^{(\beta)}_2(f,f)
=\omega^0_2(f,f)$. Indeed, the range of $p_{cl}$ is in the
domain of $|H_e|^{-1}$ by Proposition \ref{Thm_Hinvertable}
and the functions
$F(x):=e^{-\frac{\beta}{2}x}\sqrt{\frac{x}{1-e^{-\beta x}}}$
and $G(x):=\sqrt{\frac{x}{1-e^{-\beta x}}}-\sqrt{x}$
converge uniformly to $0$ on the positive half line as
$\beta\rightarrow\infty$. The explicit expression for
$p_{(\beta)}$ and the Spectral Calculus Theorem for the
functions $F(|H_e|)$ and $G(|H_e|)$ then prove the claim.
It follows that $\lim_{\beta\rightarrow\infty}
\omega^{(\beta)}(W(f))=\omega^0(W(f))$, because the
$\omega^{(\beta)}$ and $\omega^0$ are quasi-free. Hence,
$\lim_{\beta\rightarrow\infty}\omega^{(\beta)}=\omega^0$.

As $\omega^{(\beta)}$ is locally normal w.r.t.\ $\omega^0$,
it extends in a unique way to a locally normal state on
$\alg{R}$, which contains a dense, $C^*$-dynamical system
$\alg{R}_0$ (cf.\ Remark \ref{Rem_vonNeumann}), for which
$\omega$ is again a $\beta$-KMS state (by Proposition
\ref{Prop_KMSEq} and a limit argument). The
GNS-representation $\pi_{\omega}$ of $\omega$ on $\alg{R}$
restricts to the GNS-representations of $\alg{R}_0$ and of
$\alg{W}$, which all generate the same Hilbert space
$\mathcal{H}_{\omega}$. The final item then follows from
Ref.\ \cite{Sakai1991} Theorem 4.3.9.
\end{proof*}
It is known that the state $\omega^{(\beta)}$ is not pure,
but it can be purified by extending it to a so-called doubled
system \cite{Kay1985_2}. This abstract procedure finds a natural
interpretation in the setting of black hole
thermodynamics \cite{Kay1985}. Because $\omega^{(\beta)}$ is
not pure we cannot use Theorem \ref{Thm_KayUniqueness} to
obtain a uniqueness result, unlike the ground state case.

\subsection{Wick rotation in static spacetimes}\label{SSec_Wick}

In Section \ref{SSec_Ground} we have
shown the existence of unique non-degenerate $\beta$-KMS
one-particle structures for a linear scalar quantum field on a
stationary, globally hyperbolic spacetime, provided the
interaction potential is stationary and everywhere
strictly positive. In this section we will show that the
corresponding two-point distributions can also be obtained by
a Wick rotation, in case the spacetime is standard static.
The geometric backbone of the argument was already presented
in subsection \ref{SSec_Complexify}, so in this section we
may focus on the functional analytic aspects of the
technique of Wick rotation. The results we describe
correspond to those in Ref.\ \cite{Fulling+1987}, but our
presentation focusses more on the operator theoretic
language. The case of $R=\infty$, which leads to a ground
state, has already been described in some detail \cite{Wald1979},
so we will focus primarily on the case $R<\infty$.

\subsubsection{The Euclidean Green's function}\label{SSSec_GR}

For some $R>0$ consider the complexification $M^c_R$ and the
associated Riemannian manifold $M_R$ of a standard static
globally hyperbolic spacetime $M$. Because the Laplace-Beltrami
operator $\Box$ on $M$ is defined in terms of the metric and the
potential $V$ is assumed stationary, there is a natural
corresponding Euclidean Klein-Gordon operator on $M_R$, namely
$K_R:=-\Box_{g_R}+V$. Our first task is to find a preferred
Euclidean Green's function, which will be the starting point for
the Wick rotation that should lead to a two-point distribution on
the Lorentzian spacetime $M$.
\begin{definition}
A \emph{Euclidean Green's function} is a distribution
(density) $G_R$ on $M_R^{\times 2}$ which is a fundamental
solution, $(K_R)_xG_R(x,y)=(K_R)_yG_R(x,y)=\delta(x,y)$, of
positive type, $G_R(\overline{f},f)\ge 0$ for all
$f\in C_0^{\infty}(M_R)$.
\end{definition}

Just like there are many (Hadamard) two-point distributions on
$M$, there may be many Green's functions on $M_R$. The common
wisdom is to obtain a preferred one by the following method:
the partial differential operator $K_R$ can be viewed as a
positive, symmetric linear operator on the domain
$C_0^{\infty}(M_R)$ in $L^2(M_R)$. Assuming $\overline{K_R}$ is
self-adjoint and strictly positive, it has a well defined inverse.
We may then take
$G(\overline{f},f'):=\langle f,(\overline{K_R})^{-1}f'\rangle$,
whenever this is a distribution. In an attempt to substantiate
this procedure we will analyze the operator $K_R$ in some more
detail.

For a standard static spacetime $M$ we have $N^i\equiv 0\equiv w$,
so Eq.\ (\ref{Eqn_Ksplit}) simplifies to
\begin{equation}\label{Eqn_DecompL}
N^{\frac32}KN^{\frac12}=\partial_0^2+C_0,
\end{equation}
where $C_0$ is the partial differential operator
\[
C_0:=-N^{\frac12}\nabla^{(h)}_iNh^{ij}\nabla^{(h)}_jN^{\frac12}+VN^2
\]
acting on $C_0^{\infty}(\Sigma)$ in $L^2(\Sigma)$ (cf.\
Proposition \ref{Prop_StStatGround}). Recall from Section
\ref{SSec_FFStat} that the powers $\frac32$ and $\frac12$ of
$N$ to the left and right of $K$ were chosen in such a way
that $C_0$ is symmetric and at the same time the operator
$\partial_0^2$ appears without any spatial dependence. In the
case at hand that completely separates the Killing time
dependence from the spatial dependence.

In a similar manner we may split off the imaginary Killing time
dependence of $K_R$. For this we will view the circle $\mathbb{S}^1_R$
of radius $R$ as a Riemannian manifold in the canonical metric
$d\tau^2$. In analogy to the Lorentzian case
(cf.\ Sec.\ \ref{SSec_FFStat}), there is a unitary isomorphism
\[
\map{U_R}{L^2(M_R)}{L^2(\mathbb{S}^1_R)\otimes L^2(\Sigma)}:
f\mapsto\sqrt{N}f,
\]
onto the Hilbert tensor product, because
$d\mathrm{vol}_{g_R}= Nd\tau\ d\mathrm{vol}_h$. Then,
$N^{\frac32}K_RN^{\frac12}=-\partial_{\tau}^2+C_0$, with the
same operator $C_0$ on $\Sigma$ as in the Lorentzian case.
More precisely, we have
\begin{equation}\label{Eqn_Decomp}
U_RNK_RNU_R^{-1}\supset B_R\otimes I+I\otimes C_0,
\end{equation}
where the operator $B_R:=-\partial_{\tau}^2$ acts on the
dense domain $C_0^{\infty}(\mathbb{S}^1_R)$ in $L^2(\mathbb{S}^1_R)$
and the operator on the right-hand side is defined on the algebraic
tensor product of the domains of $B_R$ and $C_0$.

The properties of the operator $B_R$ are well known and we
quote them without proof:
\begin{proposition}\label{Prop_SAB}
The operator $B_R:=-\partial_{\tau}^2$ is essentially
self-adjoint on $C_0^{\infty}(\mathbb{S}^1_R)$ in $L^2(\mathbb{S}^1_R)$.
If $R$ is finite, there is a countable orthonormal basis of
eigenvectors
$\psi_n(\tau):=\frac{1}{\sqrt{2\pi R}}e^{in\tau/R}$,
$n\in\mathbb{Z}$, with eigenvalues
$\lambda_n:=\frac{n^2}{R^2}$.
\end{proposition}
This follows e.g.\ from Thm.\ II.9 in Ref.\ \cite{Reed+}
by rescaling to $R=1$. Note that for finite $R$ the
operator $B_R$ is positive, but not strictly positive.
From now on we will use $B_R$ to denote the unique
self-adjoint extension found in Proposition \ref{Prop_SAB},
to unburden our notation.

Together with the results for $C$ (Proposition
\ref{Prop_StStatGround}), Proposition \ref{Prop_SAB}
implies
\begin{theorem}\label{Thm_SA_NK_RN}
For any $R>0$ the operator $NK_RN$ is essentially
self-adjoint on $C_0^{\infty}(M_R)$ in $L^2(M_R)$, its
closure is strictly positive with $NK_RN\ge VN^2$ and
the domain of $(\overline{NK_RN})^{-\frac12}$ contains
$C_0^{\infty}(M_R)$.
\end{theorem}
\begin{proof*}
By Theorem $VIII.33$ in Ref.\ \cite{Reed+} the sum
$B_R\otimes I+I\otimes C$ is essentially self-adjoint
on the algebraic tensor product
$\mathcal{D}:=C_0^{\infty}(\mathbb{S}^1_R)\otimes C_0^{\infty}(\Sigma)$,
because both $B_R$ and $C$ are essentially self-adjoint
on the space of test-functions. By Eq.\ (\ref{Eqn_Decomp})
the operator $U_RNK_RNU_R^{-1}$ extends
$B_R\otimes I+I\otimes C$ and $U_R$ is unitary, so $NK_RN$
is already essentially self-adjoint on the smaller domain
$U_R^{-1}\mathcal{D}$. In fact, because
$\mathcal{D}\subset C_0^{\infty}(\mathbb{S}^1_R\otimes\Sigma)$ in
$L^2(\mathbb{S}^1_R\otimes\Sigma,d\tau\ d\mathrm{vol}_h)$ we have
$U_RNK_RNU_R^{-1}=B_R\otimes I+I\otimes C\ge
I\otimes C\ge I\otimes VN^2$ on $\mathcal{D}$. It follows
that $NK_RN\ge VN^2$ on $U_R^{-1}\mathcal{D}$ and hence
on $C_0^{\infty}(M_R)$. The claim on the domain of
$(\overline{NK_RN})^{-\frac12}$ then follows from Lemma
\ref{Lem_SP3} in \ref{App_SP}.
\end{proof*}

In the ultra-static case, where $N$ is constant, Theorem
\ref{Thm_SA_NK_RN} (in combination with Theorem
\ref{Thm_Kernel}) suffices to justify the procedure to
define a Euclidean Green's function by
$G_R(f,f'):=\langle(\overline{K_R})^{-\frac12}\overline{f},
(\overline{K_R})^{-\frac12}f'\rangle$. In the general case,
however, the study of the self-adjoint extensions of the
operator $K_R$ is more complicated.\footnote{Some partial
results are the following: (i) When $V=m^2>0$, $K_R$ is
essentially self-adjoint on $C_0^{\infty}(M_R)$ in
$L^2(M_R)$ if and only if its range is dense, in which
case its closure is strictly positive. For this to be the
case it is sufficient that $N^{-1}$ is bounded.
(ii) If the Riemannian manifold $(M_R,g_R)$ has a
negligible boundary \cite{Gaffney1951}, then $K_R$
is essentially self-adjoint and its closure is strictly
positive. Unfortunately, the boundedness of $N^{-1}$ does
not hold if the spacetime is the exterior region of a
black hole, while the condition in (ii) may only hold for
very special choices of $R$.}
% \begin{proposition}\label{Prop_SAK1}
% When $V=m^2>0$, $K_R$ is essentially self-adjoint on
% $C_0^{\infty}(M_R)$ in $L^2(M_R)$ if and only if its range
% is dense, in which case its closure is strictly positive.
% For this to be the case it is sufficient that $N^{-1}$ is
% bounded.
% \end{proposition}
% \begin{proof*}
% Note that $K_R\ge m^2>0$ on $C_0^{\infty}(M_R)$, so if $K_R$
% has a dense range, then its inverse is bounded, $K_R$ is
% essentially self-adjoint and its closure is strictly positive
% by Lemma \ref{Lem_SP1}. Conversely, if $K_R$ is essentially
% self-adjoint, then it is strictly positive and the dense range
% follows from the same lemma. Now the operator $N^{-1}$ maps
% the range of $NK_RN$ onto that of $K_R$. The range of the
% former is dense by Theorem \ref{Thm_SA_NK_RN} and hence so is
% the range of the latter as soon as $N^{-1}$ is bounded.
% \end{proof*}
%
% \begin{proposition}\label{Prop_SAK2}
% If the Riemannian manifold $(M_R,g_R)$ has a negligible boundary,
% then $K_R$ is essentially self-adjoint and its closure is
% strictly positive.
% \end{proposition}
Nevertheless, we can define a Euclidean Green's function
by a slight modification of the common procedure as
\begin{equation}\label{Eqn_EuclG}
G_R(f,f'):=\langle
(\overline{NK_RN})^{-\frac12}N\overline{f},
(\overline{NK_RN})^{-\frac12}Nf'\rangle,
\end{equation}
using Theorem \ref{Thm_SA_NK_RN} and the fact that
multiplication by $N$ is a continuous linear map on
$C_0^{\infty}(M)$. It is straightforward to verify that
this satisfies all the requirements to be a Euclidean
Green's function and we will see shortly that this choice
of the Euclidean Green's function will indeed allow us to
recover the KMS two-point distributions.

\subsubsection{Analytic continuation of the Euclidean Green's function}\label{SSSec_TimeDep}

We may now establish the explicit Killing time dependence
of the Euclidean Green's function and its analytic
continuation:
\begin{theorem}\label{Thm_ACGreen}
Consider a standard static globally hyperbolic spacetime $M$. For
each $R<\infty$ there is a unique continuous function $G^c_R(z,z')$
from $\mathcal{C}_R^{\times 2}$ into the distribution densities on
$\Sigma^{\times 2}$, holomorphic on the set where
$\mathrm{Im}(z-z')\not=0$, such that for all
$\chi,\chi'\in C_0^{\infty}(\mathbb{S}^1_R)$ and
$f,f'\in C_0^{\infty}(\Sigma)$ we have
\[
\langle U_R^{-1}(\chi\otimes f),G_RU_R^{-1}(\chi'\otimes f')\rangle=
\int_{S_R^{\times 2}}d\tau\ d\tau'\ \overline{\chi}(\tau)\chi'(\tau')
G_R^c(i\tau,i\tau';\overline{f},f')
\]
with $z=t+i\tau$. When $\mathrm{Im}(z-z')\in [-2\pi R,0]$ it is given
by
\[
G^c_R(z,z';\overline{f},f'):=\langle C^{-\frac12}Nf,
\frac{\cos((z-z'+i\pi R)\sqrt{C})}{2\sinh(\pi R\sqrt{C})}Nf'\rangle.
\]
\end{theorem}
\begin{proof*}
It suffices to check that the given formula for $G^c_R$ satisfies all
the desired properties, but let us first sketch a more constructive
argument to see where the formula comes from. When we try to extract
the Killing time dependence of $G_R$, as defined in
Eq.\ (\ref{Eqn_EuclG}), we may make use of the fact that the inverse
of the strictly positive operator $\overline{NK_RN}$ can be found as
a strongly converging integral of the heat kernel,
\begin{equation}\label{Eqn_SInt}
\int_0^{\infty}d\alpha\ e^{-\alpha(\overline{NK_RN})}\psi
=(\overline{NK_RN})^{-1}\psi
\end{equation}
for all $\psi\in D((\overline{NK_RN})^{-1})$.
The importance of the heat kernel (i.e.\ the exponential function)
is that it allows us to separate out the Killing time dependence.
Indeed, for all $\alpha\ge 0$ there holds
$e^{-\alpha(\overline{NK_RN})}=U_R^{-1}e^{-\alpha B_R}\otimes
e^{-\alpha C}U_R$, because of Trotter's product formula
(Ref.\ \cite{Reed+} Thm.\ VIII.31).
% Here we observe that i) the operators $B_R\otimes I$ and
% $I\otimes C$ are essentially self-adjoint (by Thm.\ $VIII.33$
% in Ref.\ \cite{Reed+} combined with Propositions \ref{Prop_SAB} and
% \ref{Prop_StStatGround}), ii) the spectral projections for
% $\overline{I\otimes C}$ and $\overline{B_R\otimes I}$ commute
% (because they are of the form $I\otimes P$, respectively
% $P\otimes I$, where $P$ is a spectral projection for $C$,
% respectively $B$) and hence iii) the bounded functions of these
% operators commute (by the Spectral Calculus Theorem,
% cf.\ Ref.\ \cite{Reed+} Sec.\ VIII.5) and, finally, iv)
% $e^{-\alpha\overline{I\otimes C}}=I\otimes e^{-\alpha C}$
% and similarly for $B_R$.
Now let $\lambda_n$, $n\in\mathbb{Z}$, denote the eigenvalues of
$B_R$ and $P_n$ the corresponding orthogonal projections. Then
we may perform the integral over the heat kernel to find
$U_R(\overline{NK_RN})^{-1}U_R^{-1}P_n
=\overline{P_n\otimes(C+\lambda_n)^{-1}}$. Summing over $n$ we
then expect the formula
\[
U_R(\overline{NK_RN})^{-1}U_R^{-1}=\sum_{n\in\mathbb{Z}}
\frac{R}{2\pi}e^{i\frac{n}{R}(\tau-\tau')}(R^2C+n^2)^{-1}
\]
where we have written $P_n$ as an integral kernel on
$(\mathbb{S}^1_R)^{\times 2}$ and we substituted the values of
$\lambda_n$. The sum over $n$ can be performed
(cf.\ Ref.\ \cite{Gradshteyn+} formula 1.445:2) in the sense of the
Spectral Calculus Theorem, leading to
\[
U_R(\overline{NK_RN})^{-1}U_R^{-1}=
\frac{\cosh((\tau-\tau'+\pi R)\sqrt{C})}{2\pi\sqrt{C}\sinh(\pi R\sqrt{C})}.
\]
The analytic continuation is then obvious.

Let us now verify that the given formula for $G^c_R$ has the
desired properties. First note that for each $z,z'$ with
$\mathrm{Im}(z-z')\in[-2\pi R,0]$ it defines a distribution
density on $\Sigma^{\times 2}$ by Theorem \ref{Thm_Kernel},
because multiplication by $N$ is a continuous linear map from
$C_0^{\infty}(\Sigma)$ to itself, $C_0^{\infty}(\Sigma)$
is in the domain of $C^{-\frac12}$, by Proposition
\ref{Prop_StStatGround}, and
\[
\frac{\cos((\tau-\tau'+\pi R)\sqrt{C})}{\sinh(\pi R\sqrt{C})}
=
(e^{(iz-iz'-2\pi R)\sqrt{C}}+e^{-i(z-z')\sqrt{C}})(I-e^{-2\pi R\sqrt{C}})^{-1}
\]
by the Spectral Calculus Theorem. Moreover, both exponential
terms in the first factor of the last expression are bounded
operators that depend holomorphically on $z,z'$ as long as
$\mathrm{Im}(z-z')\in(-2\pi R,0)$. This proves the continuity
and the holomorphicity claims. As the uniqueness of $G^c_R$
is clear from the Edge of the Wedge Theorem \cite{Berenstein+1991},
it only remains to prove that it restricts to $G_R$.

For any $f,f'\in C_0^{\infty}(\Sigma)$ the function
\[
G^c_R(i\tau,i\tau';\overline{f},f')=\frac12\langle C^{-\frac12}Nf,
(e^{-(\tau-\tau'-2\pi R)\sqrt{C}}+e^{(\tau-\tau')\sqrt{C}})
(I-e^{-2\pi R\sqrt{C}})^{-1}Nf'\rangle
\]
is continuous for $\tau-\tau'\in[-2\pi R,0]$ and holomorphic
in the interior. We may compute the derivatives in the
distributional sense, which leads to
\begin{eqnarray}
-\partial_{\tau}^2G^c_R(i\tau,i\tau';\overline{f},f')&=&
-\partial_{\tau'}^2G^c_R(i\tau,i\tau';\overline{f},f')\nonumber\\
&=&-G^c_R(i\tau,i\tau';N^{-1}CN\overline{f},f')
+\delta(\tau-\tau')\langle Nf,Nf'\rangle\nonumber\\
&=&-G^c_R(i\tau,i\tau';\overline{f},N^{-1}CNf')
+\delta(\tau-\tau')\langle Nf,Nf'\rangle.\nonumber
\end{eqnarray}
Letting $U_RK_RU_R^{-1}=N^{-1}(-\partial_{\tau}^2+C_0)N^{-1}$
act on $G^2_R(i\tau,i\tau';x,x')$ from the left and right we
find
\begin{eqnarray}
-\partial_{\tau}^2G^c_R(i\tau,i\tau';N^{-2}\overline{f},f')
+G^c_R(i\tau,i\tau';N^{-1}CN^{-1}\overline{f},f')&=&\nonumber\\
-\partial_{\tau'}^2G^c_R(i\tau,i\tau';\overline{f},N^{-2}f')
+G^c_R(i\tau,i\tau';\overline{f},N^{-1}CN^{-1}f')
&=&\delta(\tau-\tau')\langle f,f'\rangle,\nonumber
\end{eqnarray}
which shows that the restriction of $G^c_R$ to
$(\mathbb{S}^1_R)^{\times 2}$ is indeed the Euclidean Green's
function.
\end{proof*}

The case $R=\infty$ can be treated using similar methods \cite{Wald1979},
now using Ref.\ \cite{Gradshteyn+} formula 3.472:5.
The result is the distribution density-valued function
\[
G^c_{\infty}(z,z';\overline{f},f'):=\frac12\langle C^{-\frac12}Nf,
e^{-i(z-z')\sqrt{C}}Nf'\rangle.
\]
Alternatively, this expression can be obtained as the limit
\[
G^c_{\infty}(z,z';\overline{f},f')=\lim_{R\rightarrow\infty}
G^c_R(z,z';\overline{f},f')
\]
for fixed $f,f'\in C_0^{\infty}(\Sigma)$, using Lemma
\ref{Lem_Holo}.

\subsubsection{Wick rotation to fundamental solutions and thermal states}\label{SSSec_WickKMS}

Using the analytic continuation $G^c_R$ we now want to complete the Wick
rotation by considering the restriction to real values $z=t$ and $z'=t'$.
Following Ref.\ \cite{Fulling+1987} we show how the thermal two-point distribution
and the advanced, retarded and Feynman fundamental solutions are
obtained.

Both for $t>t'$ and $t<t'$ we can approach the real axis from above,
$\mathrm{Im}(z-z')>-2\pi R$, and from below, $\mathrm{Im}(z-z')<0$. This
prompts us to define the following functions on $\mathbb{R}^{\times 2}$ with
values in the distribution densities on $\Sigma^{\times 2}$:
\begin{eqnarray}\label{Eqn_G*}
\mathscr{E}^+(t,t';f,f')&:=&i\theta(t-t')\left(G^c_R(t,t';f,f')-G^c_R(t-2\pi iR,t';f,f')\right)\nonumber\\
\mathscr{E}^-(t,t';f,f')&:=&-i\theta(t'-t)\left(G^c_R(t,t';f,f')-G^c_R(t-2\pi iR,t';f,f')\right)\nonumber\\
\mathscr{E}^F_R(t,t';f,f')&:=&i\theta(t-t')G^c_R(t,t';f,f')+i\theta(t'-t)G^c_R(t-2\pi iR,t';f,f').\nonumber
\end{eqnarray}
Note that the $\mathscr{E}^{\pm}$ and $\mathscr{E}^F_R$ are given by
\begin{eqnarray}\label{Eqn_Epm}
\mathscr{E}^{\pm}(t,t';\overline{f},f')&=&\pm\theta(\pm(t-t'))\langle C^{-\frac12}Nf,
\sin\left((t-t')\sqrt{C}\right)Nf'\rangle\nonumber\\
\mathscr{E}^F_R(t,t';\overline{f},f')&=&i\langle C^{-\frac12}Nf,
\frac{\cos\left((|t-t'|+i\pi R)\sqrt{C}\right)}
{2\sin\left(\pi R\sqrt{C}\right)}Nf'\rangle.
\end{eqnarray}
They give rise to distribution densities on $M^{\times 2}$ defined by
\begin{eqnarray}\label{Eqn_Epm2}
E^{\pm}(\chi\otimes f,\chi'\otimes f')&:=&\int dt\ dt'\
\chi(t)\chi'(t')\mathscr{E}^{\pm}(t,t';\sqrt{N}f,\sqrt{N}f')\nonumber\\
E^F_R(\chi\otimes f,\chi'\otimes f')&:=&\int dt\ dt'\
\chi(t)\chi'(t')\mathscr{E}^F_R(t,t';\sqrt{N}f,\sqrt{N}f'),
\end{eqnarray}
and using Schwartz Kernels Theorem to extend the distribution to all
test-functions in $C_0^{\infty}(M)$. (Note that the factors $\sqrt{N}$
are required to account for the change in integration measure and they
can equivalently be written in terms of the unitary isomorphism $U$.)

\begin{proposition}\label{Prop_FSolnCheck}
$E^{\pm}$ and $E^F_R$ are left and right fundamental solutions for the
Klein-Gordon operator $K=-\Box+V$ and we have
$\mathscr{E}^{\pm}(t,t';f,f')=\mathscr{E}^{\mp}(t',t;f,f')
=\overline{\mathscr{E}^{\pm}(t,t';\overline{f},\overline{f'})}=\mathscr{E}^{\pm}(t,t';f',f)$.
\end{proposition}
\begin{proof*}
The first sequence of equalities follows directly from Eq.\ (\ref{Eqn_Epm})
and the fact that $L^2(\Sigma)$ carries a natural complex conjugation
which commutes with the operator $C$ and any real-valued function of $C$.
To see that the distribution densities are fundamental solutions we use
Eq.\ (\ref{Eqn_DecompL}) to find
\[
UKU^{-1}(\chi\otimes f)=\chi\otimes N^{-1}CN^{-1}f+\partial_t^2\chi\otimes N^{-2}f
\]
and we use the fact that
\[
\partial_t^2G^c_R(t,t';N^{-2}f,f')+G^c_R(t,t';N^{-1}CN^{-1}f,f')=0.
\]
(The differentiations can be carried out by going into the complex
manifold $M^c$, where $G^c_R$ is holomorphic, and then extending by
continuity to the boundary.) For the case of $E^{\pm}(t,t')$ we then
have, by Eq.\ (\ref{Eqn_Epm}):
\begin{eqnarray}
&&((K\otimes I) E^{\pm})(U^{-1}(\chi\otimes f),U^{-1}(\chi'\otimes f'))\nonumber\\
&=&E^{\pm}(U^{-1}(\chi\otimes N^{-1}CN^{-1}f),U^{-1}(\chi'\otimes f'))\nonumber\\
&&+E^{\pm}(U^{-1}(\partial_t^2\chi\otimes N^{-2}f),U^{-1}(\chi'\otimes f'))\nonumber\\
&=&\pm\int dt\ dt'\ \chi(t)\chi'(t')\theta(\pm(t-t'))\langle \sqrt{C}N^{-1}\overline{f},
\sin\left((t-t')\sqrt{C}\right)Nf'\rangle\nonumber\\
&&+\partial_t^2\chi(t)\chi'(t')\theta(\pm(t-t'))\langle C^{-\frac12}N^{-1}\overline{f},
\sin\left((t-t')\sqrt{C}\right)Nf'\rangle.\nonumber
\end{eqnarray}
We account for the factors $\theta$ by restricting the domain of integration
and then perform partial integrations, after which we are only left with
the boundary terms, which immediately yield the result. By the symmetry
properties of $\mathscr{E}^{\pm}$, $E^{\pm}$ is also a right-fundamental
solution. The proof for $E^F_R$ uses a similar computation.
\end{proof*}

It follows from the support properties of the distribution densities
$E^{\pm}$ that they are the advanced ($-$) and retarded ($+$)
fundamental solutions, so our notation is consistent. As Eq.\ (\ref{Eqn_Epm})
shows, they are independent of $R$, in line with the uniqueness of these
fundamental solutions. $E^F_R$ is the Feynman fundamental solution, as
can be inferred from the fact that the real axis of $t-t'$ is approached
by a rigid rotation from the imaginary time axis in counterclockwise
direction. It does depend on the choice of $R$ and it defines a choice
of two-point distribution as follows:
\begin{proposition}\label{Prop_KMS}
For $0<R<\infty$ the function
$G^c_R(t,t')=-i(\mathscr{E}^F_R-\mathscr{E}^-)(t,t')$ on
$\mathbb{R}^{\times 2}$ has a corresponding distribution density
$\omega^{(\beta)}_2:=-i(E^F_R-E^-)$ where we set $\beta:=2\pi R$.
$\omega^{(\beta)}_2$ is the two-point distribution density of
$\omega^{(\beta)}$ (as defined in Theorem \ref{Thm_KMSExUn}) and
\begin{eqnarray}
&&\omega^{(\beta)}_2(U^{-1}(\chi\otimes f),U^{-1}(\chi'\otimes f'))\nonumber\\
&=&\int dt\ dt'\ \chi(t)\chi'(t')G^c_R(t,t';f,f')\nonumber\\
&=&\int dt\ dt'\ \chi(t)\chi'(t')\langle C^{-\frac12}v\overline{f},
\frac{\cos((t-t'+i\pi R)\sqrt{C})}{2\sinh(\pi R\sqrt{C})}vf'\rangle.\nonumber
\end{eqnarray}
\end{proposition}
\begin{proof*}
The equality $G^c_R(t,t')=-i(\mathscr{E}^F_R-\mathscr{E}^-)(t,t')$
follows directly from the definitions of $G^c_R$, $\mathscr{E}^F_R$ and
$\mathscr{E}^-$, so it remains to check the properties of
$\omega^{(\beta)}_2$. $\omega^{(\beta)}_2$ is a bisolution to the
Klein-Gordon equation because it is $-i$ times a difference of two
fundamental solutions (Proposition \ref{Prop_FSolnCheck}).
Furthermore, comparison with Eq.'s (\ref{Eqn_Epm}, \ref{Eqn_Epm2})
shows that the anti-symmetric part of $\omega^{(\beta)}_2$ is given by
$\frac{i}{2}(E^--E^+)$. Remembering that $\partial_t=Nn^a\nabla_a$ and
that the restriction of a distribution density from $M$ to $\Sigma$
incurs a factor $N^{-1}$ we find that the initial data of
$\omega^{(\beta)}_2$ are given by
\begin{eqnarray}\label{Eqn_KMSInit}
\omega^{(\beta)}_{2,00}(\overline{f}_1,f'_1)&=&\frac{1}{2}\langle C^{-\frac12}
N^{\frac12}f_1,\coth\left(\frac{\beta}{2}C^{\frac12}\right)N^{\frac12}f'_1\rangle,
\nonumber\\
\omega^{(\beta)}_{2,10}(\overline{f},f')&=&\frac{-i}{2}\langle f,f'\rangle
=-\omega^{(\beta)}_{2,01}(\overline{f},f'),\nonumber\\
\omega^{(\beta)}_{2,11}(\overline{f}_0,f'_0)&=&\frac{1}{2}\langle C^{\frac12}
N^{-\frac12}f_0,\coth\left(\frac{\beta}{2}C^{\frac12}\right)N^{-\frac12}f'_0\rangle.
\end{eqnarray}
On the other hand, the non-degenerate $\beta$-KMS one-particle
structure, which is described in Proposition \ref{Prop_StStatGround}
for the standard static case, defines a two-point distribution
whose initial data coincide with those in Eq.\ (\ref{Eqn_KMSInit}),
as one may verify by a short computation. This proves that
$\omega^{(\beta)}_2$, as defined above, is indeed the two-point
distribution of $\omega^{(\beta)}$.
\end{proof*}

Using similar techniques one may treat the case $R=\infty$, which
leads to the two-point distribution $\omega^0_2$ of the ground
state $\omega^0$ of Theorem \ref{Thm_GroundExUn} \cite{Wald1979}.

% Because $\omega^{(\beta)}_2$ can be recovered from a Wick rotation one
% may easily deduce that it must be Hadamard, reproducing a result of
% Ref.\ \cite{Sahlmann+2000}. Indeed, $\omega^{(\beta)}_2$ is defined as the
% boundary value $G^c_R(t-i0^+,t')$ of an analytic function on
% $\mathbb{R}$ with values in the distribution densities on
% $\Sigma^{\times 2}$, so we conclude from Theorem \ref{Thm_WFBoundary}
% in \ref{App_muA} that
% \[
% WF(\omega^{(\beta)}_2)\subset \left\{(t-t',k)|\ k\le 0\right\}\times
% T^*(\Sigma^{\times 2}).
% \]
% After a change of coordinates from $(t-t',t+t')$ to $(t,t')$ this shows
% that for any $(t,x;t',x';k,k')\in WF(\omega^{(\beta)}_2)$ the light-like
% covector $k'$ must be future pointing. (Note that it also must be
% light-like, because $\omega^{\beta}_2$ is a bi-solution to the
% Klein-Gordon equation.) This proves that the two-point distribution
% (density) $\omega^{\beta}_2$ is Hadamard.

\section*{Acknowledgments}

Parts of this paper are based on a presentation given at the mathematical
physics seminar at the II.\ Institute for Theoretical Physics of the
University of Hamburg, during a visit in 2011. I thank the Institute for
its hospitality and the attendants of the seminar for their comments and
encouragement. I would also like to thank Kartik Prabhu for proof reading
much of this paper.

\appendix

\section{Some useful results from functional analysis}\label{App_SP}

\renewcommand{\thetheorem}{\Alph{section}.\arabic{theorem}}
\renewcommand{\thelemma}{\Alph{section}.\arabic{lemma}}
\renewcommand{\thedefinition}{\Alph{section}.\arabic{definition}}
\renewcommand{\theproposition}{\Alph{section}.\arabic{proposition}}

In this appendix we collect some results from functional analysis,
to make our review self-contained. Most of the proofs are omitted,
because they are elementary or make use of standard methods. For
more information we refer the reader to Refs.\ \cite{Kadison+},
\cite{Reed+} and to Ref.\ \cite{Kay1985} for strictly positive operators.
In particular these references contain a detailed formulation of
the Spectral Calculus Theorem (Ref.\ \cite{Kadison+} Sec.\ 5.6, or
Ref.\ \cite{Reed+} Thm.\ VIII.6).

If $\map{X}{\mathcal{H}_1}{\mathcal{H}_2}$ is a linear operator
between two Hilbert spaces $\mathcal{H}_i$, we denote the domain of
$X$ by $D(X)$. We wish to record the following useful relation
between operators on a Hilbert space and distributions.
\begin{theorem}\label{Thm_Kernel}
Let $\map{X}{\mathcal{H}_1}{\mathcal{H}_2}$ be a closed, densely
defined linear operator between two Hilbert spaces $\mathcal{H}_i$
and let $\map{L}{C_0^{\infty}(M)}{\mathcal{H}_1}$ be an
$\mathcal{H}_1$-valued distribution density. If the range of $L$
is contained in $D(X)$, then $f\mapsto XL(f)$ is an
$\mathcal{H}_2$-valued distribution density.
\end{theorem}
\begin{proof*}
If $X$ is a bounded operator this is immediately clear from
$\|XL(f)\|\le \|X\|\cdot\|L(f)\|$. If $X$ is a self-adjoint
operator on $\mathcal{H}_1=\mathcal{H}_2$ we may use its
spectral projections $P_{(-n,n)}$ onto the intervals $(-n,n)$
to define bounded operators $X_n:=P_{(-n,n)}X$ for
$n\in\mathcal{N}$. Each $X_nL$ defines a distribution density
and $\lim_{n\rightarrow\infty}X_nL(f)=XL(f)$ for all
$f\in C_0^{\infty}(M)$, because $L(f)\in D(X)$. From the
Uniform Bounded Principle (Ref.\ \cite{Reed+} Thm.\ III.9) we
see that $XL$ also
defines an $\mathcal{H}_2$-valued distribution density. The
general case now follows from the polar decomposition, Theorem
6.1.11 of Ref.\ \cite{Kadison+}, which allows us to write
$T=V(T^*T)^{\frac12}$, where $V$ is bounded and
$(T^*T)^{\frac12}$ is a self-adjoint operator on $\mathcal{H}_1$
with the same domain as $T$.
\end{proof*}

We now turn to injective (and therefore invertible) operators
on a Hilbert space, starting with the following four general
Lemmas:
\begin{lemma}\label{Lem_Cl1}
A densely defined, closable and injective operator $X$ in a
Hilbert space $\mathcal{H}$ has an injective closure
$\overline{X}$ if and only if $X^{-1}$ is closable.
\end{lemma}
% \begin{proof*}
% $\overline{A}$ is not injective precisely when there is a
% sequence of vectors $\psi_n\in D(A)$ such that
% $\lim_{n\rightarrow\infty}\psi_n=\psi\not=0$ and
% $\lim_{n\rightarrow\infty}A\psi_n=0$. This means that there
% are vectors $\phi_n:=A\psi_n\in D(A^{-1})$ such that
% $\lim_{n\rightarrow\infty}\phi_n=0$, but
% $\lim_{n\rightarrow\infty}A^{-1}\phi_n
% =\lim_{n\rightarrow\infty}\psi_n=\psi\not=0$, which means
% exactly that $A^{-1}$ is not closable.
% \end{proof*}
\begin{lemma}\label{Lem_Cl2}
If $X$ is a densely defined, injective operator with dense
range, then $X^*$ and $(X^{-1})^*$ are injective and
$(X^*)^{-1}=(X^{-1})^*$.
\end{lemma}
% \begin{proof*}
% Injectivity of $A^*$ follows from the dense range of $A$
% and injectivity of $(A^{-1})^*$ from the dense domain of $A$.
% Now the following statements are equivalent:
% $\phi=(A^*)^{-1}\psi$; $\psi=A^*\phi$;
% $\langle\psi,A^{-1}\eta\rangle=\langle\phi,\eta\rangle$ for
% all $\eta\in D(A^{-1})$; $\phi=(A^{-1})^*\psi$.
% \end{proof*}
\begin{lemma}\label{Lem_SP1}
A self-adjoint operator $X$ is invertible if and only if it
has a dense range on any core.
\end{lemma}
% \begin{proof*}
% This is a consequence of the following chain of equivalent
% statements for any vector $\psi\not=0$ and any core
% $\mathcal{D}$ for $X$: $X\psi=0$; $X^*\psi=0$;
% $\langle\psi,X\phi\rangle=0$ for all $\phi\in\mathcal{D}$;
% $X\mathcal{D}$ is not dense.
% \end{proof*}
\begin{lemma}\label{Lem_SP2}
If $X$ is self-adjoint and invertible, then $X^{-1}$ is
self-adjoint and invertible, where the domain of $X^{-1}$ is
the range of $X$. If $\mathcal{D}$ is a core for $X$, then
$X\mathcal{D}$ is a core for $X^{-1}$.
\end{lemma}
% \begin{proof*}
% We can define $X^{-1}$ as a linear operator on the range $R$
% of $X$, which is dense by Lemma \ref{Lem_SP1}. If $\chi$
% satisfies $\langle\chi,(X^{-1}\pm iI)\phi\rangle=0$ for all
% $\phi\in R$ and some choice of the sign, then
% $\langle\chi,(I\pm iX)\psi\rangle=0$ for all $\psi$ in the
% domain of $X$. Because $X$ is self-adjoint, this implies
% $\chi=0$ and therefore $X^{-1}$ is also self-adjoint. If
% $X^{-1}\phi=0$ for some $\phi=X\psi\in R$, then
% $\psi=X^{-1}\phi=0$ and hence $\phi=0$, so $X^{-1}$ is
% invertible. Finally, if $\mathcal{D}$ is a core for $X$,
% let $R':=X\mathcal{D}\subset R$. For any $\phi$ in the
% domain of $X^{-1}$ we have $\phi=X\psi$ and there exists a
% sequence $\psi_n\in\mathcal{D}$ such that
% $\lim_{n\rightarrow\infty}\psi_n=\psi$ and
% $\lim_{n\rightarrow\infty}X\psi_n=\phi$. Setting
% $\phi_n:=X\psi_n$ shows that
% $\lim_{n\rightarrow\infty}\phi_n=\phi$ and
% $\lim_{n\rightarrow\infty}X^{-1}\phi_n=X^{-1}\phi$, so $R'$
% is a core for $X^{-1}$.
% \end{proof*}
These Lemmas can be proved using entirely elementary methods.

As positive invertible operators are particularly useful we
make the following definition.
\begin{definition}
A densely defined operator $X$ in a Hilbert space
$\mathcal{H}$ is called \emph{strictly positive} if
and only if $X$ is self-adjoint and for any
$0\not=\phi\in D(X)$: $\langle\phi,X\phi\rangle>0$.
\end{definition}

Several equivalent characterizations can be given as follows:
\begin{lemma}\label{Lem_SPChar}
For a positive, self-adjoint operator $X$ the following
are equivalent: \textup{(i)} $X$ is strictly positive, \textup{(ii)}
$X$ is injective, \textup{(iii)} $X$ has a dense range on any core,
\textup{(iv)} $X^{-1}$ is strictly positive.
\end{lemma}
\begin{proof*}
(i) is equivalent to (iv) by Lemma \ref{Lem_SP2}, because
$\langle\phi,X^{-1}\phi\rangle=\langle X\psi,\psi\rangle$
when $\phi:=X\psi$. The implication (i)$\Rightarrow$(ii)
is immediate and (ii) is equivalent to (iii) by Lemma
\ref{Lem_SP1}. To see that (ii) implies (i) one uses the
Spectral Calculus Theorem and the fact that
$\langle\phi,X\phi\rangle=0$ implies $X^{\frac12}\phi=0$
and $X\phi=0$. If $X$ is injective, this means that
$\phi=0$.
\end{proof*}

The following estimate is often useful to find
strictly positive operators, in particular in
combination with Lemma \ref{Lem_Friedrichs} below.
\begin{lemma}\label{Lem_SP3}
Let $X$ and $Y$ be positive self-adjoint operators
with $X$ strictly positive and assume that $Y\ge X$
on a core for $Y^{\frac12}$. Then $Y$ is strictly
positive, $D(Y^{-\frac12})\supset D(X^{-\frac12})$
and $Y^{-1}\le X^{-1}$ on $D(X^{-\frac12})$.
\end{lemma}
\begin{proof*}
Let $\mathcal{D}$ denote the core for $Y^{\frac12}$
on which the estimate holds. The estimate
$\|X^{\frac12}\psi\|\le\|Y^{\frac12}\psi\|$ for
$\psi\in\mathcal{D}$ can be extended to the entire
domain $D(Y^{\frac12})$. Because $X$ is strictly
positive the same must be true for $Y$ by Lemma
\ref{Lem_SPChar}. By Lemma \ref{Lem_SP2},
$\|X^{\frac12}Y^{-\frac12}\psi\|\le\|\psi\|$ on
$D(Y^{-\frac12})$. Note in particular that the range
of $Y^{-\frac12}$ is contained in $D(X^{\frac12})$.
As $X^{\frac12}Y^{-\frac12}$ is bounded on
$D(Y^{-\frac12})$ we also find that the range of
$X^{\frac12}$, which is $D(X^{-\frac12})$, is
contained in the domain of
$(Y^{-\frac12})^*=Y^{-\frac12}$. It now follows that $(X^{\frac12}Y^{-\frac12})^*=Y^{-\frac12}X^{\frac12}$
on $D(X^{\frac12})$. As
$\|X^{\frac12}Y^{-\frac12}\|\le 1$ we must also have
$\|Y^{-\frac12}X^{\frac12}\|\le 1$, which implies that $\|Y^{-\frac12}\psi\|\le\|X^{-\frac12}\psi\|$ on
$D(X^{-\frac12})$ and the conclusion follows.
\end{proof*}

\begin{lemma}\label{Lem_Friedrichs}
Let $X\ge 0$ be a densely defined, positive operator.
Then the Friedrichs extension $\hat{X}$ is positive
and $D(X)$ is a core for $\hat{X}^{\frac12}$.
\end{lemma}

The following lemma concerns the heat kernel:
\begin{lemma}\label{Lem_Holo}
Let $X$ be a positive self-adjoint operator on
$\mathcal{H}$ and let
$\mathbb{C}_+:=\left\{z\in\mathbb{C}|\ \mathrm{Re}(z)> 0\right\}$
be the right half space. Then the function
$z\mapsto e^{-zX}$ is holomorphic on $\mathbb{C}_+$
with values in the bounded operators on $\mathcal{H}$
and for each $\psi\in\mathcal{H}$ the function
$e^{-zX}\psi$ is continuous on
$\overline{\mathbb{C}_+}$.
\end{lemma}
To close this appendix we provide some facts concerning
multiplication operators on the $L^2$ space of a
semi-Riemannian manifold:
\begin{proposition}\label{Prop_mul}
Let $(M,g)$ be an orientable semi-Riemannian manifold,
let $w\in C^{\infty}(M)$ and let $W$ be the corresponding
multiplication operator in $L^2(M,d\mathrm{vol}_g)$,
defined on $C_0^{\infty}(M)$ by $(Wf)(x)=w(x)f(x)$. If
$|w|$ is bounded, then $W$ is bounded. If $w$ is
real-valued, then $W$ is essentially self-adjoint.
$\overline{W}$ is (strictly) positive if and only if $w$
is (strictly) positive (almost everywhere).
\end{proposition}

\end{document}